\documentclass[onecolumn]{IEEEtran}

\usepackage{pdfsync,srcltx}
\usepackage{amsmath}
\usepackage{amsfonts}
\usepackage{amssymb}
\usepackage{graphicx,subfigure,psfrag}
\usepackage{url, cite}
\usepackage{pmat}

\usepackage[ieee]{jphmacros2e}

\usepackage[norefs,nocites]{refcheck}%options:norefs,nocites,msgs,chkunlbld
\makeatletter\let\NAT@parse\undefined\makeatother % enbl natbib with IEEE cls
\usepackage[numbers,sort&compress,sectionbib]{natbib}%\cite,\citet,\citep,\ldots
\newcommand{\R}{\mathbb{R}}

\newcommand{\N}{\mathbb{N}}
\newcommand{\Z}{\mathbf{Z}}

\newcommand{\G}{\mathbf{G}}
\newcommand{\V}{\mathbf{V}}
\renewcommand{\E}{\mathbf{E}}
\newcommand{\scr}{\mathcal}
\newcommand{\comment}[1]{}

\usepackage{nomencl}
\makenomenclature

\allowdisplaybreaks

\renewcommand{\baselinestretch}{1.5}

\makeatletter

\newcommand{\Rmnum}[1]{\expandafter\@slowromancap\romannumeral #1@}
\makeatother

\title{Stability Margin Scaling Laws for Distributed Formation Control as a Function of Network Structure}
\author{He Hao, Prabir Barooah, Prashant G. Mehta\thanks{He Hao and Prabir Barooah are with the  Department of Mechanical and Aerospace Engineering, University of Florida, Gainesville, FL 32611. Email: \{hehao,pbarooah\}@ufl.edu. This work was supported by the National Science Foundation through Grant CNS-0931885 and ECCS-0925534, and by the Institute for Collaborative Biotechnologies through grant DAAD19-03-D-0004. Prashant G. Mehta is with the Coordinated Science Laboratory, Department of Mechanical Science and Engineering, University of Illinois, Urbana-Champaign, IL 61801. Email: mehtapg@uiuc.edu}
}

\begin{document}
\maketitle
%%%%%%%%%%%%%%%%%%%%%%%%%%%%%%%%%%%%%%%%%%%%%%%%%%%%%%%%%%%%%%%%%%%%%%
\begin{abstract}
We consider the problem of distributed formation control of a large number of vehicles.  
An individual vehicle in the formation is assumed to be a fully actuated point mass. 
A distributed control law is examined: the control action on an individual vehicle depends on (i) its own velocity and (ii) the relative position measurements with a small subset of vehicles (neighbors) in the formation.  The neighbors are defined according to an {\em information graph}. 
\smallskip

In this paper we describe a methodology for modeling, analysis, and distributed control design of such vehicular formations whose information graph is a  $D$-dimensional lattice.  The modeling relies on an approximation based on a partial differential equation (PDE) that describes the spatio-temporal evolution of position errors in the formation. The analysis and control design is based on the PDE model.  We deduce asymptotic formulae for the closed-loop stability margin (absolute value of the real part of the least stable eigenvalue) of the controlled formation.  The stability margin is shown to approach $0$ as the number of vehicles $N\rightarrow\infty$.  The exponent on the scaling law for the stability margin is influenced by the dimension and the structure of the information graph.  We show that the scaling law can be improved by employing a higher dimensional information graph.

\smallskip

Apart from analysis, the PDE model is used for a mistuning-based design of control gains to maximize the stability margin.  Mistuning here refers to small perturbation of control gains from their nominal symmetric values.  We show that the mistuned design can have a significantly better stability margin even with a small amount of perturbation. The results of the analysis with the PDE model are corroborated with numerical computation of eigenvalues with the state-space model of the formation.
\end{abstract}

%%%%%%%%%%%%%%%%%%%%%%%%%%%%%%%%%%%%%%%%%%%%%%%%%%%%%%%%%%%%%%%%%%%%%%
%\printnomenclature

%%%%%%%%%%%%%%%%%%%%%%%%%%%%%%%%%%%%%%%%%%%%%%%%%%%%%%%%%%%%%%%%%%%%%%
\section{Introduction}\label{sec:intro}

We consider the problem of controlling a group of vehicles so that they maintain a desired formation geometry while following a desired trajectory. The desired formation geometry is specified in terms of desired relative positions between pairs of vehicles.  The desired trajectory of the formation is supplied to a subset of the vehicles, which are called reference vehicles. The problem is relevant to a number of applications such as formation control of aerial, ground, and autonomous vehicles for transportation, surveillance, reconnaissance and mine-sweeping~\cite{HT_RR_WZ_ACC:98,minesweeping_MSthesis:00,DJ_WB_MP_EW_AIAA:02,HT_DC_RAS:07}.  In many of these applications, a centralized control solution that requires all-to-all or all-to-one communication is impractical.  This motivates distributed control architectures where an individual vehicle exchanges information only with a small set of other vehicles to make control decisions.

\smallskip

Each vehicle is modeled as a fully actuated point mass. This means that (i) the dynamics of each coordinate of the vehicle's position are modeled using a double integrator, (ii) the coordinate dynamics are decoupled, and (iii) an independent force actuation is assumed for each coordinate.  A distributed control law is examined: the control action on an individual vehicle depends on (i) its own velocity and (ii) the relative position measurements with a small subset of vehicles (neighbors) in the formation.  The neighbor relationship is defined according to an {\em information graph}, which has been recognized to play an important role in closed-loop stability of the formation~\cite{FaxMurray_InfoFlowCoopCntrl_VehForm_TAC:04,BarooahHespanhaDec06a}. A node in the graph represents a vehicle, and an edge connecting two nodes represents direct information exchange between those nodes. This information exchange may occur due to one vehicle measuring the other vehicle's relative position by on-board sensors, or due to one vehicle communicating its state information to the other via a communication channel. The information graph is undirected, meaning that if vehicle $i$ can get information about vehicle $j$, then $j$ can get information about $i$.

\smallskip

The objective of this paper is to study how the stability margin (the absolute value of the real part of the least stable eigenvalue) of the closed-loop scales with the number of vehicles, structure of the $D$-dimensional information graph, and the choice of the control gains. For a specific case, when $D=1$, the stability margin of the platoons and its dependence of asymmetry in control gains was examined in our own earlier work~\cite{PB_PM_JH_TAC:09}. The extension to 2D formations appears in~\cite{HH_PB_PM_DSCC:09}. This paper is an extension of these previous works.

\medskip

In this paper, we restrict ourselves to information graphs that belong to the class of $D$-dimensional (finite) lattices.  A formal definition of lattice appears in Section~\ref{sec:results}; see Figure~\ref{fig:lattices} for a few examples. Lattices arise naturally as information graphs when the vehicles in the group are arranged in a regular pattern in space and the exchange of information occurs between pairs of vehicles that are physically close. However, lattices also allow for a flexibility to model much more general information exchange architectures. They are often used as information graph in concensus and vehicular formation problems~\cite{bamjovmitCDC08}. In this paper, we make an important distinction between the dimension of the position vector of a vehicle and the dimension of the lattice that defines the information graph. For example, a one-dimensional platoon may have a two-dimensional lattice as its information graph (see Figure~\ref{fig:2Dinfo_1D2Dspatial}).

\subsection{Related literature}

An analysis of the stability margin is important to understand the scalability of control solutions as the number of vehicles in the formation, $N$, increases.  In the formation control literature, the scalability question has been investigated primarily for a one-dimensional vehicle formation, which is usually referred to as a \emph{platoon}. An extensive literature exists on the platoon control  problem; see~\cite{SMM_BCK_subTAC:71,SS_CD_TAC:92,JH_MT_PV_CSM:94,Zhang_IntelligentCruise_TVT:99,PC_TAC:07} and references therein. The most widely studied  information exchange structures for distributed control of platoons are \emph{predecessor following control} and  \emph{bidirectional control}. In predecessor following control, every vehicle uses information from the vehicle immediately ahead. In bidirectional control, each vehicle uses information from the vehicle immediately ahead of it and the one behind it.  Scenarios in which information exchange occurs with vehicles beyond those physically closest, are studied in~\cite{SKY_SD_KRR_TAC:06,RM_JB_TAC:09}.  The focus of much of the research in this area has been on the  so-called \emph{symmetric control}, in which every vehicle uses the same control law.  Such a simplifying assumption is motivated in part by a lack of tools for analysis and design of distributed control laws. The symmetry assumption is used to simplify the design and analysis. References that studied non-symmetric control design include~\cite{KhatirDavison_NonIdenticalK_ACC:04,RM_JB_TAC:09}.

\smallskip

For platoons, the distributed control architectures with symmetric control are known to scale poorly, both in terms of closed-loop stability margin and sensitivity to external disturbances.  In a symmetric bidirectional architecture, the least stable closed-loop eigenvalue approaches zero as $N$ increases~\cite{PB_PM_JH_TAC:09}.  This progressive loss of stability margin causes the closed-loop performance to become arbitrarily sluggish as the number of vehicles, $N$, increases.  Small stability margin can also lead to long transients due to initial conditions, which can result in control saturation~\cite{jovfowbamdanACC04}.  It is worthwhile to point out that the stability margin for a platoon is known to scale poorly as a function of $N$ even with the centralized LQR control~\cite{MJ_BB_TAC:05}. In addition to the loss of stability margin, the sensitivity of the closed-loop platoon to external disturbances increases without bound as a function of $N$. This effect is also referred to as \emph{string instability}~\cite{LP_TAC:74,SwaroopHedrick_stringstability_TAC:96} or \emph{slinky-type effect}~\cite{Zhang_IntelligentCruise_TVT:99}. String instability is observed for both symmetric predecessor following and symmetric bidirectional control~\cite{Seiler_disturb_vehicle_TAC:04,PB_JH_CDC:05}. Non-symmetric control design within the bidirectional architecture was proposed in~\cite{PB_PM_JH_TAC:09} that helps improve the closed-loop stability margin. A non-symmetric control design within the framework of predecessor architecture was proposed in~\cite{KhatirDavison_NonIdenticalK_ACC:04}, which ameliorates string instability at the expense of control gains that increase without bound as $N$ increases.

\smallskip

%PGM Dec 10: this paragraph may NOT be strictly necessary to include at this time.
Control of platoons with inter-vehicle communication that allows for information exchange with vehicles that are not just nearest neighbors was considered in~\cite{SKY_SD_KRR_TAC:06,RM_JB_TAC:09}.  It was concluded in~\cite{SKY_SD_KRR_TAC:06} that to eliminate string instability with symmetric control, the number of vehicles that each vehicle communicates with has to grow without bound as $N$ increases. It was shown in~\cite{RM_JB_TAC:09} that heterogeneity in control gains does not significantly alter string instability if certain constraints are imposed on integral absolute error and high frequency response of the loop transfer function.

\smallskip

Bamieh~\textit{et. al.} studied controlled vehicle formations with a $D$-dimensional torus as the information graph~\cite{bamjovmitCDC08}.  Scaling laws with symmetric control are obtained for certain performance measures that quantify the sensitivity of the closed-loop to stochastic disturbance.  It is shown in~\cite{bamjovmitCDC08} that the scaling of these performance measures with $N$ is strongly dependent on the dimension $D$ of the information graph.  In~\cite{AP_PS_KH_TAC:02}, Pant \emph{et. al.} introduced the notion of mesh-stability for two-dimensional formations with a ``look-ahead'' information exchange structure, which refers to a particular kind of directed information flow. The scenario considered in our paper, with undirected information graphs, does not fall under the look-ahead information exchange structure. 

\subsection{Contributions of this paper}

In this paper we describe a methodology for modeling, analysis, and distributed control design of vehicular formations whose information graph belongs to the class of $D$-dimensional lattices.  The approach is to use a partial differential equation (PDE) based continuous approximation of the (spatially) discrete platoon dynamics. Just as a PDE can be discretized using a finite difference approximation, we can carry out the procedure in reverse: the spatial difference terms in the discrete model are approximated by spatial derivatives. The resulting PDE yields the original set of ordinary differential equations upon discretization. This approach is motivated by earlier work on PDE modeling of one-dimensional platoons~\cite{PB_PM_JH_TAC:09}.  The PDE model is used for analysis of stability margin and for mistuning-based design of distributed control laws.

\smallskip

There are two contributions of this work that are summarized below.

\smallskip

First, we obtain scaling laws of the stability margin of the closed-loop formation with symmetric control.
We show that the stability margin scales as $O(\frac{1}{n_1^{2}})$ where $n_1$ is the number of vehicles along a certain axis of the information graph. By choosing the structure of the information graph in such a way that $n_1$ increases slowly in relation to $N$, the reduction of the stability margin as a function of $N$ can be slowed down. In fact, by holding $n_1$ to be a constant independent of the number of vehicles $N$, the stability margin can be bounded away from zero even as the number of vehicles increase without bound.  It turns out, however, that keeping $n_1$ fixed while $N$ increases causes the number of reference vehicles to increase.  When the information graph is a square $D$-dimensional lattice (equal number of nodes on each side of the lattice), the stability margin scales as $O(\frac{1}{N^{2/D}})$  in the limit of large  $N$.  This formula is a generalization of the estimate given in~\cite{PB_PM_JH_TAC:09} for a one-dimensional formation.

\smallskip

The second contribution of this work is a procedure to design \emph{asymmetric} control gains so that the stability margin scaling law is significantly improved over that with symmetric control. For the case of square information graphs, we show that an arbitrarily small asymmetry in the proportional control gains from their nominal symmetric values results in stability margin scaling as $O(\frac{1}{N^{1/D}})$. In contrast to the $O(\frac{1}{N^{2/D}})$ scaling seen in the symmetric case, this is an order of magnitude improvement. The resulting control design is called a \emph{mistuning}-based design since the control gains are changed only slightly, i.e., mistuned, from their values in the nominal, symmetric case.  Mistuning-based approaches have been used for stability augmentation in several applications~\cite{Shapiro:1998,Bendiksen:2000,RGM:2003,PM_GH_AB:07}, and recently for distributed control of one-dimensional platoons~\cite{PB_PM_JH_TAC:09}.

\smallskip

% The third contribution of the work is the approach used in deriving the results mentioned above. We derive a partial differential equation (PDE) based
% continuous approximation of the (spatially) discrete formation
% dynamics. Just as a PDE can be discretized using a finite difference
% approximation, we carry out a reverse procedure: spatial difference
% terms in the discrete model are approximated by spatial derivatives.
% The resulting PDE yields the original set of ordinary differential
% equations upon discretization.
The advantage of using a PDE-based analysis is that the PDE reveals, better than the state-space model does, the mechanism of
loss of stability and suggests the mistuning-based approach to
ameliorate it. Numerical computations of eigenvalues of the state-space model of the formation is used to confirm the scaling laws with symmetric as well as mistuned control. Although the PDE model approximates the (spatially) discrete formation dynamics in the limit $N\to \infty$, numerical calculations show that the conclusions drawn from the PDE-based analysis holds even for small number of vehicles. %Thus, the results of this paper are useful in the  analysis and design of control architectures for realistic vehicle formations whose sizes may not be too large.

\smallskip

The remainder of this paper is organized as follows. Section~\ref{sec:results} presents the problem statement and the main results of this paper. Section~\ref{sec:problem} describes the state-space and PDE models of the formation control problem. Analysis and control design results together with their numerical verification appear in Sections~\ref{sec:instability-analysis} and~\ref{sec:mistuning}, respectively. In Section~\ref{sec:remarks}, we present time-domain simulations to illustrate these results, and comment on various aspects of the proposed design and analysis methodology.

%%%%%%%%%%%%%%%%%%%%%%%%%%%%%%%%%%%%%%%%%%%%%%%%%%%%%%%%%%%%%%%%%%%%%%
\section{Problem statement and main results}\label{sec:results}
%%%%%%%%%%%%%%%%%%%%%%%%%%%%%%%%%%%%%%%%%%%%%%%%%%%%%%%%%%%%%%%%%%%%%%
\subsection{Problem statement}\label{sec:statement}
We consider the formation control of $N$ identical vehicles. The position of each vehicle is a $D_s$-dimensional vector (with $D_s=1,2$ or $3$); $D_s$ is referred to as the \emph{spatial dimension} of the formation.  Let $p_i^{(\mrm{d})} \in \R$ be the $\mrm{d}$-th coordinate of the $i$-th vehicle's position, whose dynamics are modeled by a double integrator:
\begin{align}\label{eq:vehicle-dynamics}
  \ddot{p}_i^{(\mrm{d})} = u_i^{(\mrm{d})},\ \ \ \ \mrm{d}=1,\dots,D_s,
\end{align}
where $u_i^{(\mrm{d})} \in \R$ is the control input (acceleration or deceleration command). The underlying assumption is that each of the $D_s$ coordinates of a vehicle's position can be independently actuated.
We say that the vehicles are \emph{fully actuated}.
The spatial dimension $D_s$ is $1$ for a platoon of vehicles moving in a straight line, $D_s=2$ for a formation of  ground vehicles and $D_s=3$ for a formation of  aerial vehicles flying in the three dimensional space. 

The control objective is to make the group of vehicles track a pre-specified desired trajectory while maintaining a desired formation geometry. The desired formation geometry is specified by a desired relative position vector $\Delta_{i,j} \eqdef p_i^{*}(t) - p_j^{*}(t)$ for \emph{every} pair of vehicles $(i,j)$, where $p_i^*(t)$ is the desired trajectory of the vehicle $i$. The desired inter-vehicular spacings have to be specified in a mutually consistent fashion, i.e. $\Delta_{i,j} =\Delta_{i,k}+\Delta_{k,j}$ for every triple $i,j,k$.  Desired trajectory of the formation is specified in the form of a few fictitious ``reference vehicles'', each of which perfectly tracks its own desired trajectory. The reference vehicles are generalization of the fictitious leader and follower vehicles in one-dimensional platoons~\cite{PB_PM_JH_TAC:09, MJ_BB_TAC:05,SMM_BCK_subTAC:71}. A subset of vehicles can measure their relative positions with respect to the reference vehicles, and these measurements are used in computing their control actions. In this way, desired trajectory information of the formation is specified only to a subset of the vehicles in the group. In this paper we consider the desired trajectory of the formation to be of a constant-velocity type, so that $\Delta_{i,j}$'s don't change with time.

Next we define an {\em information graph} that  makes it convenient to describe distributed control architectures.

\begin{definition} An \emph{information graph} is an undirected graph $\G = (\V,\E)$, where the set of \emph{nodes} $\V=\{1,2,\dots, N, N+1, \dots, N + N_r\}$ consists of $N$ real vehicles and $N_r$ reference vehicles.  The set of edges $\E \subset \V \times \V$ specify which pairs of nodes (vehicles) are allowed to exchange information to compute their local control actions.  Two nodes $i$ and $j$ are called \emph{neighbors} if $(i,j) \in \E$, and the set of neighbors of $i$ are denoted by $\scr{N}_i$. \frqed
\end{definition}

%PGM Dec 10 check?
Note that information exchange may or may not involve an explicit communication network. For example, if vehicle $i$  measures the relative position of vehicle $j$ with respect to itself by using a radar and uses that information to compute its control action, we consider it as ``information exchange'' between $i$ and $j$. If a vehicle $i$ has access to desired trajectory information then there is an edge between $i$ and a reference vehicle.

\nomenclature{$N$}{Number of vehicles in the formation}%

In this paper we consider the following \emph{distributed} control law, whereby the control action at a vehicle depends on i) its own velocity and ii) the {\em relative position measurements} with its neighbors in the information graph:
\begin{align}\label{eq:control-lawd}
	u_{i}^{(\mrm{d})} & = \sum_{j \in \scr{N}_i}-k_{(i,j)}^{(\mrm{d})}(p_{i}^{(\mrm{d})}-p_{j}^{(\mrm{d})}-\Delta_{i,j}^{(\mrm{d})}) - b_i^{(\mrm{d})}(\dot{p}_{i}^{(\mrm{d})}- v^{*(\mrm{d})}), \quad i=1,\dots,N,
\end{align}
where $v^{*(\mrm{d})}$ is the $d$-th component of the desired velocity of the formation, $k_{(\cdot)}^{(\mrm{d})}$ is the proportional gain and $b_{(\cdot)}^{(\mrm{d})}$ is the derivative gain.  Note that all the variables in~\eqref{eq:control-lawd} are scalars. It is assumed that  vehicle $i$ knows its own neighbors (the set $\scr{N}_i$), desired spacing $\Delta_{i,j}^{(\mrm{d})}$, and the desired velocity $v^{*(\mrm{d})}$.

%%%%%%%%%%%%%%%%%%%%%%%%%%%%%%%%%%%%%%%%%%%%%%%%%%%%%%%%%%%%%%%%%%%%%
\begin{figure}
	  \psfrag{x1}{$x_1$}
	  \psfrag{x2}{$x_2$}
	  \psfrag{O}{$O$}
	  \psfrag{X1}{$X_1$}
	  \psfrag{X2}{$X_2$}
	  \psfrag{l1}{$1$}
	  \psfrag{l2}{$2$}
	   \psfrag{v1}{$v^{*(1)}\;t$}
	   \psfrag{v2}{$v^{*(2)}\; t$}
	   \psfrag{d1}{$\Delta_{5,2}^{(1)}$}
	   \psfrag{d2}{$\Delta_{2,7}^{(1)}$}
	   \psfrag{d3}{$\Delta_{5,7}^{(1)}$}
	   \psfrag{d4}{$\Delta_{2,5}^{(2)}$}
	   \psfrag{d5}{$\Delta_{7,2}^{(2)}$}
	   \psfrag{d6}{$\Delta_{7,5}^{(2)}$}
	   \psfrag{d7}{$\Delta_{6,5}^{(1)}$}
\subfigure[The desired formation geometry of a 1D spatial platoon with $6$ vehicles and $3$ reference vehicles.]{\includegraphics[scale = 0.25]{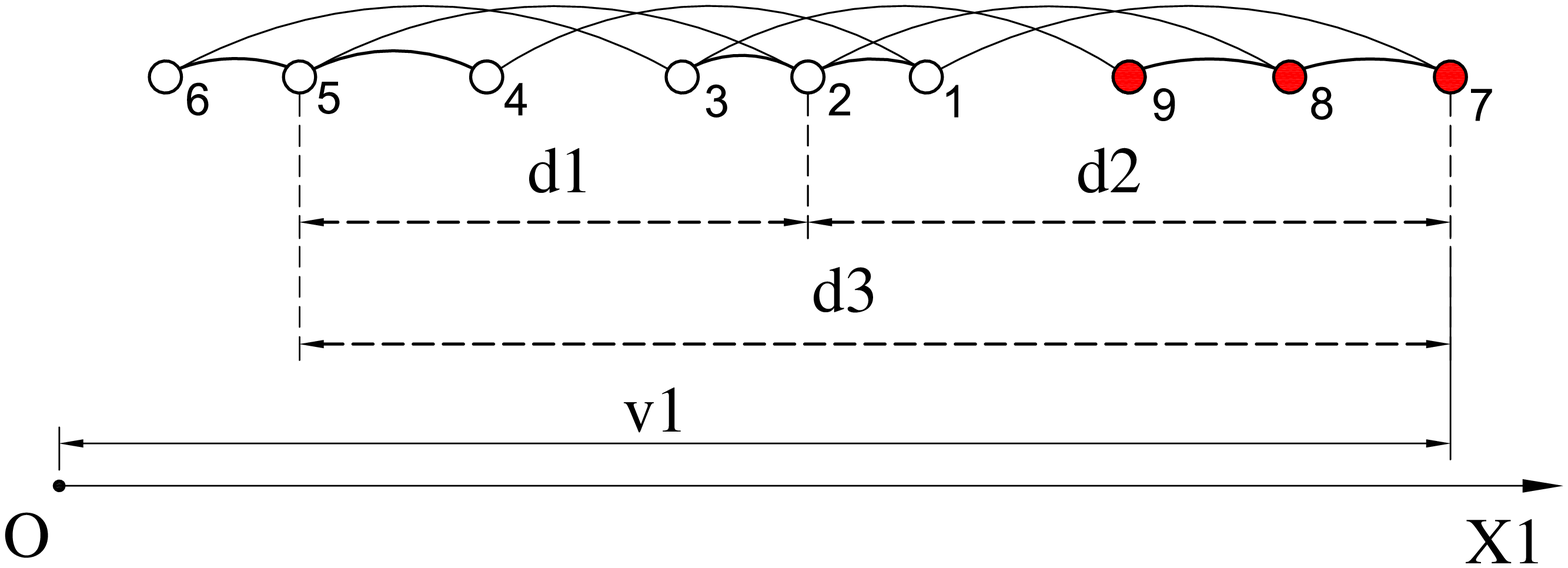} } \quad 
\subfigure[The desired formation geometry of a  2D spatial vehicle formation with $6$ vehicles and $3$ reference vehicles.]{ \includegraphics[scale = 0.22]{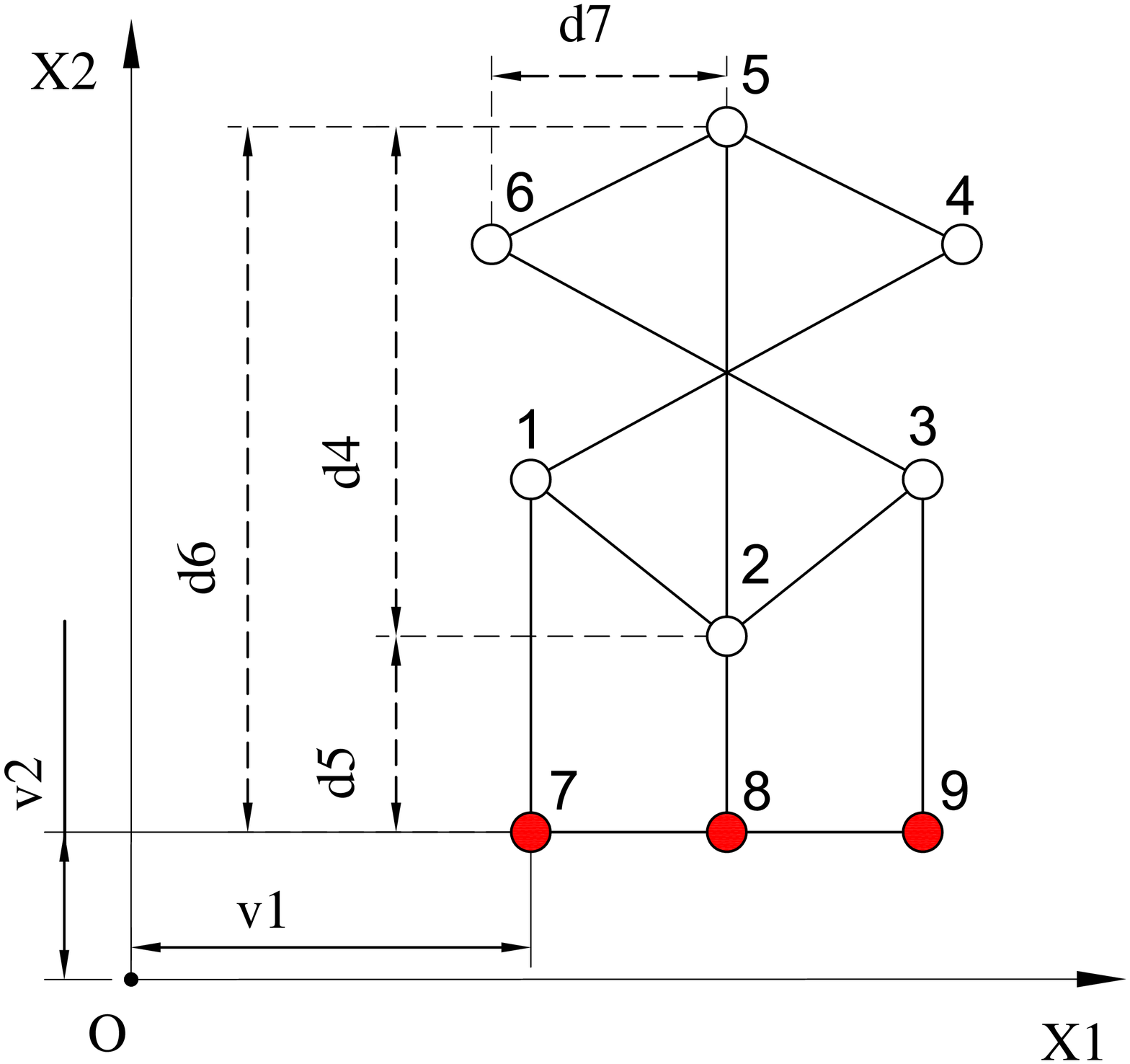}} \quad
\subfigure[The information graph for both the 1D platoon and the 2D formation shown in (a) and (b).]{\includegraphics[scale = 0.25]{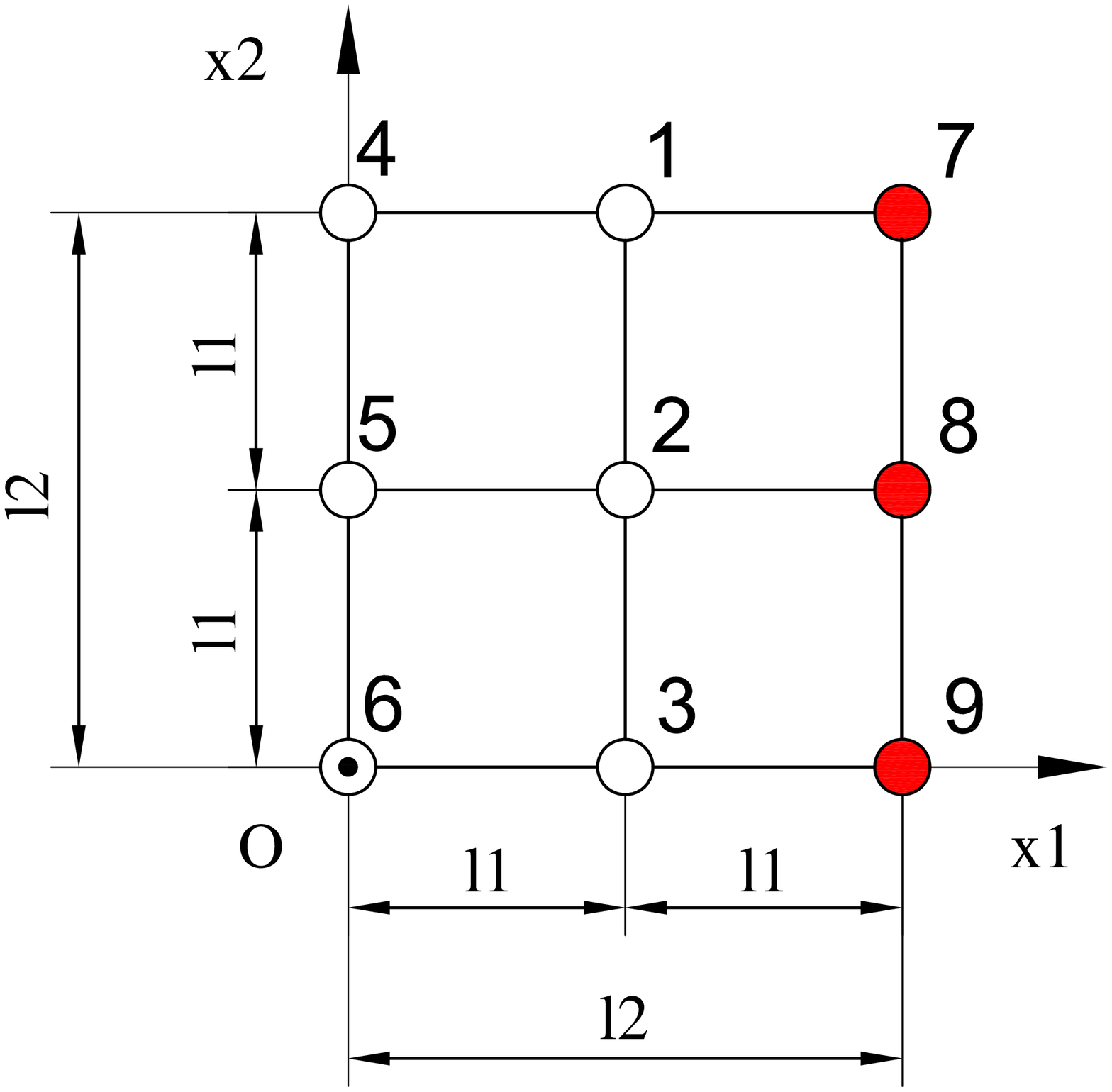} } 
\caption{(a, b): Two distinct spatial formations that have the same associated information graph (c). Red (filled) circles represent reference  vehicles and black (unfilled) circles represent real vehicles. Dashed lines (in (a), (b)) represent desired relative positions, while solid lines represent edges in the information graph. }
\label{fig:2Dinfo_1D2Dspatial}
\end{figure}

\begin{example}
Consider the two formations shown in Figure~\ref{fig:2Dinfo_1D2Dspatial}~(a) and~(b). Their spatial dimensions are $D_s=1$ and $D_s=2$, respectively.  The information graph, however, is the same in both cases:
\begin{align*}
\V=\{1,2,\dots,9\}, \ \E = \{(1,2), (1,4), (1,7), (2,3), (2,5), (2,8), (3,6), (3,9), (4,5), (5,6),(7,8),(8,9)\}.
 \end{align*}
A drawing of the information graph appears in Figure~\ref{fig:2Dinfo_1D2Dspatial}~(c).  Although the information graph is the same, the desired spacings $\Delta_{i,j}$'s are different in the two formations.  For example, $\Delta_{2,5}^{(1)} \neq 0$ in the one-dimensional formation shown in Figure~\ref{fig:2Dinfo_1D2Dspatial} (a)  whereas  $\Delta_{2,5}^{(1)}=0$ in the two-dimensional formation shown in Figure~\ref{fig:2Dinfo_1D2Dspatial} (b).
\end{example}

%Note that the desired formation geometry can be specified in terms of the desired spacing vectors between every pair of vehicles $i,j$, subject to being mutually consistent. However, since the control law~\eqref{eq:control-lawd} is distributed, only some of the vehicle pairs (that have edges between them in the information graph $\G$) uses this information.
%For the formation shown in Figure~\ref{fig:2Dinfo_1D2Dspatial}(a), the desired spacing between pairs of physically adjacent vehicles is specified to be $\Delta^{(1)}$, which completely specifies the desired spacing between every pair of vehicles. However, since there is no edge between vehicles $4$ and $3$, the error between their actual spacing  and desired spacing will not affect their control actions.

In this paper we restrict ourselves to a specific class of information graph, namely a finite rectangular lattice:

\begin{definition}[$D$-dimensional lattice] A $D$-dimensional lattice, specifically a $n_1 \times n_2 \times \dots \times n_D$ lattice, is a graph with $n_1 n_2 \dots  n_D$ nodes.  In the $D$-dimensional space $\R^D$, the coordinate of $i$-th node is $\vec{i} \eqdef [i_1,\dots,i_D]^T$, where $i_1 \in \{0,1,\dots,(n_1-1)\}$, $i_2 \in \{0,1,\dots,(n_2-1)\}$, $\dots$ and $i_D \in \{0,1,\dots,(n_D-1)\}$.  An edge exists between two nodes $\vec{i}$ and $\vec{j}$ if and only if $\|\vec{i} - \vec{j} \| = 1$, where $\| \cdot\|$ is the Euclidean norm in $\R^D$.  A $n_1 \times n_2 \times \dots \times n_D$ lattice is denoted by $\mbf{Z}_{n_1 \times n_2 \times \dots \times n_D}$.  With a slight abuse of notation, ``the $i$-th node'' is used to denote the node on the lattice with coordinate $\vec{i}$. \frqed
\end{definition}

\begin{figure}
\begin{center}
	  \psfrag{O}{$O$}
	  \psfrag{x1}{$x_1$}
	  \psfrag{x2}{$x_2$}
	  \psfrag{x3}{$x_3$}
 \subfigure[A 1D $4$ lattice.]{\includegraphics[scale = 0.2]{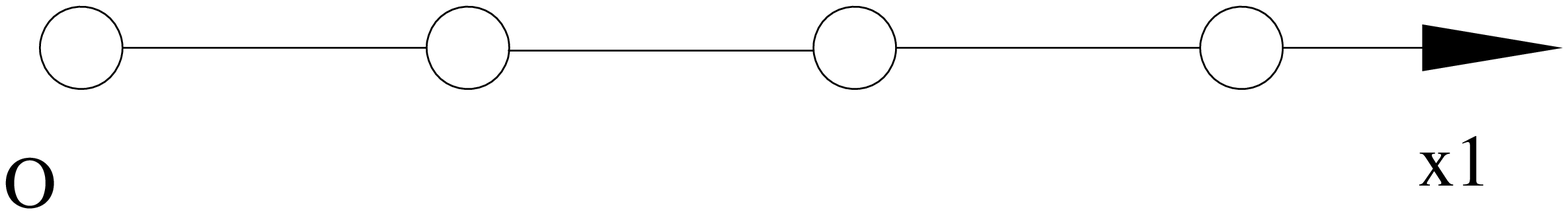}} \ \ \ \
 \subfigure[A 2D $4 \times 4$ lattice.]{\includegraphics[scale = 0.2]{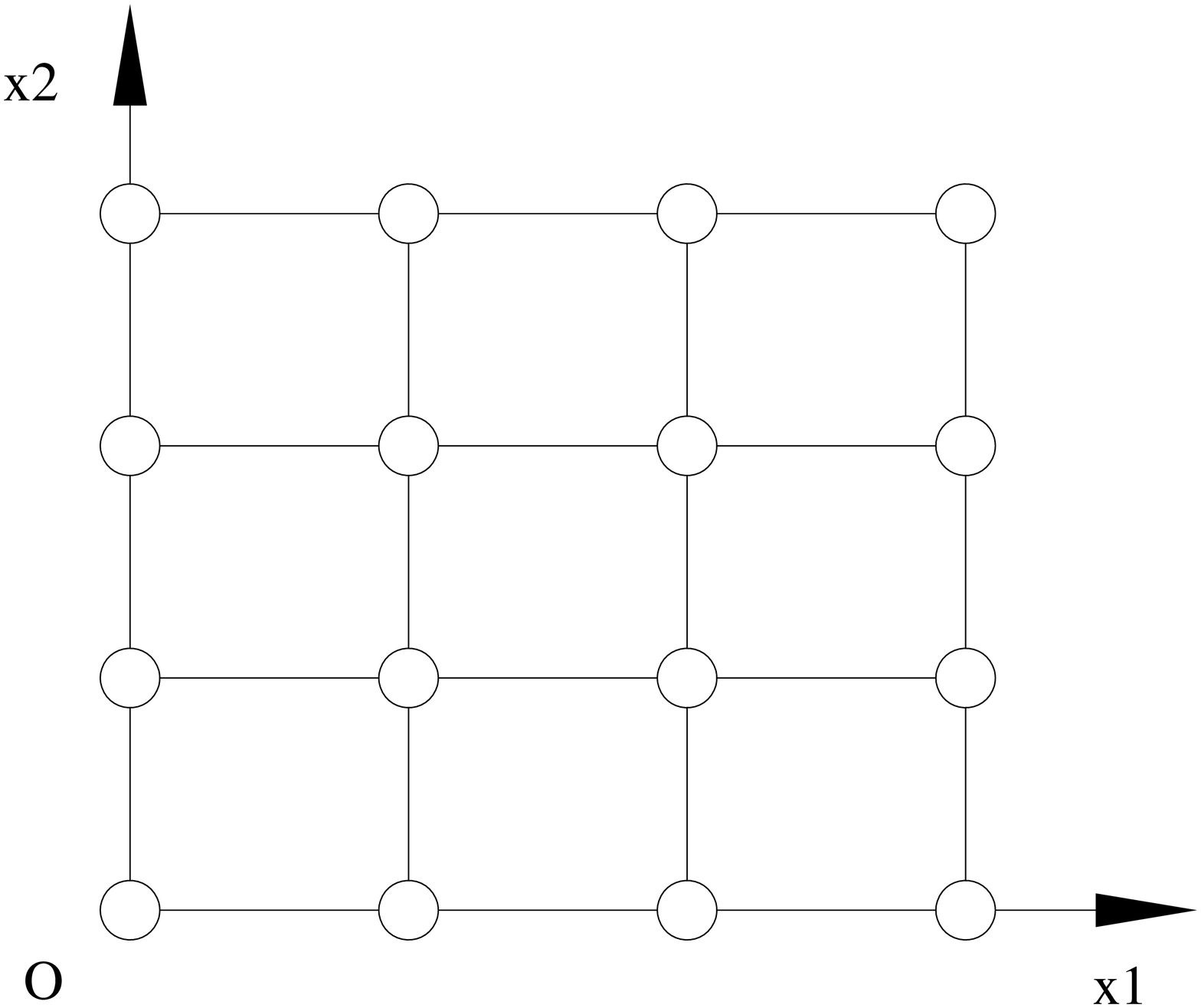}}  \ \ \ \
 \subfigure[A 3D $2 \times 3 \times 3$ lattice.]{\includegraphics[scale = 0.25]{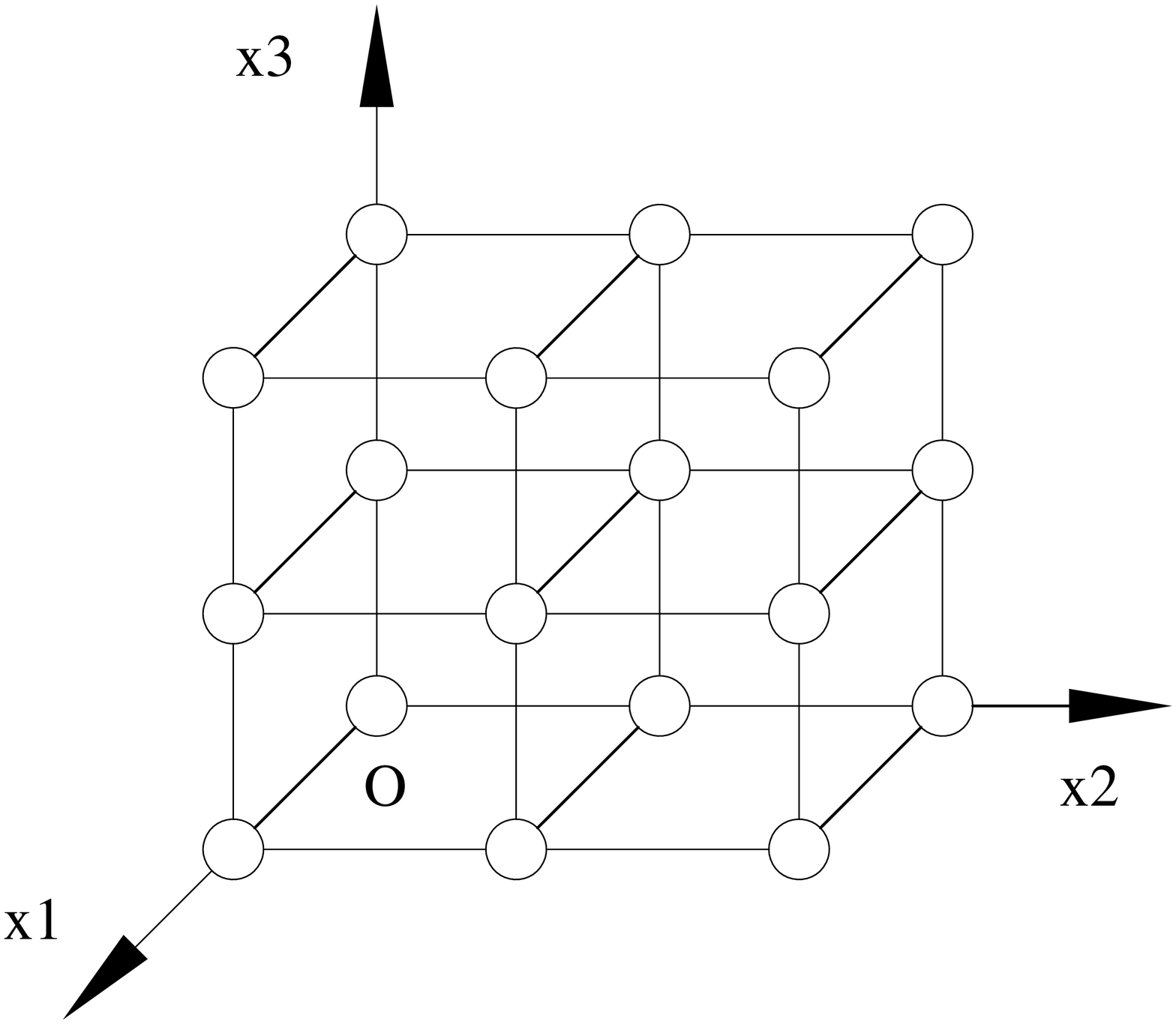}}\ \ \
 \end{center}
 \caption{Examples of 1D, 2D and 3D lattices. }
  \label{fig:lattices}
\end{figure}

Figure~\ref{fig:lattices} depicts three examples of lattices. A $D$-dimensional lattice is drawn in $\R^D$ with a Cartesian reference frame whose axes are denoted by $x_1,x_2,\dots,x_D$. Note that these coordinate axes may not be related to the coordinate axes in the physical space $\R^{D_s}$.

In this paper an information graph $\G$ is always a lattice $\mbf{Z}_{n_1 \times n_2 \dots \times n_D }$, where $n_1 n_2 \dots n_D = N+N_r$. For a given $N$, the choice of $N_r,D,n_1,n_2,\dots,n_D$ serves to determine the specific choice of the information graph within the class. An information graph is said to be \emph{square} if  $n_1-1=n_2=\hdots=n_D$.

For the ease of exposition and notational simplicity, we make the following two assumptions regarding the reference vehicles and the distributed control architecture~\eqref{eq:control-lawd}:

\begin{assumption}\label{as:k-same}
For each $(i,j)\in\E$, the gain $k_{(i,j)}^{(\mrm{d})}$ does not depend on $\mrm{d}$, and for each $i \in \V$, $b_i^{(\mrm{d})}$ does not depend on $\mrm{d}$.   \frqed
\end{assumption}

\begin{assumption}\label{as:L1}
The reference vehicles are arranged so that a node $i$ in the information graph corresponds to a reference vehicle if and only if  $i_1=n_1-1$. \frqed
\end{assumption}

Assumption~\ref{as:k-same} means that the local control gains do not explicitly depend upon the coordinate $d$.  Such an assumption is not restrictive because of the fully actuated assumption.  If the local control gains are allowed to depend upon $d$ then one could repeat the analysis of this paper separately for each value of $d$.  Note that the assumption does not mean that the control gains are spatially homogeneous; for example, the control gains $k_{(i,j)}^{(1)}\neq k_{(i,j)}^{(2)}$ for the same $(i,j) \in \E$.

Assumption~\ref{as:L1} means that all reference vehicles are assumed to be arranged on a single ``face'' of the lattice, and every vehicle on this face is a reference vehicle. Assumption~\ref{as:L1} implies that $N = (n_1-1)n_2\dots n_D$ and $N_r = n_2\dots n_D$.  Other arrangements of reference vehicles do not significantly change the main conclusions of this paper. Some of these extensions are discussed in Sec.~\ref{sec:remarks}.

As a result of the Assumption~\ref{as:k-same}, we can rewrite~\eqref{eq:control-lawd} as
\begin{align}\label{eq:control-law}
    u_{i} & = \sum_{j \in \scr{N}_i}-k_{(i,j)}(p_{i}-p_{j}-\Delta_{i,j}) - b_i (\dot{p}_{i}- v^*),
\end{align}
where the superscript $(\mrm{d})$ has been suppressed.  %The symbols $p_{i}, \dot{p}_{i}$, and $u_{i}$ denote position, velocity and control, respectively, along one of the $D_s$ coordinate directions.

\begin{remark}\label{rem:Ds-vs-D}
The dimension $D$ of the information graph is distinct from the spatial dimension $D_s$.  Figure~\ref{fig:2Dinfo_1D2Dspatial} shows an example of two formations in space, one with $D_s=1$ and the other with $D_s=2$. The information graph for both the formations is the same $3 \times 3$ two-dimensional lattice, i.e.,  $D=2$.  On account of the fully actuated dynamics and Assumption~\ref{as:k-same}, the spatial dimension $D_s$ plays no role in the results of this paper.  The dimension of the information graph $D$, on the other hand, will be shown to play a crucial role.
\end{remark}

\begin{remark}
Analysis of the control law~\eqref{eq:control-law} is relevant even when there are additional dynamic elements in the controller. There are several reasons for this.
First, a dynamic controller cannot have a zero at the origin, for that will result in a pole-zero cancellation, causing the steady-state
errors to grow without bound as $N$ increases~\cite{PB_JH_CDC:05}. Second, a
dynamic controller cannot have an integrator either. If it
does, the closed-loop platoon dynamics become unstable for a
sufficiently large value of $N$~\cite{PB_JH_CDC:05}. Thus, any allowable
dynamic compensator must essentially act as a static gain at low frequencies. The results of~\cite{PB_JH_CDC:05,SKY_SD_KRR_TAC:06} indicate that the low frequency behavior is the dominant factor in the control of large networks of agents with double integrator dynamics. Hence, the issues that arise with the control law~\eqref{eq:control-law} are also relevant to the
case where additional dynamic elements appear in the control law.
\end{remark}

%%%%%%%%%%%%%%%%%%%%%%%%%%%%%%%%%%%%%%%%%%%%%%%%%%%%%%%%%%%%%%%%%%%%%%
%%%%%%%%%%%%%%%%%%%%%%%%%%%%%%%%%%%%%%%%%%%%%%%%%%%%%%%%%%%%%%%%%%%%%% %%%%%%%%%%%%%%%%%%%%%%%%%%%%%%%%%%%%%%%%%%%%%%%%%%%%%%%%%%%%%%%%%%%%%%
\subsection{Main result \Rmnum{1}: Stability margin with symmetric control and $D$-dimensional information graph}\label{sec:result-nominal}

\begin{definition}\label{def:stabiltity_margin}
The \emph{stability margin} is the absolute value of the real part of the least stable eigenvalue of the closed-loop system. \frqed
\end{definition}
%The first main result of this paper is an asymptotic formula that relates the stability margin of the closed-loop formation with a \emph{symmetric} control.
%Recall that the \emph{stability margin} $S$ of the formation is defined as the absolute value of the real part of the least stable eigenvalue of the closed-loop state matrix $A$ (when the dynamics are expressed in the state-space form).  The formula is asymptotic in the sense that it holds when $N$ is large.

\begin{definition}
The control law~\eqref{eq:control-law} is \emph{symmetric} if all the vehicles use the same control gains: $k_{(i,j)}=k_{0},$ for all $(i,j) \in \E$ and $b_i=b_0$ for all $i \in \V$, where $k_0$ and $b_0$ are positive constants.
\end{definition}

The first main result gives an asymptotic formula for controlled formation with symmetric control:

\begin{theorem}\label{thm:symmetric}
Consider an $N$-vehicle formation with vehicle dynamics~\eqref{eq:vehicle-dynamics} and control law~\eqref{eq:control-lawd}, with Assumptions~\ref{as:k-same} and~\ref{as:L1}. With symmetric control, the stability margin of the closed-loop is given by the formula
\begin{align}\label{eq:less_stable_eigenvalue}
	S=\frac{\pi^2 k_0}{4b_0}\frac{1}{(n_1-1)^{2}} + O(\frac{1}{n_1^4}),
\end{align}
that holds when $n_1 \to \infty$. % is large (in particular, when $n_1 \gg 1+\frac{\pi\sqrt{k_0}}{b_0}$).
\frqed
\end{theorem}
We remark that the stability margin depends only upon $n_1$ -- the number of vehicles along the $x_1$ axis.  The $x_1$ axis is special because it is normal to the face with the reference vehicles; see Assumption~\ref{as:L1}.  In the PDE model, the boundary condition is of the Dirichlet type on this face (see~\eqref{eq:BC}). Analogous estimates also hold with different arrangement of the reference vehicles (see Section~\ref{sec:remarks} for details).

\smallskip

\paragraph{Square information graph} %An important special case is that of a square information graph, which refers  to the case when the number of vehicles (excluding the reference ones) in each direction of the lattice is the same. Under Assumption~\ref{as:L1}, this means $n_1 -1 = n_2 = \dots = n_D$\footnote{Note that this is a slight abuse of notation since the number of nodes along each of the $D$ coordinate axes in $\R^D$ (in which the graph is drawn) is not the same, since that would have led to $n_1 = n_2 = \dots = n_D$.}. Since $L_1$ is defined as $L_1 = n-1$, for a square information graph we have $N = (n_1 -1)n_2\dots n_D = L_1^D$.  The following corollary follows immediately from the lemma above.

For a square information graph, $N = (n_1 -1)n_2\dots n_D = (n_1-1)^D$, and we have the following corollary:
\begin{corollary}\label{cor:symmetric-square}
Consider an $N$-vehicle formation with vehicle dynamics~\eqref{eq:vehicle-dynamics} and control law~\eqref{eq:control-lawd}, with Assumptions~\ref{as:k-same} and~\ref{as:L1}. When the information graph is a square $D$-dimensional lattice, the closed-loop stability margin with symmetric control is given by the asymptotic formula
\begin{align}\label{eq:stability-margin-symmetric-square}
	S=\frac{\pi^2 k_0}{4b_0}\frac{1}{{N^{2/D}}} + O(\frac{1}{{N^{4/D}}}).
\end{align}\frqed
\end{corollary}
The special case of Corollary~\ref{cor:symmetric-square} for $D=1$ was established in~\cite{PB_PM_JH_TAC:09}.

The result from Corollary~\ref{cor:symmetric-square} shows that for a constant choice of symmetric control gains $k_0$ and $b_0$, the stability margin approaches $0$ as $N\rightarrow\infty$.  The dimension $D$ of the information graph determines the scaling. Specifically, the stability margin scales as $O(1/{N^2})$ for 1D information graph, as $O(1/{N})$ for 2D information graph, and as $O(1/{N^{2/3}})$ for 3D information graph.
Thus, \emph{for the same control gains, increasing the dimension of the information graph improves the stability margin significantly}.  In practice, this may require a communication network with long range connections in the physical space.  Note that an information graph is only a drawing of the connectivity.  A neighbor in the information graph need not be physically close.

% \begin{figure}[t]
% \centering
% \includegraphics[scale = 0.366]{corollary1_prediction.eps}
%  \caption{ stability margin predicted by Corollary~\ref{cor:symmetric-square} with that numerically computed from the state matrix $A$ in~\eqref{eq:closedloop-wholeplatoon} for a vehicle formation with $2$-dimensional \emph{square} information graph.}\label{fig:corollary-1}
% \end{figure}

%In many settings, it is often convenient to have the dimension of the information graph to be equal to the spatial dimension: $D_s = D$. For example, consider a formation of ground vehicles arranged in a two-dimensional grid.  Here, the spatial dimension $D_s=2$. If the control action depends only upon relative position measurements with the physically nearest vehicles, then the information graph is also a  $2$-dimensional lattice. So, $D=2$.  In this case, the Corollary~\ref{cor:symmetric-square} implies that the stability margin scales as $O(1/{N})$. %PB: this para did not add anything, so I removed it

%We note that the stability margin can be trivially made independent of $N$ by choosing $k_0 = c N^{\frac{2}{D}}$, which increases the gains without bound as $N$ increases. This choice of control gains has the obvious disadvantage associated with any high gain feedback.

\medskip

\begin{remark}\label{rem:LQR}
It was shown in~\cite{MJ_BB_TAC:05} that the closed-loop stability margin for a circular platoon approaches zero as $O(1/N^2)$ even with the centralized LQR controller. It is interesting to note that distributed control (with an information graph of dimension $D>1$) yields a better scaling law for the stability margin than centralized LQR control.  
%In contrast to the centralized control in~\cite{MJ_BB_TAC:05}, the control law~\eqref{eq:control-law} is distributed because the number of neighbors with a $D$-dimensional information graph is at most $2D$.
%note that a decentralized controller can improve the asymptotic decay of the stability margin over centralized LQR control, as long as the information graph has dimension higher than $1$.
%\frqed
\end{remark}

\paragraph{Non-square information graph}
It follows from Theorem~\ref{thm:symmetric} that by choosing the structure of the information graph in such a way that $n_1$ increases slowly in relation to $N$, the loss of the stability margin as a function of $N$ can be slowed down. In fact, when $n_1$ is held at a constant value independent of $N$, it follows from Theorem~\ref{thm:symmetric} that the stability margin is a constant independent of the total number of vehicles. More generally, consider an information graph with $n_1 = O(N^c)$, where $c\in[0,1]$ is a fixed constant.  Using Theorem~\ref{thm:symmetric}, it follows that  $S =  O(1/N^{2c})$ as $N\rightarrow\infty$. If $c<\frac{1}{D}$, the resulting reduction of $S$ with $N$ is slower than that obtained for a square lattice; cf.~Corollary~\ref{cor:symmetric-square}. This shows that within the class of $D$ dimensional lattices (for a fixed $D$), certain information graphs provide better scaling of the stability margin than others. The price one pays for  improving stability margin by reducing $n_1$ is an increase in the number of reference vehicles. This is because the number of reference vehicles $N_r$ is related to $n_1$ by $N_r = N/(n_1-1)$ (see  Assumption~\ref{as:L1}).

It is important to stress that not all non-square graphs are advantageous. For example, if $n_1 = O(N)$ and $n_2$ through $n_D$ are $O(1)$, it follows from Theorem~\ref{thm:symmetric} that the stability margin is $S = O(1/N^2)$. This is the same trend as in a 1-D information graph. In this case, we can say that the $D$ dimensional information graph effectively behaves as a one dimensional graph.

\begin{figure}[t]
	  \psfrag{x1}{$x_1$}
	  \psfrag{x2}{$x_2$}
	  \psfrag{l1}{\scriptsize $n_1=O(1)$}
	  \psfrag{l2}{\scriptsize $n_2=O(N)$}
	  \psfrag{O}{$O$}
	  \begin{center}
 \subfigure[Non-square information graph, $S = O(1)$]{\includegraphics[scale = 0.28]{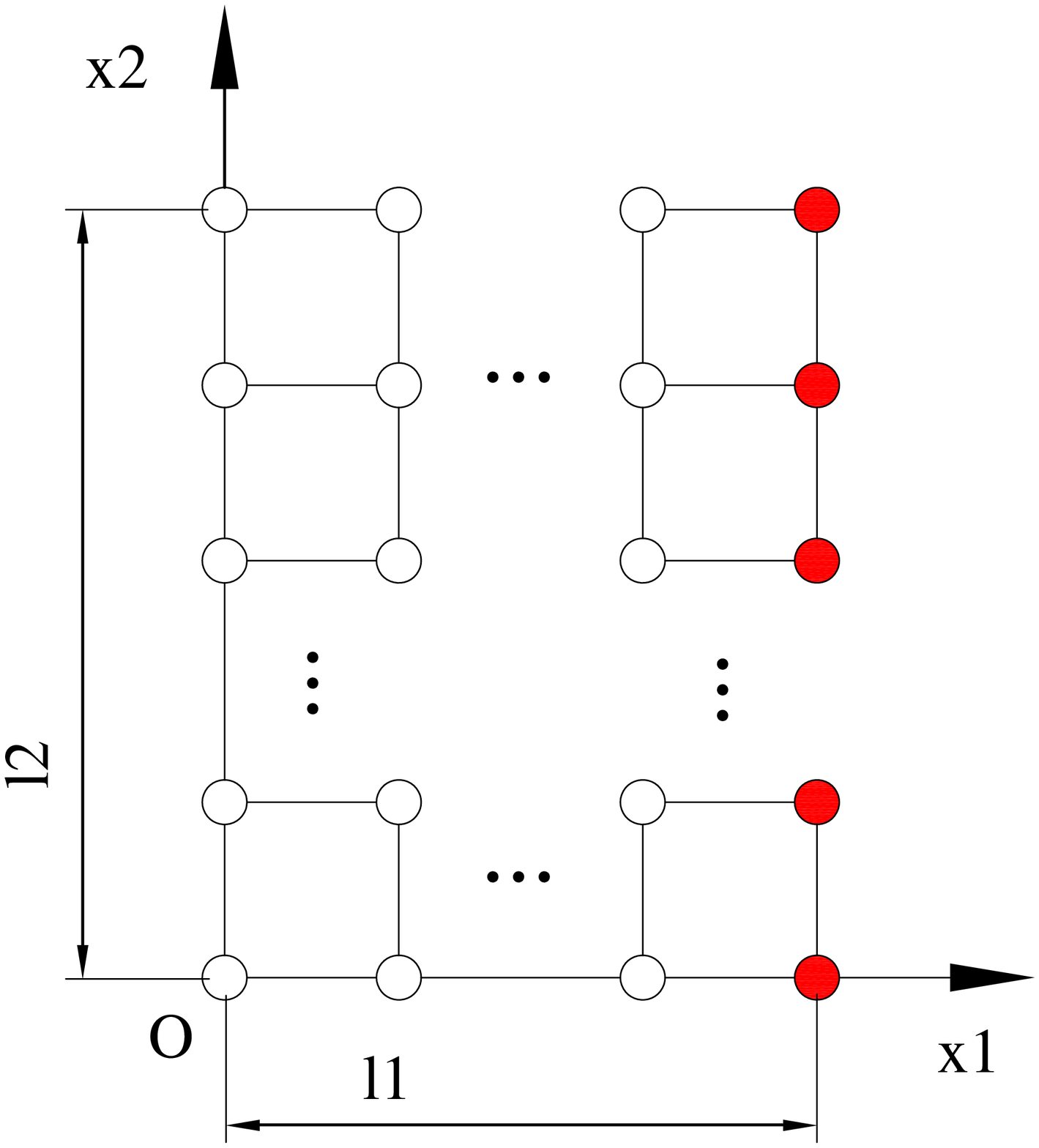}} \
	  \psfrag{l3}{\scriptsize $n_1=O(N)$}
	  \psfrag{l4}{\scriptsize $n_2=O(1)$}
 \subfigure[Non-square information graph, $S = O(1/N^2)$]{\includegraphics[scale = 0.26]{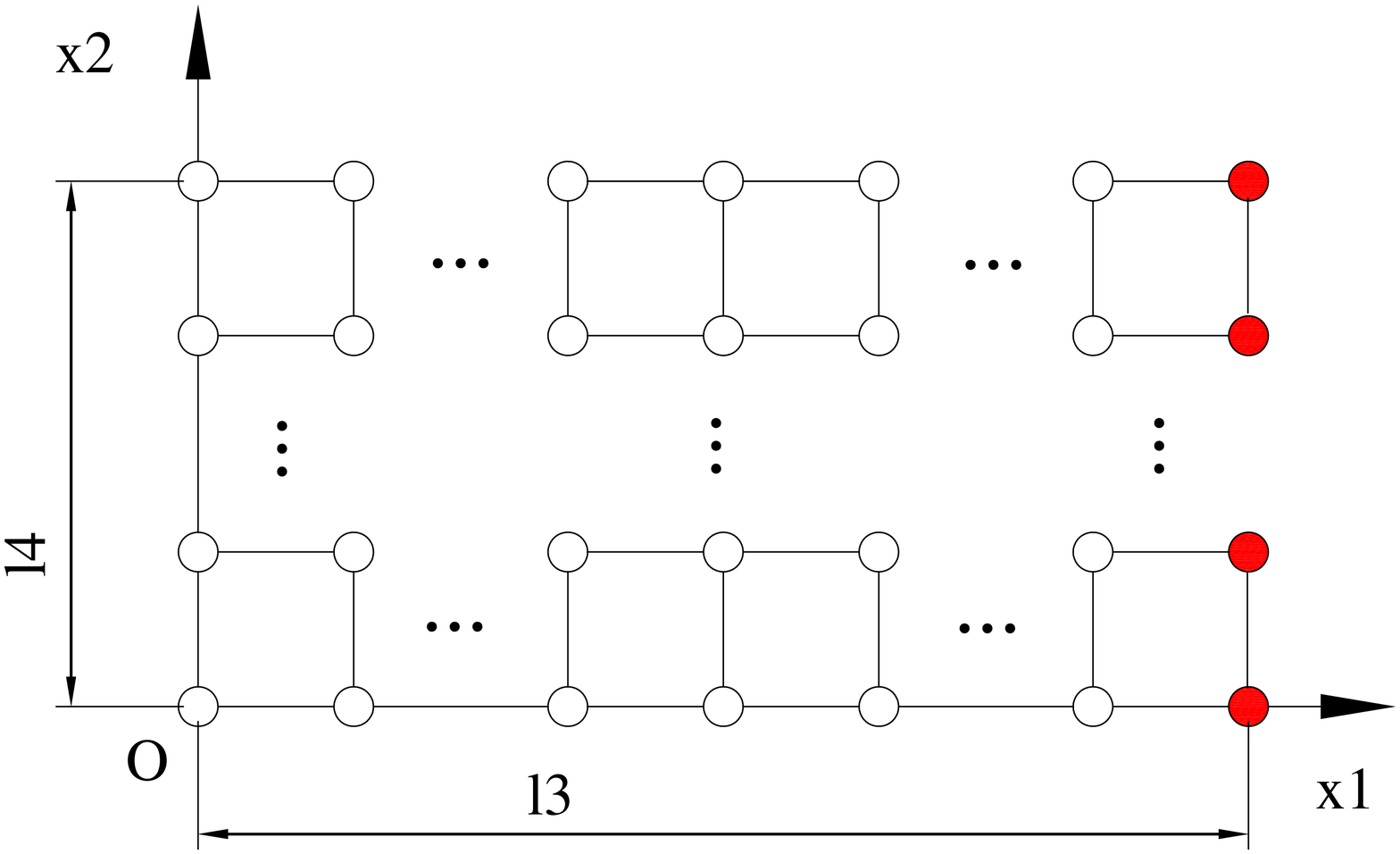}}\
	\psfrag{l5}{\scriptsize $n_1= O(\sqrt{N})$}
	\psfrag{l6}{\scriptsize $n_2= O(\sqrt{N})$}
 \subfigure[``Approximately'' square information graph, $S = O(1/N)$]{\includegraphics[scale = 0.25]{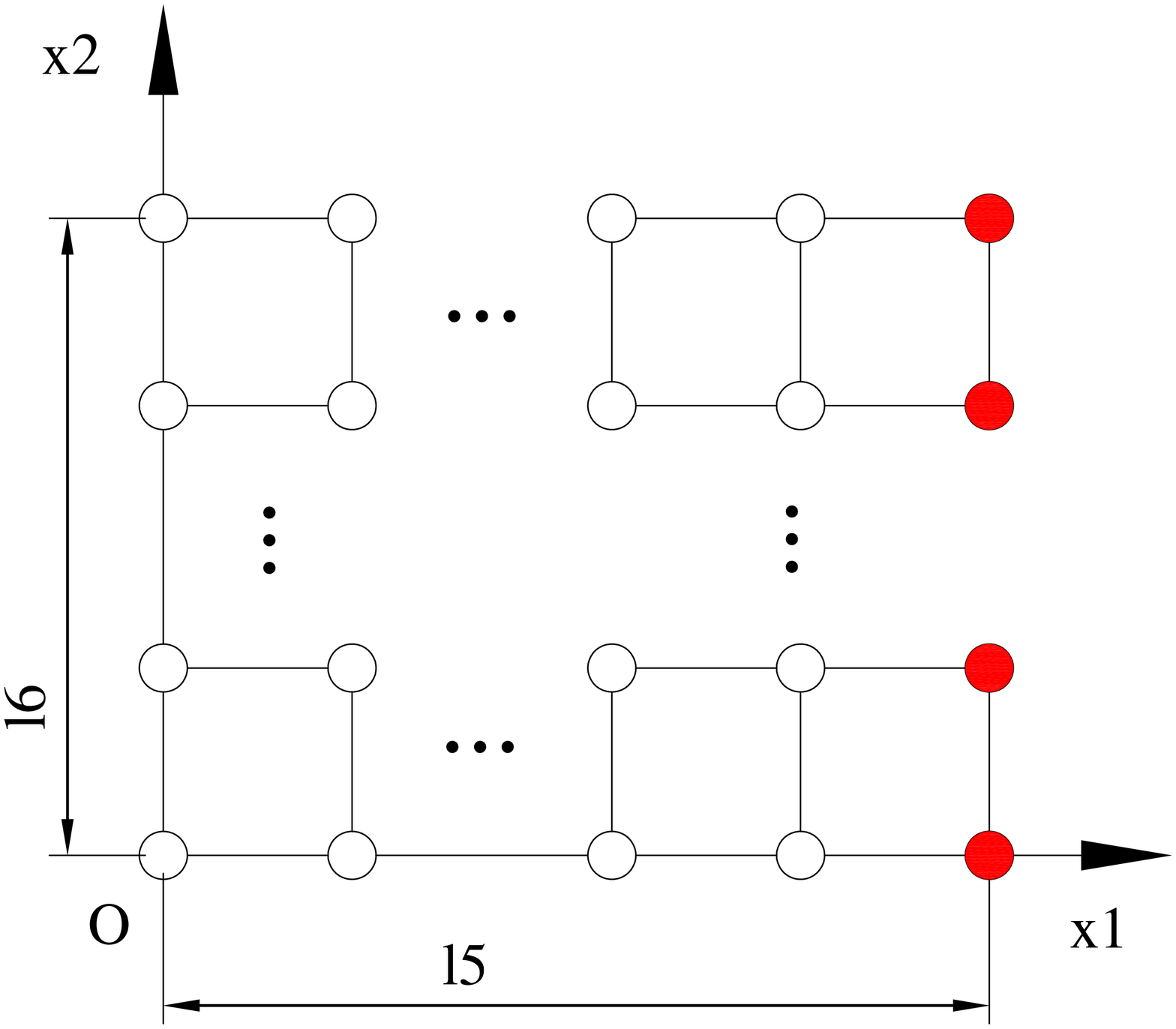}}
\end{center}
 \caption{(a) A $2$-dimensional information graph in which the first dimension is held constant, resulting in a stability margin that is independent of $N$, $S = O(1)$. (b) A $2$-dimensional information graph that is "asymptotically" 1D (as $N \to \infty$) since the size of the first dimension increases linearly with $N$, resulting in a stability margin scaling law $S =O(1/N^2)$, which is the same as that with an 1D information graph. (c) A $2$-dimensional information graph in which both sides are of length $O(\sqrt{N})$, for which we have  $S=O(1/N)$, the same behavior as that of a square 2D graph.}\label{fig:graphs_aspect-ratio}
\end{figure}

Figure~\ref{fig:graphs_aspect-ratio} shows a few examples of information graph that are relevant to the discussion above. The 2D information graph shown in Figure~\ref{fig:graphs_aspect-ratio} (a) has $n_1 = O(1)$ and $n_2 = O(N)$, whereas the one in Figure~\ref{fig:graphs_aspect-ratio} (b) has $n_1 = O(N)$ and $n_2 = O(1)$. The graph shown in Figure~\ref{fig:graphs_aspect-ratio} (c)  is approximately square, both $n_1$ and $n_2$ are $O(\sqrt{N})$.

\begin{figure}[h]
  \centering
  \psfrag{N}{$\ N$} \psfrag{S}{$S$}
  \psfrag{A}{$n_1 -1= 5$ (SSM)}
  \psfrag{B}{$n_1 -1= N/5$ (SSM)}
  \psfrag{C}{$n_1 -1= \sqrt{N}$ (SSM)}
  \psfrag{D}{$n_1 -1= 5$ (Theorem~\ref{thm:symmetric})}
  \psfrag{E}{$n_1 -1= N/5$ (Theorem~\ref{thm:symmetric})}
  \psfrag{F}{$n_1 -1= \sqrt{N}$ (Corollary~\ref{cor:symmetric-square})}
\includegraphics[scale = 0.4]{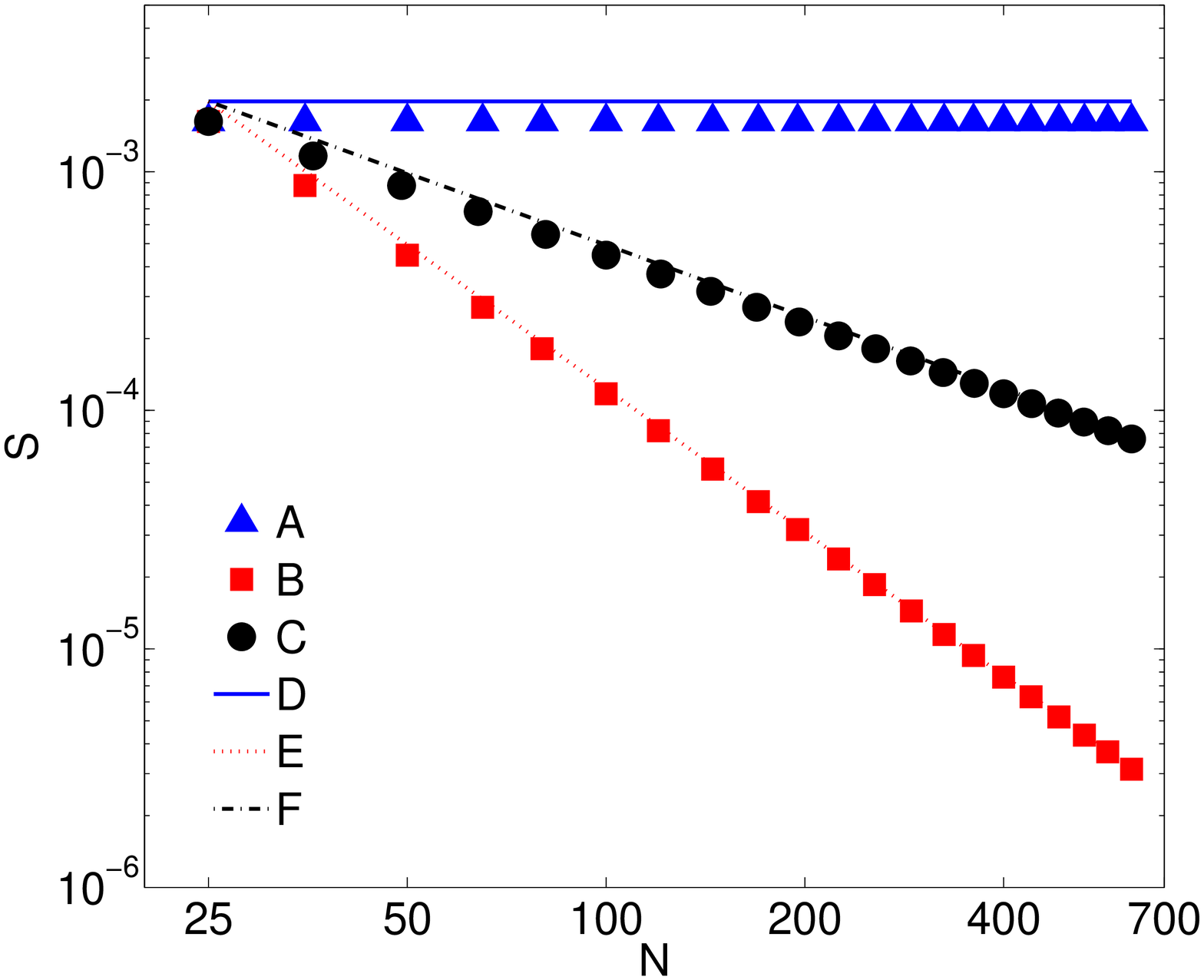}
 \caption{ Stability margin predicted by Theorem~\ref{thm:symmetric}  for a vehicle formation with  information graphs of various ``shapes'' as shown in Figure~\ref{fig:graphs_aspect-ratio}. The legend "SSM'' means computed from the "state space model''~\eqref{eq:closedloop-wholeplatoon}, which is presented in Section~\ref{sec:problem}.  For the first case,  $n_1-1=5$  and $n_2 = N/5$. Theorem~\ref{thm:symmetric} predicts that in this case $S=O(1)$ even as $N \to \infty$. In the second case, $n_2=5$ and $n_1-1=N/5$, which leads to $S=O(1/N^2)$.  The third case is that of a square information graph, $n_1-1=n_2=\sqrt{N}$, which leads to $S=O(1/N)$. Theorem~\ref{thm:symmetric} and corollary~\ref{cor:symmetric-square} predicts the stability margin quite accurately in each of the cases. The control gains used in all the calculations are $k_0=0.01$ and $b_0=0.5$. }\label{fig:Eig-trend-early}
\end{figure}

Figure~\ref{fig:Eig-trend-early} provides numerical corroboration of the discussion above. The stability margin as a function of $N$ for three distinct 2D information graphs (that are described in Figure~\ref{fig:graphs_aspect-ratio}) are shown in this figure. The stability margin is computed by computing the eigenvalues of the closed-loop state matrix; the state space model is described in~\eqref{eq:closedloop-wholeplatoon} in Section~\ref{sec:problem}.  The control gains used are $k_0=0.01$, $b_0=0.5$. The plots show that the formula~\eqref{eq:less_stable_eigenvalue} in Theorem~\ref{thm:symmetric}  makes an excellent prediction of the trend of stability margin. The asymptotic nature of the result in Theorem~\ref{thm:symmetric} (and Corollary~\ref{cor:symmetric-square}) is seen from the plot: the prediction becomes more and more accurate as $N$ increases.

%%%%%%%%%%%%%%%%%%%%%%%%%%%%%%%%%%%%%%%%%%%%%%%%%%%%%%%%%%%%%%%%%%%%%%

%%%%%%%%%%%%%%%%%%%%%%%%%%%%%%%%%%%%%%%%%%%%%%%%%%%%%%%%%%%%%%%%%%%%%%
% \begin{figure}
%   \centering
%   	{\includegraphics[scale = 0.4]{aspect_ratio.eps}}
% 	\caption{Numerical corroboration of stability margin for three specific cases of $L_1$ as shown in Figure~\ref{fig:2Dlattice-asymptotic-1D}. For the first case, $L_1=\Theta(1)=5$, $L_2=N/L_1$ and $N$ ranges from $25$ to $625$, the stability margin is a constant. For the second case,  $L_1=\Theta(N)=N/L_2$, $L_2=5$ and $N$ ranges from $25$ to $625$, the stability margin has a trend of $O(1/N^2)$. For the third case,  $L_1=L_2= \Theta(N^{1/2})$, and $N$ ranges from $25$ to $625$, the stability margin has a trend of $O(1/N)$. In all three cases, the gains used are $k_0=0.01$, $b_0=0.5$.}
%   \label{fig:aspect_ratio}
% \end{figure}
%%%%%%%%%%%%%%%%%%%%%%%%%%%%%%%%%%%%%%%%%%%%%%%%%%%%%%%%%%%%%%%%%%%%%%
\subsection{Main result \Rmnum{2}: Stability margin with non-symmetric control and $D$-dimensional information graph}\label{sec:result-mistuning}
%This is NOT necessary atleast as a prelude!  most of this has been said before.  Limitations are obvious.
%The results in the previous section shows that with symmetric control, to improve the asymptotic decay rate of the stability margin with $N$, one has to either increase the gains without bound as $N$ increases, or increase the dimension (real or effective) of the information graph. The first option is not attractive for obvious reasons. The second option may require long-range communication. For example, for a platoon of vehicles located in 1D, to make the information graph a square lattice in 2D, vehicles need to communicate with vehicles that are $\sqrt{N}$ vehicles away (see Figure~\ref{fig:2Dinfo_1D2Dspatial}). So the second option also has limitations, especially if distributed control with short range communication is desired.
%

The second main result of this work is that for a fixed information graph, the scaling law for stability margin can be improved by choosing a non-symmetric control law. We call the resulting design a \emph{mistuning}-based design because it relies on small changes from the symmetric control.  The improvement is achieved by making small perturbations to the proportional gains alone $k_{(i,j)}$; changing derivative gains alone do not have the same disproportionate effect, it only effects the $O(1/n_1^2)$ term in the stability margin. The mistuning-based design and the resulting scaling law is summarized with the aid of the following theorem:

%%%%%%%%%%%%%%%%%%%%%%%%%%%%%%%%%%%%%%%%%%%%%%%%%%%%%%%%%%%%%%%%%%%%%
%%%%%%% main mistuning result %%%%%%%%%%%%%%%%%%%%%
\begin{theorem}\label{thm:mistuned}
Consider an $N$-vehicle formation with vehicle dynamics~\eqref{eq:vehicle-dynamics} and control law~\eqref{eq:control-lawd} under Assumptions~\ref{as:k-same} and~\ref{as:L1}, with nominal symmetric control gains $k_0$ and $b_0$. Now consider the problem of maximizing the stability margin by designing the proportional control gains $k_{(i,j)}$, where the gains are required to satisfy $| k_{(i,j)} - k_0 | \leq  \varepsilon$ for every $(i,j)\in\E$, with $\varepsilon \in ( 0, \; k_0)$ being an arbitrary and small pre-specified constant. For vanishingly small values of $\varepsilon$, the optimal control gains of  the $i$-th vehicle ($i=1, \dots, N$) are given by:
\begin{align}\label{eq:optimal-mistuned-gains}
	k_{(i,i^{1+})}=k_0+\varepsilon, \ \ k_{(i,i^{1-})}=k_0-\varepsilon, \ \ k_{(i,j)} = k_0 \text{ for all other neighbors } j,
\end{align}
where $i^{1+}$ denotes $i$'s neighbor in the positive $x_1$ direction (in the drawing of the information graph) relative to node $i$ and $i^{1-}$ denotes $i$'s neighbor in the negative $x_1$ direction.  The resulting stability margin is given by
\begin{align}\label{eq:stability-margin-mistuned}
	S =\frac{2 \varepsilon}{b_0}\frac{1}{n_1-1} + O(\frac{1}{n_1^2}),
\end{align}
The formula is asymptotic in the sense that it holds when $n_1,\dots,n_D \to \infty$ and $\epsilon \to 0$. \frqed
\end{theorem}
\nomenclature{$k_0$}{The symmetric proportional control gain}%

\begin{figure}[t]
  \centering
  \psfrag{k1}{$k_{(i,i^{1+})}$}
  \psfrag{k2}{$k_{(i,i^{2+})}$}
  \psfrag{k3}{$k_{(i,i^{1-})}$}
  \psfrag{k4}{$k_{(i,i^{2-})}$}
  \psfrag{k5}{$k_0+\varepsilon$}
  \psfrag{k6}{$k_0-\varepsilon$}
  \psfrag{k7}{$k_0$}

  \psfrag{O}{$O$}
  \psfrag{x1}{$x_1$}
  \psfrag{x2}{$x_2$}
  \includegraphics[scale = 0.3]{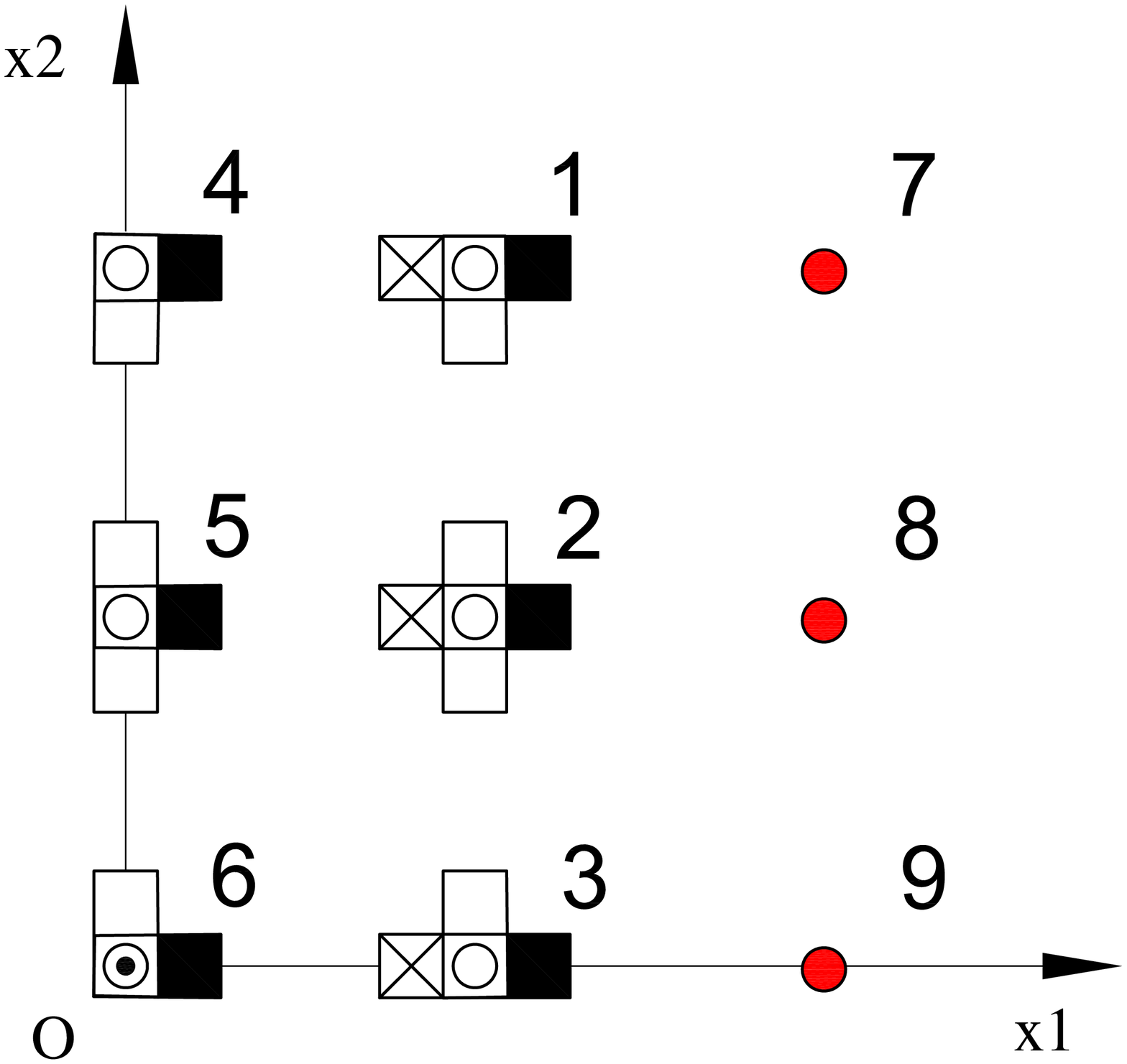}
\hspace{1 cm}
  \includegraphics[scale = 0.35, clip = true, trim  = 0in 0in 0in 0in]{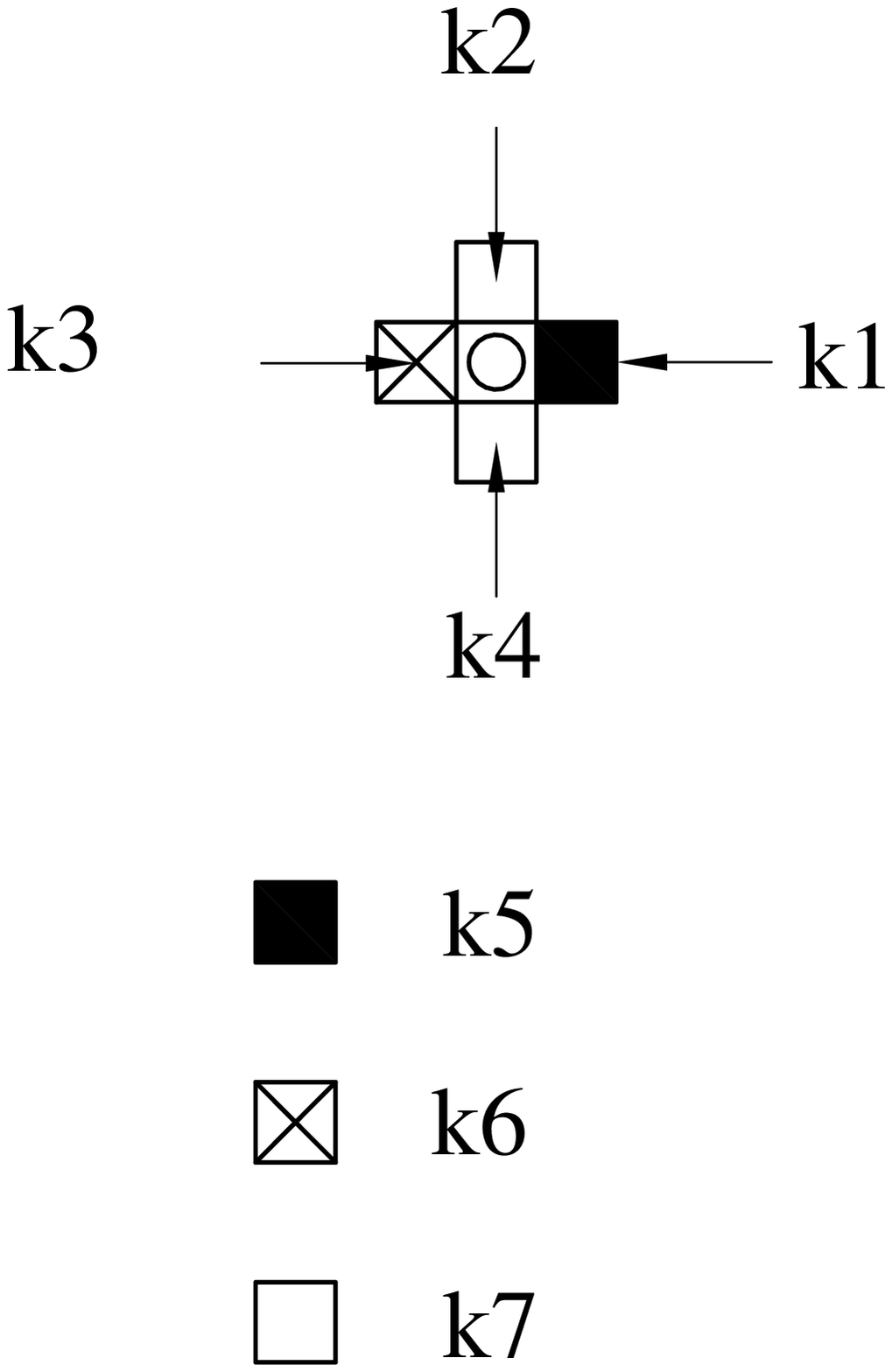}
  \caption{Optimal mistuned proportional control gains for a formation of 6 vehicles with $3$ reference vehicles whose information graph is a $3 \times 3$ lattice. In general, each vehicle in the $2$-dimensional information graph has $4$ proportional gains (as shown on the top of the right figure), $k_{(i,i^{1+})}$, $k_{(i,i^{1-})}$, $k_{(i,i^{2+})}$ and $k_{(i,i^{2-})}$. The gain $k_{(i,i^{1+})}$ is the proportional gain with respect to the neighbor in the positive $x_1$ direction of vehicle $i$. The other three proportional gains have similar interpretations. Notice that the optimal mistuned control gains are achieved by simply making $k_{(i,i^{1+})}$ larger than the nominal gain $k_0$ by $\varepsilon$ and making $k_{(i,i^{1-})}$ smaller than the nominal gain $k_0$ by $\varepsilon$. The other proportional gains remain the same as the nominal gain.}
  \label{fig:kopt-mistuned}
\end{figure}

% \begin{figure}[t]
%   \centering
%   	{\includegraphics[scale = 0.4]{stability_margin.eps}}
% 	\caption{Stability margin improvement with mistuned control (asymmetric) over symmetric control for a vehicle formation with $2$-dimensional square information graph. The control gains for symmetric control are $k_0=0.01$, $b_0=0.5$, and the mistuning amount for mistuned control is $\varepsilon=0.001$ ( i.e., $\pm 10\%$ mistuning). The optimal mistuned gains for each vehicle $i$ ($i=1,\dots, N$) are $k_{(i,i^{1+})}=k_0+\varepsilon=0.011$, $k_{(i,i^{1-})}=k_0-\varepsilon=0.009$ and $k_{(i,j)}=k_0=0.01$ for others, they have the same arrangements as those shown in Figure~\ref{fig:kopt-mistuned}.}
%   \label{fig:mistuning-improvement-square}
% \end{figure}

%%%%%%%%%%%%%%%%%%%%%%%%%%%%%%%%%%%%%%%%%%%%%%%%%%%%%%%%%%%%%%%%%%%%%%
%%%%%%% mistuning result for square lattice %%%%%%%%
For the special case of a square information graph, we have the following corollary.
\begin{corollary}\label{cor:mistuning-square}
For a vehicular formation of $N$ vehicles with square information graph  and mistuned control design described in Theorem~\ref{thm:mistuned}, the  stability margin is given by
\begin{align}\label{eq:stability-margin-asymmetric-square}
    S=\frac{2\varepsilon}{b_0}\frac{1}{N^{1/D}}+ O(\frac{1}{N^{2/D}}),
\end{align}
where $\varepsilon$ is defined in Theorem~\ref{thm:mistuned}.\frqed
\end{corollary}

\medskip

We note that the additional information needed by each vehicle $i$ to implement the mistuned control comprises of (i) the parameter $\varepsilon$ and (ii) the knowledge of which one of its neighbors is the neighbor $i^{1+}$ and which is $i^{1-}$. The special case of Corollary~\ref{cor:mistuning-square} for $D=1$ was established in~\cite{PB_PM_JH_TAC:09}.

\medskip

Comparing Theorems~\ref{thm:symmetric} and~\ref{thm:mistuned} (similarly, Corollaries~\ref{cor:symmetric-square} and~\ref{cor:mistuning-square}), we see that the effect of mistuning is to introduce a square root in the stability margin formula. Thus, even for a small  $\varepsilon$, mistuning can improve the closed-loop stability margin by a large amount, especially when $N$ is large. Numerical verification of  the conclusion of Theorem~\ref{thm:mistuned} is presented in Section~\ref{sec:mistuning}; see, in particular, Figure~\ref{fig:mistuned-lseig-compare}.
Figure~\ref{fig:kopt-mistuned} depicts the optimal mistuned control gains for the case where the information graph is a $3 \times 3$ lattice.

%%%%%%%%%%%%%%%%%%%%%%%%%%%%%%%%%%%%%%%%%%%%%%%%%%%%%%%%%%%%%%%%%%%%%%
\section{Closed-loop dynamics: State-space and PDE models}\label{sec:problem}
%%%%%%%%%%%%%%%%%%%%%%%%%%%%%%%%%%%%%%%%%%%%%%%%%%%%%%%%%%%%%%%%%%%%%%
\subsection{State-space model of the controlled vehicle formation}\label{sec:ss_model}
The dynamics of the $i$-th vehicle is obtained by combining the open loop dynamics~\eqref{eq:vehicle-dynamics} with the control law~\eqref{eq:control-law}, which yields
\begin{align}\label{eq:state_blah_blah_1}
	\ddot{p}_i  = \sum_{j \in \scr{N}_i}-k_{(i,j)}(p_{i}-p_{j}-\Delta_{i,j}) - b_{i}(\dot{p}_{i}- v^*), \ \ i=1,\dots,N.
\end{align}
Let $p_i^*(t)$ denote the desired trajectory of the $i$-th vehicle.  The trajectory is uniquely determined from the trajectories of the reference vehicles and the desired formation geometry. For example, suppose the trajectory of a reference vehicle $r$ is $v^{*}t$. If the $d$-th coordinate of the desired gap between a vehicle $i$ and the reference vehicle $r$ is $\Delta_{i,r}^{(d)}$,  then the $d$-th coordinate of the desired trajectory of $i$ is $p^{*(d)}(t) =v^{*(d)}t + \Delta_{i,r}^{(d)} $.

To facilitate analysis, we define the following coordinate transformation:
\begin{align}\label{eq:state_error}
	\tilde{p}_i  \eqdef p_i-p_i^*  \ \ \ \ \   \Rightarrow  \ \ \ \ \     \dot{\tilde p}_i=\dot{p}_i-v^*.    
\end{align}
Substituting~\eqref{eq:state_error} into~\eqref{eq:state_blah_blah_1}, we have
\begin{align}\label{eq:new_dynamics}
	\ddot{\tilde{p}}_{i}  = \sum_{j \in \scr{N}_i}-k_{(i,j)}(\tilde{p}_{i}-\tilde{p}_{j}) - b_{i}\dot{\tilde{p}}_{i}.
\end{align}
Since the trajectory of a reference vehicle is assumed to be equal to its desired trajectory, $\tilde{p}_{i}  =0$ if $i$ is a reference vehicle. To express the closed-loop dynamics of the formation compactly, we define:
\begin{align*}%\label{eq:tall_vector}
	\tilde{\mbf p}   \eqdef[\tilde{p}_{1},\tilde{p}_{2},\cdots,\tilde{p}_{N}]^{T}, \ \
	\tilde{ \mbf v}  \eqdef \dot{\tilde {\mbf p}} = [\dot{\tilde p}_{1},\dot{\tilde p}_{2},\cdots,\dot{\tilde p}_{N}]^{T}
\end{align*}
Using~\eqref{eq:new_dynamics}, the state-space model of the vehicle formation can now be written
compactly as:
\begin{align}\label{eq:closedloop-wholeplatoon}
\matt{\mbf{\dot {\tilde p}} \\ \mbf{\dot{\tilde v}}} & =
\mbf A \matt{\mbf{\tilde p} \\ \mbf{\tilde v}} \Leftrightarrow \dot{\mbf{\psi}}=\mbf{A} \mbf{\psi}
\end{align}
where $\psi \eqdef [\mbf{ \tilde p}; \mbf{{\tilde v}}]$ is the state vector and $\mbf A$  the closed-loop state matrix.%
\nomenclature{$\mbf A$}{State matrix of the state-space model}%

\begin{example}[Example 1 contd.]
Consider the 1D and the 2D spatial formations depicted in Figure~\ref{fig:2Dinfo_1D2Dspatial} (a) and (b), respectively.  The information graph for both these formations is the same and drawn in Figure~\ref{fig:2Dinfo_1D2Dspatial}(c). We will now show that the closed-loop dynamics of both the formations are the same; cf.~Remark~\ref{rem:Ds-vs-D}. Specifically, let us examine the dynamics~\eqref{eq:state_blah_blah_1} for the vehicle $i=2$. For the 1D formation ($D_s=1$), we have
\begin{align}\label{eq:2th_vehicle}
	\ddot{p}_2^{(1)}  =&-k_{(2,1)}^{(1)} (p_{2}^{(1)} -p_{1}^{(1)} -\Delta_{2,1}^{(1)} )- k_{(2,3)}^{(1)} (p_{2}^{(1)} -p_{3}^{(1)} -\Delta_{2,3}^{(1)} )-k_{(2,5)}^{(1)} (p_{2}^{(1)} -p_{5}^{(1)} -\Delta_{2,5}^{(1)} )\notag \\&-k_{(2,8)}^{(1)} (p_{2}^{(1)} -p_{8}^{(1)} -\Delta_{2,8}^{(1)} ) - b_{2}^{(1)} (\dot{p}_{2}^{(1)} - v^{*(1)} ).
\end{align}
For the purpose of illustration, we focus on the third term on the right hand side of the above equation, and note that the desired trajectories are defined with respect to reference vehicle $7$ (it can be defined with respect to any reference vehicle):
\begin{align}\label{eq:pstar-etc-1}
	p_2^{*(1)}=v^{*(1)} t + \Delta_{2,7}^{(1)}, \ \ \ p_5^{*(1)}=v^{*(1)} t + \Delta_{5,7}^{(1)}.
\end{align}
Using the notation in Eq.~\eqref{eq:state_error}, the third term in the right hand side of~\eqref{eq:2th_vehicle}  can now be expressed as
\begin{align*}
-k_{(2,5)}^{(1)}(p_{2}^{(1)}-p_{5}^{(1)}-\Delta_{2,5}^{(1)}) &=-k_{(2,5)}^{(1)}(\tilde{p}_{2}^{(1)}+p_2^{*(1)}-\tilde{p}_{5}^{(1)}-p_5^{*(1)}-\Delta_{2,5}^{(1)}) = -k_{(2,5)}^{(1)}(\tilde{p}_{2}^{(1)}-\tilde{p}_{5}^{(1)}),
\end{align*}
where the first equality follows from~\eqref{eq:pstar-etc-1} and $\Delta_{2,7}^{(1)}-\Delta_{5,7}^{(1)}=\Delta_{2,5}^{(1)}$, which follows from the definition $\Delta_{i,j} = p^*_i - p_j^*$. By evaluating the other terms in a similar manner, we obtain
\begin{align}\label{eq:p2-dynamics-1D}
	\ddot{\tilde{p}}_{2}^{(1)}= -k_{(2,1)}^{(1)}(\tilde{p}_{2}^{(1)}-\tilde{p}_{1}^{(1)})-k_{(2,3)}^{(1)}(\tilde{p}_{2}^{(1)}-\tilde{p}_{3}^{(1)})-k_{(2,5)}^{(1)}(\tilde{p}_{2}^{(1)}-\tilde{p}_{5}^{(1)})-k_{(2,8)}^{(1)}(\tilde{p}_{2}^{(1)}-\tilde{p}_{8}^{(1)})-b_{2}^{(1)}\dot{\tilde{p}}_{2}^{(1)}.
  \end{align}

\medskip

In case of  the formation with spatial dimension $D_s=2$, we examine the dynamics of the second component of the position vector of vehicle $2$:
\begin{align}\label{eq:2th_vehicle-2}
\ddot{p}_2^{(2)}  = & -k_{(2,1)}^{(2)} (p_{2}^{(2)}-p_{1}^{(2)}-\Delta_{2,1}^{(2)})- k_{(2,3)} ^{(2)}(p_{2}^{(2)}-p_{3}^{(2)}-\Delta_{2,3}^{(2)})-k_{(2,5)}^{(2)} (p_{2}^{(2)}-p_{5}^{(2)}-\Delta_{2,5}^{(2)}) \notag\\
            & -k_{(2,8)}^{(2)} (p_{2}^{(2)}-p_{8}^{(2)}-\Delta_{2,8}^{(2)}) - b_{2}^{(2)} (\dot{p}_{2}^{(2)}- v^{*(2)}).
\end{align}
For this formation, the desired trajectories are also defined with respect to reference vehicle $7$,
\begin{align}\label{eq:pstar-etc-2}
	p_2^{*(k)}= v^{*(k)} t +\Delta_{2,7}^{(k)}, \ \ \ p_5^{*(k)}=v^{*(k)} t +\Delta_{5,7}^{(k)}, \quad k =1,2,
\end{align}
so that the third term on the right hand side of~\eqref{eq:2th_vehicle-2} can be expressed as
\begin{align*}
k_{(2,5)}^{(2)}(p_{2}^{(2)}-p_{5}^{(2)}-\Delta_{2,5}^{(2)})  &  = -k_{(2,5)}^{(2)}(\tilde{p}_{2}^{(2)}+p_2^{*(2)}-\tilde{p}_{5}^{(2)}-p_5^{*(2)}-\Delta_{2,5}^{(2)}) =-k_{(2,5)}^{(2)}(\tilde{p}_{2}^{(2)}-\tilde{p}_{5}^{(2)}),
\end{align*}
where the second equality follows from~\eqref{eq:pstar-etc-2}  and $\Delta_{2,7}^{(2)}-\Delta_{5,7}^{(2)}=\Delta_{2,5}^{(2)}$, which follows from the definition $\Delta_{i,j} = p^*_i - p_j^*$. Repeating this procedure for each of the terms, one obtains:
\begin{align}\label{eq:p2-dynamics-2D}
	\ddot{\tilde{p}}_{2}^{(2)}= -k_{(2,1)}^{(2)}(\tilde{p}_{2}^{(2)}-\tilde{p}_{1}^{(2)})-k_{(2,3)}^{(2)} (\tilde{p}_{2}^{(2)}-\tilde{p}_{3}^{(2)})-k_{(2,5)}^{(2)} (\tilde{p}_{2}^{(2)}-\tilde{p}_{5}^{(2)})-k_{(2,8)}^{(2)} (\tilde{p}_{2}^{(2)}-\tilde{p}_{8}^{(2)})-b_{2}^{(2)}\dot{\tilde{p}}_{2}^{(2)}.
\end{align}
Under Assumption~\ref{as:k-same} that the gains are independent of $d$,~\eqref{eq:p2-dynamics-2D} has  the same structure as~\eqref{eq:p2-dynamics-1D}. The same holds for all the vehicles, which shows that the closed-loop dynamics~\eqref{eq:closedloop-wholeplatoon} depends only on the information graph. \frqed
\end{example}

\medskip

Our goal is to analyze the closed-loop stability margin with increasing number of vehicles $N$ and to devise ways to improve it by appropriately choosing the controller gains. While in principle this can be done by numerically computing the eigenvalues of the matrix $\mbf{A}$, such a computation does not clearly reveal the dependence of stability margin on $N$, control gains, graph structure etc. For this purpose, we approximate the dynamics of the spatially discrete formation by a partial differential equation (PDE) model that is valid for large values of $N$. The PDE model is used for analysis and control design.

\subsection{PDE model of the controlled vehicle formation}\label{sec:pde_model}
For a given choice of the information graph, the $i$-th vehicle has the coordinate $\vec{i} = [i_1,i_2,\dots,i_D]^T$ in $\R^D$. We interpret $\tilde{p}_i$ as a function of the coordinate $\vec{i}$.  In the following, we consider a continuous approximation of this function to write a PDE model.
%
%
%Via a continuous approximation, we can imagine $\tilde{p}_i$ as the value of a continuous function defined over $\R^D$ evaluated at the coordinate $\vec{i}$. Our approach is to construct dynamics of such a continuous function of space so that its value at $\vec{i}$ is identical to $\tilde{p}_i$ at every time $t$.
%
%The properties of the coupled ODE system~\eqref{eq:closedloop-wholeplatoon} can then be examined in terms of the properties of the PDE that defines the dynamic evolution of this function.

%%%%%%%%%%%%%%%%%%%%%%%%%%%%%%%%%%%%%%%%%%%%%%%%%%%%%%%%%%%%%%%%%%%%%
\begin{figure}
\begin{center}
	\psfrag{x1}{$x_1$}
	\psfrag{x2}{$x_2$}
	\psfrag{i}{$i$}
	\psfrag{i1}{$i^{1+}$}
	\psfrag{i2}{$i^{2+}$}
	\psfrag{i3}{$i^{1-}$}
	\psfrag{i4}{$i^{2-}$}
	{\includegraphics[scale = 0.2]{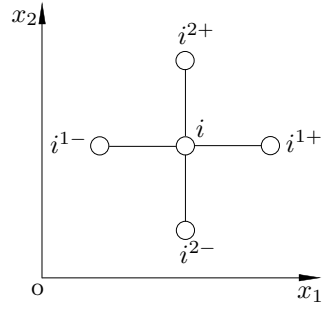}}
\end{center}
\caption{A pictorial representation of the $i$-th vehicle and its four nearby neighbors in a 2D information graph. $i^{1+}$ stands for the neighbor of the $i$-th vehicle in the $x_1$ positive direction relative to vehicle $i$, and $i^{1-}$ stands for the neighbor of the $i$-th vehicle in the $x_1$ negative direction relative to vehicle $i$. And  $i^{2+}$ and  $i^{2-}$ can be interpreted in the same way.}
\label{fig:neighbor}
\end{figure}

%For notational simplicity, we present details of the derivation for $D=2$. The generalization to other values of $D$ is straightforward, and the PDE model for the general case is summarized following the derivation for $D=2$.
For the $i$-th node with coordinate $\vec{i} = [i_1,\dots,i_D]^T$, we use $i^{d+}$ and $i^{d-}$ to denote the nodes with coordinates $[i_1,\dots,i_{d-1},i_d+1,i_{d+1},\dots,i_D]^T$ and $[i_1,\dots,i_{d-1},i_d-1,i_{d+1},\dots,i_D]^T$, respectively.
%Similarly, $i^{d-}$ denotes the node whose coordinate is
%$[i_1,\dots,i_{d-1},i_d-1,i_{d+1},\dots,i_D]^T$.
For $D=2$, a node $i$ in the interior of the graph and its four neighbors, i.e., $i^{1+}$, $i^{1-}$,$i^{2+}$, and $i^{2-}$, are shown in Figure~\ref{fig:neighbor}.   The dynamics~\eqref{eq:new_dynamics} can now  be expressed as:
\begin{align}\label{eq:2D1}
	\ddot{\tilde{p}}_{i}  =-\sum_{d=1}^{D} k_{(i,i^{d+})}(\tilde{p}_i - \tilde{p}_{i^{d+}})  -\sum_{d=1}^{D} k_{(i,i^{d-})}(\tilde{p}_i - \tilde{p}_{i^{d-}}) - b_{i}\dot{\tilde{p}}_{i},
\end{align}%
\nomenclature{$i^{1+}$}{The nearest neighbor of node $i$ in the positive $x_1$ direction with respect to $i$}%
\nomenclature{$i^{1-}$}{The nearest neighbor of node $i$ in the negative $x_1$ direction with respect to $i$}%
\nomenclature{$i^{2+}$}{The nearest neighbor of node $i$ in the positive $x_2$ direction with respect to $i$}%
\nomenclature{$i^{2-}$}{The nearest neighbor of node $i$ in the positive $x_2$ direction with respect to $i$}%
We define,
%for every $d \in \{1,\dots,D\}$,
\begin{align}\label{eq:define_kplusminus}
	k_i^{d,f+b} \eqdef & k_{(i,i^{d+})}+ k_{(i,i^{d-})} & k_i^{d,f-b} \eqdef & k_{(i,i^{d+})} - k_{(i,i^{d-})}, & d \in \{1,\dots,D\}.
\end{align}%
where the superscripts $f$ and $b$ denote \emph{front} and \emph{back}, respectively. Substituting~\eqref{eq:define_kplusminus} into~\eqref{eq:2D1}, we have
\begin{align}\label{eq:prePDExx}
	\ddot{\tilde{p}}_{i}+b_{i}\dot{\tilde{p}}_{i} =&-\sum_{d=1}^{D}\frac{k_i^{d,f+b}+ k_i^{d,f-b}}{2}(\tilde{p}_{i}-\tilde{p}_{i^{d+}}) - \sum_{d=1}^{D}\frac{k_i^{d,f+b}- k_i^{d,f-b}}{2}(\tilde{p}_{i}-\tilde{p}_{i^{d-}})
\end{align}

\nomenclature{$\vec{i}$}{The coordinate of the node $i$ in the information graph}%
\nomenclature{$\vec{x}$}{Stand for $(x_1,x_2,\dots,x_D)$, is the argument of all the continuous approximation functions}%

\begin{figure}[t]
	\begin{center}
	\psfrag{O}{$O$}
	\psfrag{x1}{$x_1$}
	\psfrag{x2}{$x_2$}
	\psfrag{c1}{$c_1$}
	\psfrag{c2}{$c_2$}
	\psfrag{l1}{$1$}	
	\psfrag{f}{$ $}	
	
\subfigure[Original lattice]{\includegraphics[scale = 0.24]{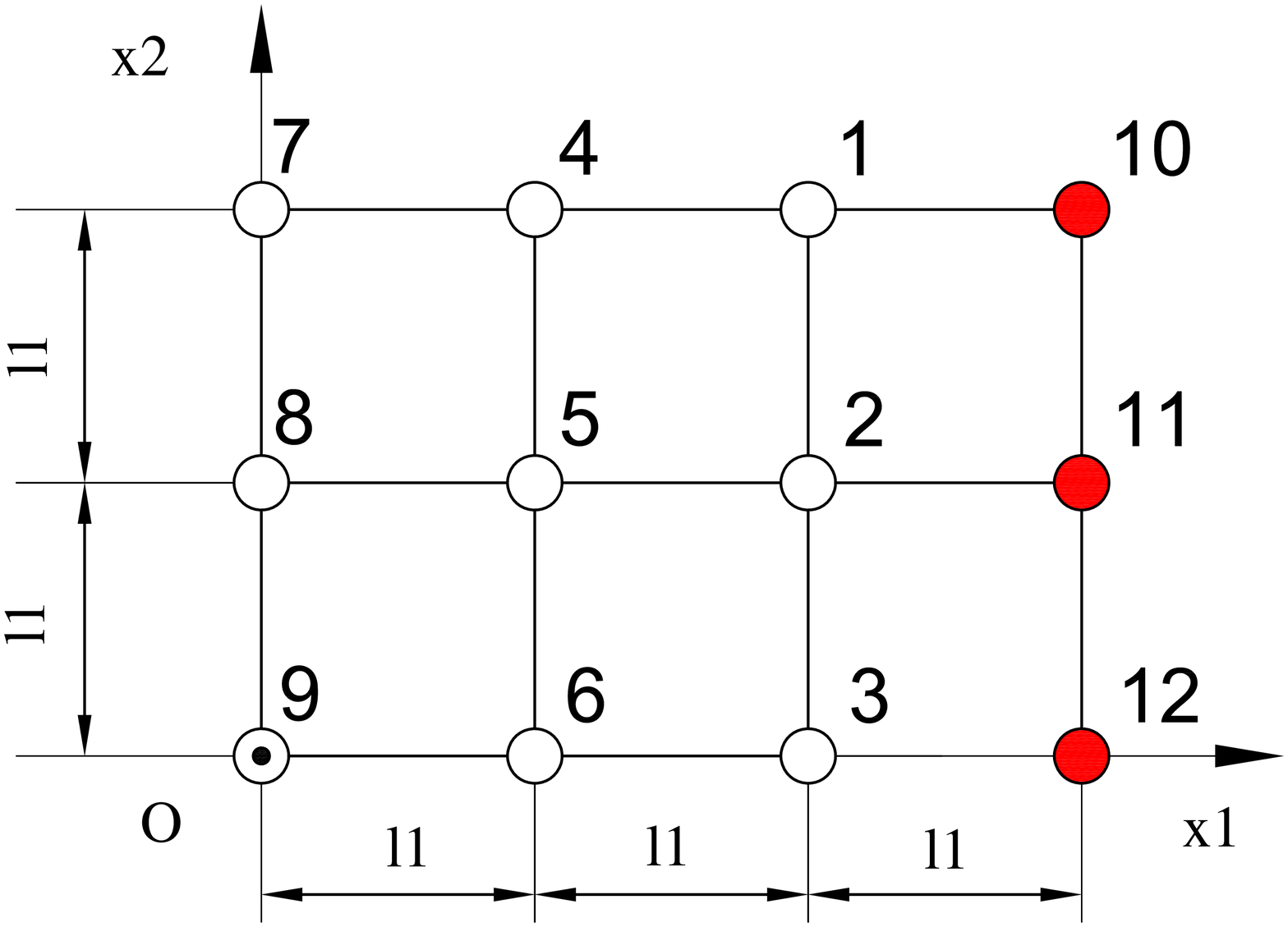}}
\subfigure[Redrawn lattice]{\includegraphics[scale = 0.25]{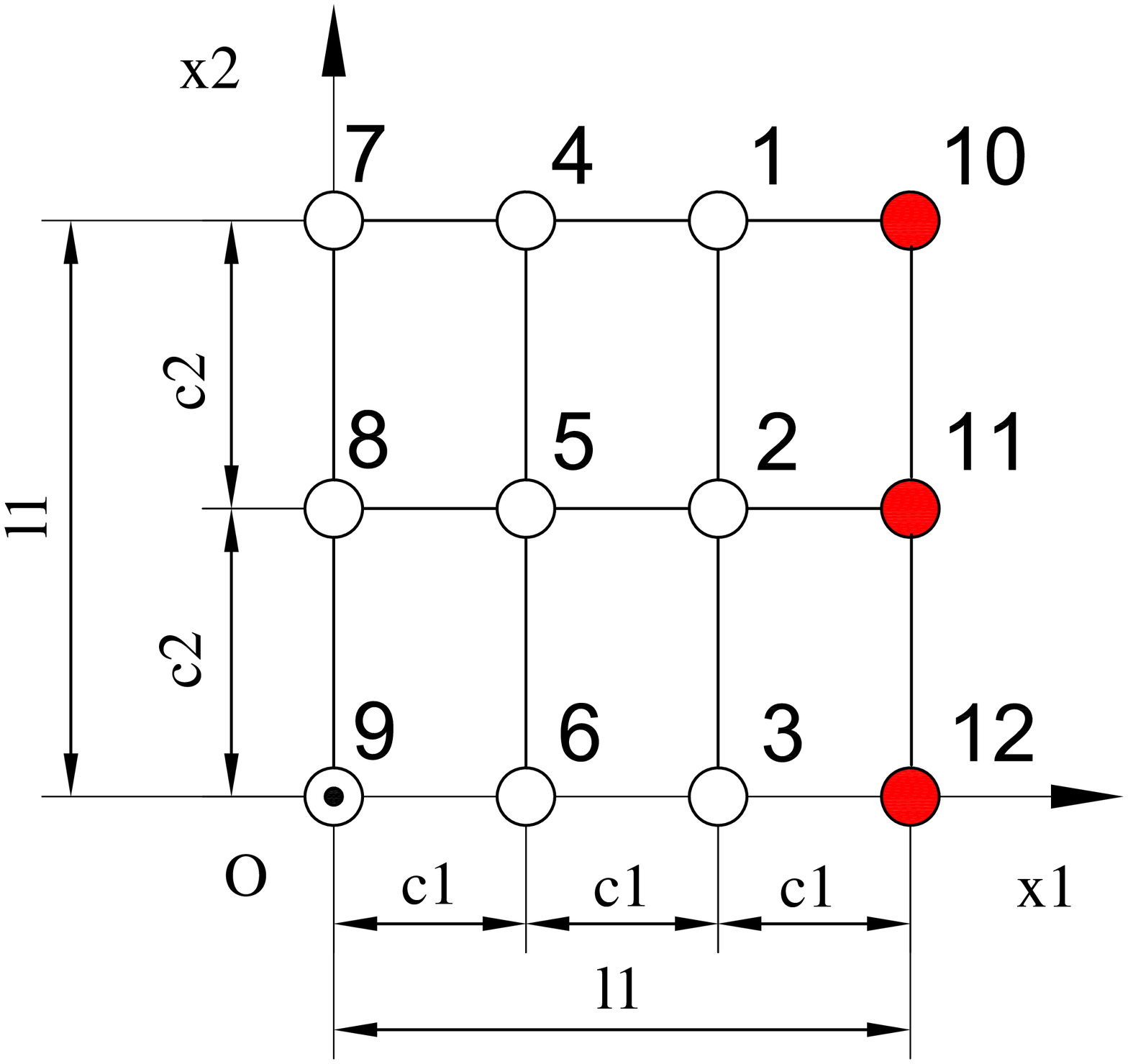}}
\subfigure[function approximation]{\includegraphics[scale = 0.25]{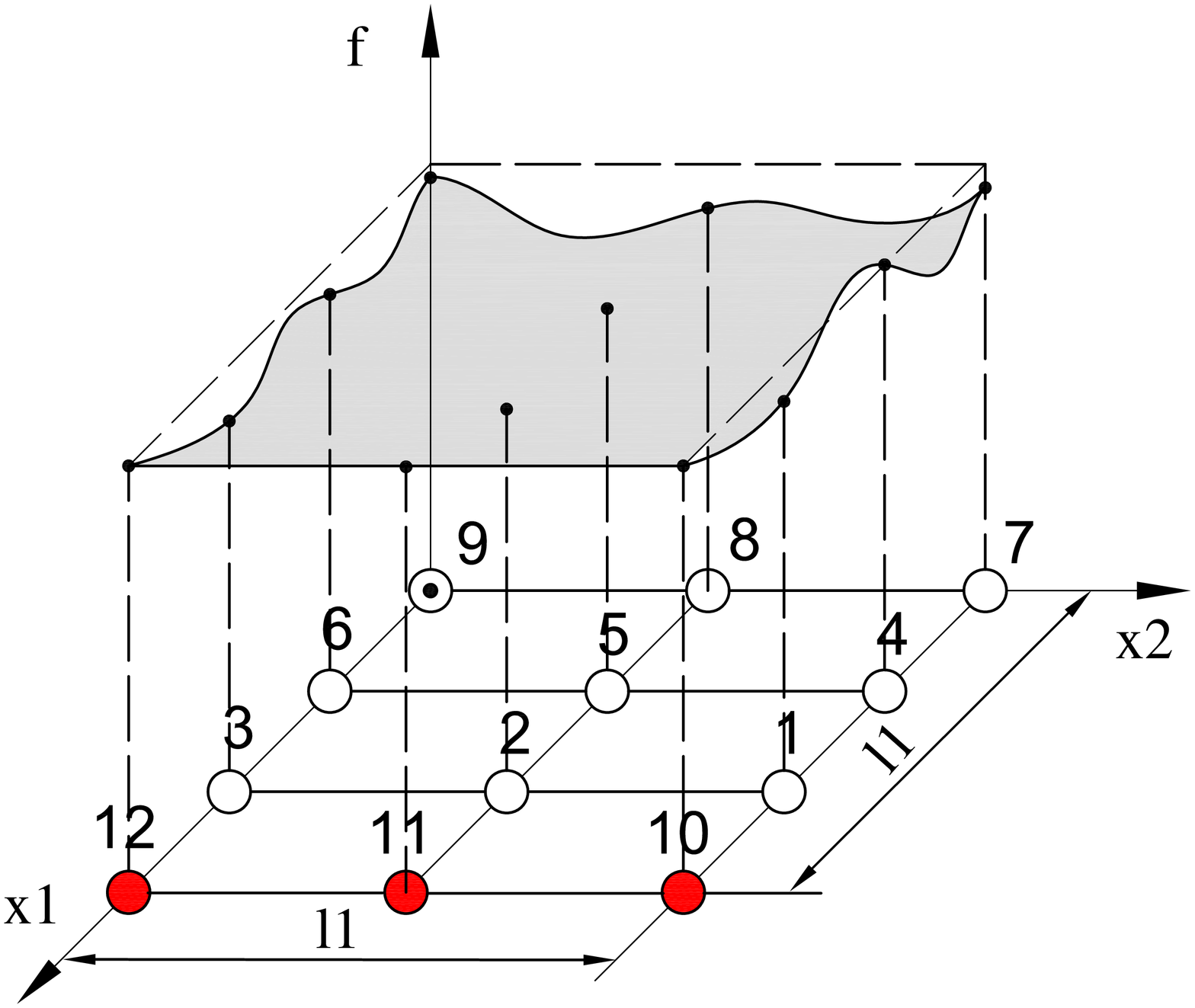}}
\end{center}
\caption{ Original lattice, its redrawn lattice and a continuous approximation of a discrete value function  defined on the redrawn lattice. (a) is a 2D information graph for a formation with $3\times3$ vehicles and $3$ reference vehicles. (b) shows a redrawn information graph of (a), so that it lies in the unit 2-cell $[0, 1]^2$. (c) gives a pictorial representation of continuous approximation of a discrete function whose values are well defined on the nodes in the redrawn lattice as shown in (b).}
\label{fig:2Dlattice-redraw}
\end{figure}

To proceed further, we first redraw the information graph in such a way so that it always lies in the unit $D$-cell $[0,1]^D$, irrespective of the number of vehicles. Note that in graph-theoretic terms, a  graph is defined only in terms of its node and edge sets. A drawing of a graph in an Euclidean space, also called an embedding~\cite{Diestel_GT:05}, is merely a convenient visualization tool. For the rest of this section, we will consider the following drawing (embedding) of the lattice $\mbf{Z}_{n_1 \times \dots \times n_D}$ in the Euclidean space $\R^D$. The Euclidean coordinate of the $i$-th node, whose ``original'' Euclidean position was $[i_1,\dots,i_D]^T$, is  now drawn at position $[i_1c_1, i_2c_2,\dots, i_Dc_D]^T$, where
\begin{align}\label{eq:cd_def}
c_d \eqdef & \frac{1}{n_d-1},  \ d=1,\dots,D.
\end{align}
Figure~\ref{fig:2Dlattice-redraw} shows an example, where the original lattice, shown in Figure~\ref{fig:2Dlattice-redraw} (a), is redrawn to fit into $[0,1]^2$, which is shown in Figure~\ref{fig:2Dlattice-redraw} (b).

%%%%%%%%%%%%%%%%%%%%%%%%%%%%%%%%%%%%%%%%%%%%%%%%%%%%%%%%%%%%%%%%%%%%%

The starting point for the PDE  derivation is to consider a function
$\tilde{p}(\vec{x},t): [0 , 1]^D \times [0, \; \infty ) \to \R$ defined over the unit D-cell in $\R^D$  that satisfies:
\begin{align}\label{eq:p_approx}
  \tilde{p}_{i}(t) = \tilde{p}(\vec{x},t)|_{\vec{x} = [i_1c_1, i_2c_2,\dots, i_Dc_D]^T}
\end{align}
Figure~\ref{fig:2Dlattice-redraw}(c) pictorially depicts the approach: functions that are defined at discrete points (the vertices of the lattice drawn in $[0, 1]^D$) will be approximated by functions that are defined everywhere in $[0,1]^D$. The original functions are thought of as samples of their continuous approximations. We formally introduce the following scalar functions  $k_{d}^f,k_{d}^b,b:[0,1]^D\to \R$ (for $d\in \{1,\dots,D\}$) defined according to the stipulation:
\begin{align}\label{eq:gain_approx}
k_{(i,i^{d+})} &=k_{d}^f(\vec{x})|_{\vec{x} = [i_1c_1, i_2c_2,\dots, i_Dc_D]^T}, \notag \\
k_{(i,i^{d-})} &=k_{d}^b(\vec{x})|_{\vec{x} = [i_1c_1, i_2c_2,\dots, i_Dc_D]^T}, \\
b_{i}&=b(\vec{x})|_{\vec{x} = [i_1c_1, i_2c_2,\dots, i_Dc_D]^T}. \notag
\end{align}
In addition, we define functions $k_{d}^{f+b},k_{d}^{f-b}:[0,1]^D\to \R$ as
\begin{align}\label{eq:f+b-etc}
  k_{d}^{f+b}(\vec{x}) & \eqdef k_{d}^{f}(\vec{x}) + k_{d}^{b}(\vec{x}),  &  k_{d}^{f-b}(\vec{x}) & \eqdef k_{d}^{f}(\vec{x}) - k_{d}^{b}(\vec{x}).
\end{align}
Due to~\eqref{eq:gain_approx}, these satisfy
\begin{align*}
k_i^{d,f+b} &=k_{d}^{f+b}(\vec{x})|_{\vec{x} = [i_1c_1, i_2c_2,\dots, i_Dc_D]^\text{T}}, &
k_i^{d,f-b} &=k_{d}^{f-b}(\vec{x})|_{\vec{x} = [i_1c_1, i_2c_2,\dots, i_Dc_D]^\text{T}}.
\end{align*}
To obtain a PDE model from~\eqref{eq:prePDExx}, we first rewrite it as
\begin{align}\label{eq:prePDE2}
	\ddot{\tilde{p}}_{i}+b_{i}\dot{\tilde{p}}_{i} =& \sum_{d=1}^{D} k_i^{d,f-b}c_d\frac{(\tilde{p}_{i^{d+}}-\tilde{p}_{i^{d-}})}{2c_d} +\sum_{d=1}^{D}\frac{k_i^{d,f+b}}{2}c_d^2\frac{(\tilde{p}_{i^{d+}}-2\tilde{p}_{i}+\tilde{p}_{i^{d-}})}{c_d^2}
\end{align}%
and then use the following finite difference approximations for every $d \in \{ 1,\dots,D\}$:
\begin{align*}
\Big[ \frac{\tilde{p}_{i^{d+}}-\tilde{p}_{i^{d-}}}{2c_d} \Big ] &=\Big [\frac{\partial \tilde{p}(\vec{x},t)}{\partial x_d} \Big ]_{\vec{x} = [i_1c_1, i_2c_2,\dots, i_Dc_D]^\text{T}}, \\
\Big [ \frac{\tilde{p}_{i^{d+}}-2\tilde{p}_{i}+\tilde{p}_{i^{d-}}}{c_d^2} \Big ]
&=\Big [\frac{\partial^2 \tilde{p}(\vec{x},t)}{\partial {x_d}^2} \Big ]_{\vec{x} = [i_1c_1, i_2c_2,\dots, i_Dc_D]^\text{T}}.
\end{align*}
We emphasize that $x_1,\dots, x_D$ above are the coordinate directions in the Euclidean space in which the information graph is drawn, which are unrelated to the coordinate axes of the  Euclidean space that the vehicles physically occupy. Substituting the expression~\eqref{eq:cd_def} for $c_d$,~\eqref{eq:prePDE2} is seen as a finite difference approximation of the following PDE:
\begin{align}\label{eq:PDE}
	\Big(\frac{\partial^2}{\partial t^2}  + b(\vec{x})\frac{\partial}{\partial t} \Big)\tilde{p}(\vec{x},t) & = \sum_{d=1}^{D}\Big (\frac{k_d^{f-b}(\vec{x})}{n_d-1}\frac{\partial}{\partial x_d}+\frac{k_d^{f+b}(\vec{x})}{2{(n_d-1)}^2}\frac{\partial ^2}{\partial {x_d}^2} \Big ) \tilde{p}(\vec{x},t),
\end{align}
The boundary conditions of PDE~\eqref{eq:PDE} depend on the arrangement of reference vehicles in the information graph. If there are reference vehicles on the boundary, the boundary condition is of Dirichlet type.  If there are no reference vehicles, the boundary condition is of the Neumann type.
% Under Assumption~\ref{as:L1}, the boundary conditions are of the Dirichlet type on that face of the unit cell where the reference vehicles are
% \begin{align}\label{eq:BC}
% \tilde{p}(1,x_2,\dots,x_Dt) &= 0.
% \end{align}
% All other boundary conditions are of the Neumann type. For example, in case of $D=2$, we have Nuemann boundary conditions on three of the four boundaries:
% \begin{align*}
% \frac{\partial \tilde{p}}{\partial x_1}(0,x_2,t) &= 0, & \frac{\partial \tilde{p}}{\partial x_2}(x_1,0,t) &= 0, & \frac{\partial \tilde{p}}{\partial x_2}(x_1,1,t) &= 0.
% \end{align*}
Under Assumption~\ref{as:L1}, the boundary conditions are of the Dirichlet type on that face of the unit cell where the reference vehicles are, and Neumann on all other faces:
\begin{equation}\label{eq:BC}
  \begin{split}
\tilde{p}(1,x_2,\dots,x_D,t) &= 0, \quad \frac{\partial \tilde{p}}{\partial x_1}(0,x_2,\dots,x_D,t) = 0, \\
\frac{\partial \tilde{p}}{\partial x_d}(\vec{x},t) &= 0,\quad \vec{x}  = [x_1,\dots,x_{d-1},0 \text { or } 1,x_{d+1},\dots,x_D]^T, \quad (d>1).
 \end{split}
\end{equation}
If other arrangements of reference vehicles are used, the boundary conditions may be different. For future use, we rewrite the PDE~\eqref{eq:PDE} as
\begin{align}\label{eq:PDE_D}
	\Big(\frac{\partial^2}{\partial t^2}  +b(\vec{x})\frac{\partial}{\partial t} \Big)\tilde{p}(\vec{x},t) =\scr{L}(\frac{\partial}{\partial x_d},\frac{\partial ^2}{\partial {x_d}^2}) \tilde{p}(\vec{x},t),
\end{align}
where the linear operator $\scr{L}$ is defined as
\begin{align}\label{eq:laplacian}
\scr{L}(\frac{\partial}{\partial x_d},\frac{\partial ^2}{\partial {x_d}^2}) \eqdef \sum_{d=1}^{D} \frac{k_d^{f-b}(\vec{x})}{n_d-1}\frac{\partial}{\partial x_d}+\frac{k_d^{f+b}(\vec{x})}{2(n_d-1)^2}\frac{\partial ^2}{\partial {x_d}^2}.
\end{align}
It can be verified in a straightforward manner that the PDE~\eqref{eq:PDE} yields the original set of coupled ODEs~\eqref{eq:2D1} upon discretization.

%%%%%%%%%%%%%%%%%%%%%%%%%%%%%%%%%%%%%%%%%%%%%%%%%%%%%%%%%%%%%%%%%%%%%%
\section{Stability margin with symmetric control}\label{sec:instability-analysis}
%%%%%%%%%%%%%%%%%%%%%%%%%%%%%%%%%%%%%%%%%%%%%%%%%%%%%%%%%%%%%%%%%%%%%%
\subsection{PDE-based analysis of stability margin}
Recall that in case of symmetric control we have
\begin{align*}
	k_{(i,j)} & =k_0, \quad \forall (i,j)\in \E, & b_i=b_0, \quad \forall i \in\V,
\end{align*}
where $k_0$ and $b_0$ are positive scalars. In this case, using the notation in~\eqref{eq:define_kplusminus} and~\eqref{eq:gain_approx}, we have
\begin{align*}
	k_d^{f+b}(\vec{x}) &= 2k_0, & k_d^{f-b}(\vec{x}) &=0, & b(\vec{x}) &=b_0, \quad d = 1,\dots,D.
\end{align*}
The PDE~\eqref{eq:PDE_D} simplifies to a damped wave equation:
\begin{align}\label{eq:PDE_D_simplified}
	\Big(\frac{\partial^2}{\partial t^2} + b_0\frac{\partial}{\partial t} \Big)\tilde{p}(\vec{x},t) =\scr{L}_0(\frac{\partial ^2}{\partial {x_d}^2})\tilde{p}(\vec{x},t),
\end{align}
where $\scr{L}_0(\vec{x})$ is the Laplacian operator:
\begin{align}\label{eq:new_laplacian}
\scr{L}_0(\frac{\partial ^2}{\partial {x_d}^2})=a_{1}^2\frac{\partial ^2}{\partial {x_1}^2}+a_{2}^2\frac{\partial ^2}{\partial {x_2}^2}+\dots  +a_{D}^2\frac{\partial ^2}{\partial {x_D}^2},
\end{align}
where
\begin{align}\label{eq:wave-speed-def}
a_{d}^2 \eqdef  \frac{k_0}{(n_d-1)^2} , \quad d=1,\dots, D,
\end{align}
are the \emph{wave-speeds}. The closed-loop eigenvalues of the PDE model require consideration of the boundary value problem
\begin{align}   \label {eq:eigen_problem}
\scr{L}_0(\frac{\partial ^2}{\partial {x_d}^2})\phi(\vec{x}) = -\lambda \phi(\vec{x}),
\end{align}
For the given boundary condition of~\eqref{eq:BC}, the eigenvalues (different from the eigenvalue of PDE) and eigenfunctions of $\scr{L}_0$ are respectively given by
\begin{align}\label{eq:lamda-symmetric}
 %\begin{split}
	\lambda_{\vec{l}}  & = \Big(\frac{(2l_1-1)\pi}{2} \Big)^2 a_{1}^2 +(l_2\pi)^2 a_{2}^2+\dots+ (l_D\pi)^2 a_{D}^2 \notag \\
                & = \pi^2 k_0\Big( \frac{(2l_1-1)^2}{4(n_1-1)^2} + \frac{l_2^2}{(n_2-1)^2} +\dots+ \frac{l_D^2}{(n_D-1)^2} \Big), \notag \\
\phi_{\vec{l}}(\vec{x}) &=\cos \big( \frac{(2l_1-1)\pi x_1}{2} \big ) \cos(l_2\pi x_2)\cdots \cos(l_D\pi x_D).
%\end{split}
\end{align}
where $l_1 \in\{ 1,2,\dots$ \} and $l_2,\dots,l_D \in \{ 0,1,2,\dots \}$.
We use the notation $\vec{l}=(l_1,\dots,l_D)$ to denote the wave vector and $\lambda_{\vec{l}}$, $\phi_{\vec{l}}(\vec{x})$ to denote the associated eigenvalue and eigenfunction given by~\eqref{eq:lamda-symmetric}. After taking a Laplace transform of both sides of  the PDE~\eqref{eq:PDE_D_simplified} with respect to $t$, we get $\scr{P} \eta(\vec{x},s)=0$ where $\scr{P}\eqdef s^2+b_0s-\scr{L}_0$ and $\eta(\vec{x},s)=\sum \phi_{\vec{l}}(\vec{x}) \alpha_{\vec{l}}(s)$ is the Laplace transform of $\tilde{p}(\vec{x},t)$ with $\alpha_{\vec{l}}(s)$ being its weights. Note that $\phi_{\vec{l}}$ is also the $\vec{l}$-th basis of the null space of operator $\scr{P}$. The eigenvalues of the PDE turn out to be the roots of the characteristic equation:
\begin{align}\label{eq:s-quad}
s^2 + b_0 s + \lambda_{\vec{l}} = 0,
\end{align}
where $s$ as the Laplace variable and $\lambda_{\vec{l}}$ is an eigenvalue of~\eqref{eq:eigen_problem}.  
%For a given wave vector $\vec{l}$, and corresponding eigenvalue $\lambda_{\vec{l}}$. 
The two roots of~\eqref{eq:s-quad} are
\begin{align}\label {eq:eigenvalue}
 s_{\vec l}^{\pm} \eqdef \frac{-b_0\pm\sqrt{b_0^2- 4\lambda_{\vec{l}}}}{2}.
\end{align}
We call $s_{\vec l}^{\pm}$ the $\vec{l}$-th pair of eigenvalues. %, with the understanding that $l$ really refers to the  wave vector $\vec{l}$.
If the discriminant in~\eqref{eq:eigenvalue} is positive, both the eigenvalues are real-valued. In this case, $s_{\vec{l}}^{+}$ is closer to the origin than $s_{\vec{l}}^{-}$; so we call $s_{\vec{l}}^{+}$ the $\vec{l}$-th {\em less-stable} eigenvalue.  The \emph{least stable} eigenvalue is the one among them that is closest to the imaginary axis, and the stability margin is the absolute value of its real part:
\begin{align}\label{eq:s-min}
  s_\mrm{min} & = \min_{\vec{l}} s_{\vec{l}}^+, & S \eqdef |Re(s_\mrm{min})|.
\end{align}
Provided each of the $n_d$'s are large so that the PDE~\eqref{eq:PDE} with the boundary condition~\eqref{eq:BC} is an accurate approximation of the (spatially) discrete formation dynamics~\eqref{eq:closedloop-wholeplatoon} under Assumption~\ref{as:L1}, the least stable eigenvalue of the PDE~\eqref{eq:PDE_D} provides information on the stability margin (see Definition~\ref{def:stabiltity_margin}) of the closed-loop formation dynamics. We are now ready to prove the Theorem~\ref{thm:symmetric} that was stated in Section~\ref{sec:results}.
\begin{proof-theorem}{\ref{thm:symmetric}}
Consider the eigenvalue problem for PDE~\eqref{eq:PDE_D_simplified} with mixed Dirichlet and Neumann boundary conditions~\eqref{eq:BC}. Since the less stable eigenvalues are given by $s^+_{\vec{l}} = \frac{1}{2}(-b_0 + \sqrt{b_0^2 - 4\lambda_{\vec{l}}})$. If the discriminant $b_0^2 - 4\lambda_{\vec{l}}$ is positive, both of the eigenvalues are real-valued. In this case, $s^+_{\vec{l}}$ is closer to the origin than $s^-_{\vec{l}}$; so we call $s^+_{\vec{l}}$ the $\vec{l}$-th \emph{less-stable} eigenvalue. It follows from~\eqref{eq:eigenvalue} that the least stable among them is the one that  is obtained by minimizing $\lambda_{\vec{l}}$ over the $D$-tuples $(l_1,\dots,l_D)$.  Using~\eqref{eq:lamda-symmetric}, this minimum is achieved at $l_1=1,l_2=\dots=l_D=0$, where $\lambda(1,0,\hdots,0) = 0.25 \pi^2k_0/(n_1-1)^2$. Therefore,
 \begin{align*}%\label{eq:smin-L0}
  s_\text{min}= \min_{(l_1,\dots,l_D)} s^+ =  \frac{b_0}{2}(-1 + \left( 1 - \frac{\pi^2k_0}{b_0^2(n_1-1)^2}\right)^{1/2}) = -\frac{\pi^2k_0}{4b_0 (n_1-1)^2} + O(\frac{1}{n_1^4})
 \end{align*}
where the last equality holds when $n_1 \gg 1+\frac{\pi\sqrt{k_0}}{b_0}$. Due to the definition of stability margin~\eqref{eq:s-min}, the result follows immediately from the equation above. 
\frQED
\end{proof-theorem}

%%%%%%%%%%%%%%%%%%%%%%%%%%%
\begin{figure}[h]
\psfrag{Real}{Real}
\psfrag{Imaginary}{Imaginary}
\psfrag{SSM}{SSM}
\psfrag{PDE}{PDE}
\begin{center}
{\includegraphics[scale = 0.4]{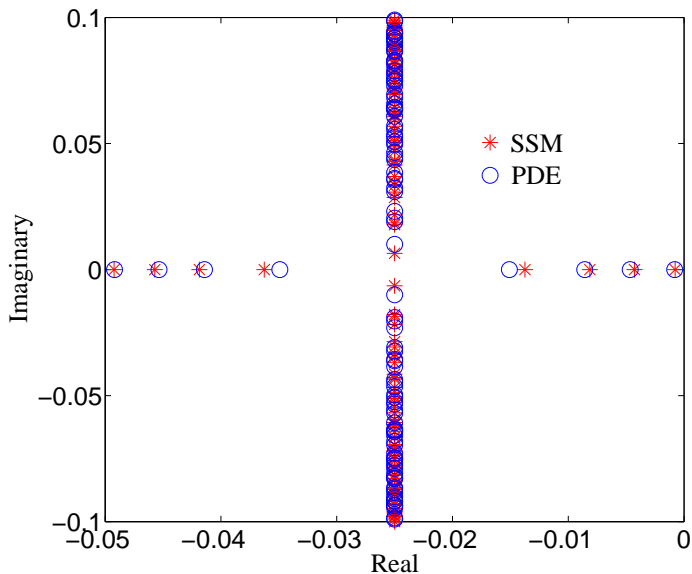}}
\end{center}
\caption{Numerical comparison of closed-loop eigenvalues with symmetric control predicted by the  state-space model (SSM)~\eqref{eq:closedloop-wholeplatoon} and PDE model~\eqref{eq:PDE_2D} with mixed Dirichlet-Neumann boundary conditions. Eigenvalues shown are for a square information graph with $26\times 25$ nodes ($625$ vehicles), and the control gains used are $k_0=0.01$, $b_0=0.05$. Only a few eigenvalues are compared in the figure. PDE eigenvalues are computed using a Galerkin method~\cite{Canuto}.}\label{fig:Eig_comparison}
\end{figure}
%%%%%%%%%%%%%%%%%%%%%%%%%%%

%%%%%%%%%%%%%%%%%%%%%%%%%%%%%%%%%%%%%%%%%%%%%%%%%%%%%%%%%%%%%%%%%%
\subsection{Numerical comparison of eigenvalues between SSM and PDE}
\begin{figure}[h]
\psfrag{N}{$\ N$}
\psfrag{S}{$S$}
\psfrag{SSM}{SSM}
\psfrag{PDE}{PDE}
\psfrag{Corollary 1}{Corollary 1}
\begin{center}
{\includegraphics[scale = 0.4]{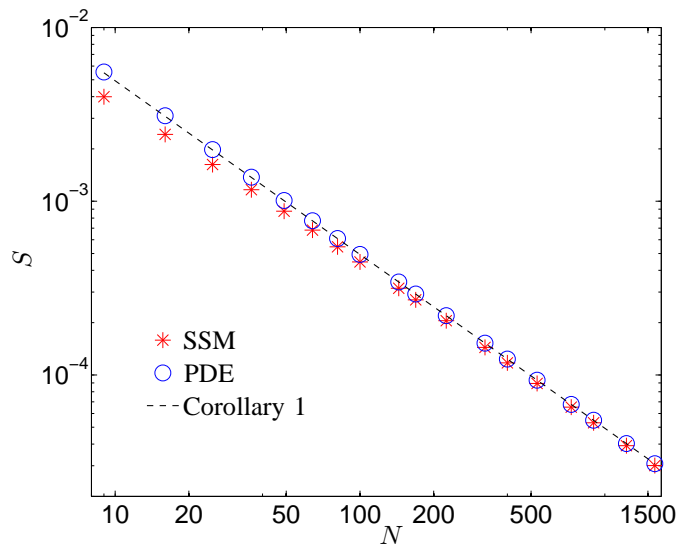}}
\end{center}
\caption{The stability margin of the closed-loop formation dynamics with symmetric control ($k_0=0.01$ and $b_0=0.5$) as a function of number of vehicles: the legends of SSM, PDE and Corollary~\ref{cor:symmetric-square} stand for the stability margin computed from the  state space model, from the PDE model, and from the asymptotic formula~\eqref{eq:stability-margin-symmetric-square} in Corollary~\ref{cor:symmetric-square}.
}\label{fig:Eig-trend}
\end{figure}

We now present numerical computations that corroborates the PDE-based analysis.  We consider a $26 \times 25$ square two-dimensional information graph with symmetric control. The gains are
\begin{align*}
	k_{(i,i^{1+})}=k_{(i,i^{1-})}=k_{(i,i^{2+})}=k_{(i,i^{2-})}=k_0=0.01,\ b_i=b_0=0.05.
\end{align*}
The associated PDE model is given by
\begin{align}\label{eq:PDE_2D}
	(\frac{\partial^2}{\partial t^2}+b_0\frac{\partial}{\partial t})\tilde{p}(\vec{x},t)=\Big (\frac{k_0}{(n_1-1)^2} \frac{\partial^2}{\partial x_1^2} +\frac{k_0}{(n_2-1)^2} \frac{\partial^2}{\partial x_2^2}\Big )\tilde{p}(\vec{x},t).
\end{align}
The eigenvalues of the state matrix $\mbf A$ in~\eqref{eq:closedloop-wholeplatoon} are compared against the eigenvalues of the PDE~\eqref{eq:PDE_2D} with mixed Neumann-Dirichlet boundary conditions in Figure~\ref{fig:Eig_comparison}. The eigenvalues of the PDE are computed numerically using a Galerkin method with Fourier basis~\cite{Canuto}.  The comparison in Figure~\ref{fig:Eig_comparison} shows that the PDE eigenvalues match the state-space model eigenvalues well, especially the ones close to the imaginary axis.  Figure~\ref{fig:Eig-trend} shows, as a function of $N$,  the stability margin computed from the PDE and the state-space model. The prediction from the asymptotic formula~\eqref{eq:stability-margin-symmetric-square} in Corollary~\ref{cor:symmetric-square} is also shown. We see from Figure~\ref{fig:Eig-trend} that the least stable eigenvalue of the closed-loop is well captured by both the PDE model as well as the asymptotic formula~\eqref{eq:stability-margin-symmetric-square} that is derived from analysis of the PDE.

%%%%%%%%%%%%%%%%%%%%%%%%%%%
%%%%%%%%%%%%%%%%%%%%%%%%%%%

% The numerical plots confirm the prediction of the analysis. The least stable eigenvalue with symmetric control approaches $0$ as $O(\frac{1}{N^{2/D}})$ for a vehicle formation with $D$-dimensional square information graph, irrespective of the fixed constant values of the gains $k_0$ and $b_0$.  This means that the closed-loop system has an arbitrary small stability margin as a function of $N$.  Moreover, this conclusion is independent of the control gains within the class of symmetric control.  In the following section, a mistuning design method is proposed to partially ameliorate this progressive loss of stability margin.

%%%%%%%%%%%%%%%%%%%%%%%%%%%%%%%%%%%%%%%%%%%%%%%%%%%
%%%%%%%%%%%%%%%%%%%%%%%%%%%%%%%%%%%%%%%%%%%%%%%%%%%
%%%%%%%%%%%%%%%%%%%%%%%%%%%%%%%%%%%%%%%%%%%%%%%%%%%
\section{Mistuning-based control design}\label{sec:mistuning}
With symmetric control, one obtains an $O(\frac{1}{n_1^2})$ scaling law for the least stable eigenvalue
because the coefficient of the $\frac{\partial ^2}{\partial {x_1}^2}$ term  in the PDE~\eqref{eq:PDE_D_simplified} is $O(\frac{1}{n_1^2})$ and the coefficient of the $\frac{\partial}{\partial x_1}$ term  is $0$. Any asymmetry
between the forward and the backward gains will lead
to non-zero $k_d^{f-b}(\vec{x})$ and the presence of $O(\frac{1}{n_1})$ term as coefficient
of $\frac{\partial}{\partial x_1}$. By a judicious choice of asymmetry, there is
thus a potential to improve the stability margin from $O(\frac{1}{n_1^2})$
to $O(\frac{1}{n_1})$. The subsequent analysis shows that this is indeed so, and a control design is proposed to achieve the $O(\frac{1}{n_1})$ trend. One should also note that this insight into the control design problem is difficult to obtain from the  examination of the state matrix $\mbf{A}$.

\subsection{Reducing loss of stability by mistuning}
In this section, we consider the problem of designing the control gain functions $k_d^{f}(x)$ and $k_d^{b}(x)$ so as to improve the stability margin over symmetric control.
Specifically, we consider the problem of minimizing the least-stable eigenvalue $s_\mrm{min}$ of the PDE~\eqref{eq:PDE_D} by changing the control gains slightly (mistuned) from their values in the symmetric case. We begin by considering the forward and backward position feedback gain profiles
\begin{align}\label{eq:kf-kf-tilde}
	k_d^f(\vec{x})&=k_0+\varepsilon \tilde{k}_d^f(\vec{x}), &  k_d^b (\vec{x})=k_0+\varepsilon \tilde{k}_d^b(\vec{x}),
\end{align}
where $\varepsilon>0$ is a small parameter signifying the amount of mistuning
and $\tilde{k}_d^f(\vec{x}),\ \tilde{k}_d^b(\vec{x})$ are functions defined over $[0,1]^D$ that capture gain  perturbation from the nominal value $k_0$. Define
\begin{align}\label{eq:ks-km-def}
\tilde{k}_d^s(\vec{x}) & \eqdef \tilde{k}_d^f(\vec{x})+ \tilde{k}_d^b(\vec{x}), &
\tilde{k}_d^m(\vec{x}) & \eqdef \tilde{k}_d^f(\vec{x})- \tilde{k}_d^b(\vec{x}).
\end{align}
Due to the definition of $k_d^{f+b}$ and  $k_d^{f+b}$ in~\eqref{eq:f+b-etc}, we have
\begin{align*}
k_d^{f+b}(\vec{x})&=2k_0+\varepsilon \tilde{k}_d^s(\vec{x}), & k_d^{f-b}(\vec{x}) & =\varepsilon \tilde{k}_d^m(\vec{x}).
\end{align*}
The mistuned version of the PDE~\eqref{eq:PDE_D} is thus given by
\begin{align}\label{eq:mistuned_PDE_D}
	\Big (\frac{\partial^2}{\partial t^2}+b_{0}\frac{\partial}{\partial t}  \Big)\tilde{p}(\vec{x},t)=
&\sum_{d=1}^{D}{\Big (  \frac{k_0}{(n_d-1)^2} \frac{\partial^2}{\partial x_d^2}} \Big) \tilde{p}(\vec{x},t)+  \varepsilon \sum_{d=1}^{D}{ \Big ( \frac{\tilde{k}_d^{s}(\vec{x})}{2(n_d-1)^2} \frac{\partial^2}{\partial x_d^2}+ \frac{\tilde{k}_d^{m}(\vec{x})}{n_d-1}\frac{\partial}{\partial x_d}} \Big )\tilde{p}(\vec{x},t).
\end{align}
We study the problem of improving the stability margin by judicious
choice of $\tilde{k}_d^{s}(\vec{x})$ and $\tilde{k}_d^{m}(\vec{x})$ while keeping the gains $\tilde{k}_d^f(\vec{x})$ and $\tilde{k}_d^b(\vec{x})$ within certain pre-specified bounds. The results of our investigation, described in the following sections, provide a systematic
framework for designing control gains in the formation by introducing
small changes to the symmetric design.

\medskip

To design the ``mistuning'' profiles
$\tilde{k}_d^{s}(\vec{x})$ and $\tilde{k}_d^{m}(\vec{x})$ to minimize the least stable eigenvalue $s_\mrm{min}$, we first obtain an explicit asymptotic formula for the eigenvalues when $\varepsilon$ is small. The result is presented in the following theorem. The proof appears in the Appendix.
%%%%%%%%%%%%%%%%%%%%%%%%%%%%%%%%%%%%%%%%%%%%%%%%%%%%%%%%%%%%%%%%%%%%%%%
\begin{theorem}\label{thm:theorem1}
	Consider the eigenvalue problem of the mistuned PDE~\eqref{eq:mistuned_PDE_D} with mixed Dirichlet and Neumann boundary condition~\eqref{eq:BC}. The least stable eigenvalue is given by the following formula that is valid when $\varepsilon\rightarrow 0$ and $n_1,n_2,\dots, n_D \rightarrow\infty$:
\begin{align}\label{eq:less_stable_eig_mistuning_D}
   s_\mrm{min}  &=s_\mrm{min}^{(0)}  -   \varepsilon \frac{\pi}{2 b_0 (n_1-1)} \int_{0}^{1} \tilde{k}_1^{m}(\vec{x})\sin \big(\pi x_1 \big ) \ dx_1  -  \varepsilon \frac{\pi^2}{4b_0(n_1-1)^2}\int_{0}^{1} \tilde{k}_1^{s}(\vec{x}) \cos^2(\frac{\pi}{2} x_1) \ dx_1 + O(\varepsilon^2),
\end{align}
where $s_\mrm{min}^{(0)}$ is the least stable eigenvalue without mistuning, i.e., of PDE~\eqref{eq:PDE_D_simplified} with the same boundary conditions. \frqed
\end{theorem}
%%%%%%%%%%%%%%%%%%%%%%%%%%%%%%%%%%%%%%%%%%%%%%%%%%%%%%%%%%%%%%%%%%%%%%%%%
 
\medskip

It follows from Theorem~\ref{thm:theorem1} that to minimize the least stable eigenvalue, one needs to choose only $\tilde{k}^m_1(\vec{x})$ carefully; all other $\tilde{k}^m_d$'s and all $\tilde{k}^s_d$'s can be set to $0$. The reason is that only $\tilde{k}^m_1(\vec{x})$ and $\tilde{k}^s_1(\vec{x})$ affect the least stable eigenvalue, and the term involving $\tilde{k}_1^s(\vec{x})$ is of order $1/(n_1-1)^2$, whereas the term involving $\tilde{k}_1^m(\vec{x})$ is of order $1/(n_1-1)$. For large $n_1$ the effect of the function  $\tilde{k}_1^m(\vec{x})$ on the least stable eigenvalue will be far greater than that of $\tilde{k}_1^s(\vec{x})$. Therefore, we choose
\begin{align*}
\tilde{k}^s_d(\vec{x}) & \equiv 0 \equiv \tilde{k}^m_d(\vec{x}) \quad \text{ for } \;\; d=2,\dots,D, & \text{ and } & & \tilde{k}^s_1(\vec{x}) & \equiv 0.
\end{align*}
This means that the perturbations to the ``front'' and ``back'' gains satisfy $\tilde{k}_d^f(\vec{x}) = \tilde{k}_d^b(\vec{x})=0$ for $d=2,\dots,D$. For  $d=1$, the choice $\tilde{k}_1^s(\vec{x}) \equiv 0$ leads to
\begin{align*}
\tilde{k}_1^f(\vec{x})=-\tilde{k}_1^b(\vec{x}) \Leftrightarrow \tilde{k}_{1}^{m}(\vec{x})=2\tilde{k}_1^f(\vec{x}).
\end{align*}
The most beneficial gains can now be readily obtained from Theorem~\ref{thm:theorem1}. To minimize the least stable eigenvalue with $\tilde{k}^s_1(\vec{x}) \equiv 0$, we should choose $\tilde{k}^m_1(\vec{x})$ to make the integral $\int_{0}^{1} \tilde{k}^m_1(\vec{x})\sin(\pi x_1) dx_1$  as large as possible, which is achieved by setting $\tilde{k}^m_1(\vec{x})$ to be the largest possible value everywhere in the unit cell. This result is summarized in the next Corollary.

%%%%%%%%%%%%%%%%%%%%%%%%%%%%%%%%%%%%%%%%%%%%%%%%%%%%%%%%%%%%%%%%%%%%%%%%%%%%%%%%%%%%%% corollary on mistuing gain design
\begin{corollary}\label{cor:opt_profile}
Consider the problem of minimizing the least-stable eigenvalue of the PDE~\eqref{eq:mistuned_PDE_D} with mixed Dirichlet and Neumann boundary condition~\eqref{eq:BC} in the limit as $\varepsilon\rightarrow 0$ by choosing $\tilde{k}_1^f(\vec{x}),\ \tilde{k}_1^b(\vec{x})\in L^\infty([0,1])$ with the constraint that $\|\tilde{k}_1^f(\vec{x})\|_{\infty}= \|\tilde{k}_1^b(\vec{x})\|_{\infty} = 1$, where $\| \cdot\|_\infty$ denotes the sup-norm. The solution to this optimization problem is given by
\begin{align*}% \label{eq:opt_profile}
	\tilde{k}_1^f(\vec{x}) & =1,\ \tilde{k}_1^b(\vec{x})=-1, \quad \forall \vec{x} \in [0,1]^D. \tag*{\frqed}
\end{align*}
\end{corollary}
%%%%%%%%%%%%%%%%%%%%%%%%%%%%%%%%%%%%%%%%%%%%%%%%%%%%%%%%%%%%%%%%%%%%%%%
The proof of Theorem~\ref{thm:mistuned} now follows in a straightforward manner from Corollary~\ref{cor:opt_profile}.
\begin{proof-theorem}{\ref{thm:mistuned}}
Note that ensuring $|k_{(i,j)} - k_0|<\varepsilon$ in the formation is equivalent to keeping $|k_d^f - k_0|\leq \varepsilon$ and  $|k_d^b - k_0| \leq \varepsilon$ for $d=1,\dots,D$ in the PDE domain; cf.~\eqref{eq:gain_approx}. This is equivalent to keeping $\|\tilde{k}^f_d\|_\infty \leq 1$ and $\|\tilde{k}^b_d\|_\infty \leq 1$ for each $d$; cf.~\eqref{eq:kf-kf-tilde}. In this case, the optimal gains are those given in Corollary~\ref{cor:opt_profile}. It follows from~\eqref{eq:gain_approx} that the optimal gains for the vehicles are
\begin{align*}
  k_{(i,i^{1+})} & = (k_0 + \varepsilon \tilde{k}_1^f(\vec{x}))|_{\vec{x} = [i_1c_1,\dots,i_D c_D]^T} = k_0 + \varepsilon , \quad \forall \ i \in \V\\
  k_{(i,i^{1-})} & = (k_0 + \varepsilon \tilde{k}_1^b(\vec{x}))|_{\vec{x} = [i_1c_1,\dots,i_D c_D]^T} = k_0 - \varepsilon , \quad \forall \ i \in \V\\
  k_{(i,i^{d+})} & = k_{(i,i^{d-})} = k_0, \quad d>1, \forall \ i \in \V.
\end{align*}
The resulting least stable eigenvalue is, from Theorem~\ref{thm:theorem1},
\begin{align*}
  s_\mrm{min} & = -\varepsilon \frac{\pi}{b_0 (n_1-1)}\int_0^1 \sin(\pi x_1)dx_1 + s_\mrm{min}^{(0)} = -\varepsilon \frac{2}{b_0(n_1-1)} + O(\frac{1}{n_1^2}),
\end{align*}
since $s_\mrm{min}^{(0)} = O(1/n_1^2)$. The result follows upon taking absolute value of $s_\mrm{min}$.\frQED
\end{proof-theorem}

%The result shows that even with an arbitrarily small amount of
%mistuning $\epsilon$, one can improve the closed-loop platoon damping
%by a large amount, especially for large values of $N$. The least stable
%eigenvalue asymptotes to 0 as $O(\frac{1}{n_1})$ in the mistuned
%case as opposed to $O(\frac{1}{n_1^2})$ in the symmetric case.

\subsection{Comparison of eigenvalues between mistuned SSM and PDE}
\begin{figure}[t]
\psfrag{N}{$\ N$}
\psfrag{S}{$S$}
\psfrag{Nominal SSM}{Nominal SSM}
\psfrag{Nominal PDE}{Nominal PDE}
\psfrag{Mistuned SSM}{Mistuned SSM}
\psfrag{Mistuned PDE}{Mistuned PDE}
\psfrag{Corollary 3}{Corollary 3}
\begin{center}
{\includegraphics[scale = 0.4]{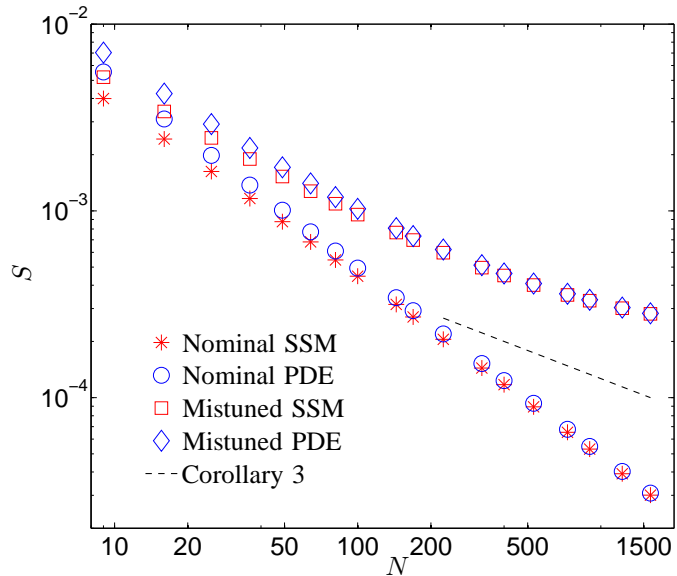}}
\end{center}
\caption{Stability margin improvement by mistuning for a vehicle formation with 2D square information graph. The nominal control gains are $k_0=0.01$, $b_0=0.5$, and the mistuned gains used are the ones shown in Figure~\ref{fig:kopt-mistuned}, with $\varepsilon=0.001$. The symbol $N$ on the $x$-axis is the number of vehicles and $S$ on the $y$-axis is the stability margin. The legends ``Nominal SSM'' and ``Nominal PDE'' stand for the stability margin computed from the state-space model and the PDE model, respectively, with symmetric control. The legends ``Mistuned SSM'' and ``Mistuned PDE'' stand for the stability margin computed from the state-space model and PDE model, respectively, with mistuned control. We see that the (i) the PDE model predicts the stability margin quite accurately, and (ii) the stability margin is improved significantly by mistuning control design even with  $\pm 10\%$ variation from the symmetric gains, especially for large $N$.}\label{fig:mistuned-lseig-compare}
\end{figure}

Figure~\ref{fig:mistuned-lseig-compare} depicts the numerically obtained mistuned and nominal eigenvalues for both the PDE and state-space model for a 2D square information graph. The nominal control gains are $k_0=0.01$, $b_0=0.5$, and the mistuned gains used are the ones shown in Figure~\ref{fig:kopt-mistuned}, with $\varepsilon=0.001$.
%%%%%%%%%%%%%%%%%%%%%%%%%%%
The figure shows that
\begin{enumerate}
\item the closed-loop poles match the PDE eigenvalues accurately
over a range of $N$;
\item the mistuned eigenvalues show large improvement over the
nominal case even though the controller gains differ from
their nominal values only by $\pm 10\%$. The improvement is
particularly noticeable for large values of $N$, while being
significant even for small values of $N$.
\end{enumerate}
For comparison, the figure also depicts the asymptotic eigenvalue
formula given in Theorem~\ref{thm:mistuned}. The improvement in the stability margin with mistuning is remarkable since the gains are changed from their symmetric values by only $\pm 10\%$. Another interesting aspect of the result in Corollary~\ref{cor:mistuning-square} is that the improvement from $O(1/N^{2/D})$ to $O(1/N^{1/D})$ can be achieved by \emph{arbitrarily small changes} to the nominal gains. In addition, the optimal mistuned gain profile is quite simple to implement. For a vehicle formation with arbitrary dimensional information graph and with a maximum variation of $\pm 10\%$ from the symmetric gains, the optimal gains are obtained by letting  $k_{(i, i^{1+})}$ be $10$ percent larger than the nominal gain $k_0$ and letting $k_{(i, i^{1-})}$ be $10$ percent smaller than the nominal gain.

%%%%%%%%%%%%%%%%%%%%%%%%%%%%%%%%%%%%%%%%%%%%%%%%%%%%%%%%%%%%%%%%%%%%%%
\section{Discussion}\label{sec:remarks}
%%%%%%%%%%%%%%%%%%%%%%%%%%%%%%%%%%%%%%%%%%%%%%%%%%%%%%%%%%%%%%%%%%%%%%
\subsection{Relationship between the stability margins of the coupled-ODE and PDE models }\label{sec:ssm_vs_PDE}
In this paper, all the analysis and control design are based on the stability margin of the PDE model, which is an approximation of the coupled-ODE model under the assumption that each $n_i$ ($i \in \{1,2,\dots,D\}$) is very large. This raises the question: how large is the difference between the stability margin of the PDE (continuous problem) and the coupled-ODE (discrete problem) model? In this section, we provide an analysis on the difference between the stability margins of the continuous and the discrete problems, which we call the \emph{stability margin approximation error}. The results are summarized in the following lemma.
 
\begin{lemma}\label{lem:ssm_vs_pde}
Consider an $N$-vehicle formation with vehicle dynamics~\eqref{eq:vehicle-dynamics} and control law~\eqref{eq:control-lawd}, under Assumptions~\ref{as:k-same} and~\ref{as:L1}. With symmetric control (respectively, mistuning design), the stability margin approximation error between the PDE model~\eqref{eq:PDE} with boundary condition~\eqref{eq:BC} and the discrete model is $O(1/n_1^{3})$ (respectively, $O(1/n_1^{2})+O(\varepsilon^2)$).\frqed
\end{lemma}

In particular, for a square information graph, the stability margin approximation error bounds for symmetric control and mistuning design are $O(1/N^{3/D})$ and $O(1/N^{2/D})+O(\varepsilon^2)$ respectively.

\begin{figure}[h]
	\psfrag{0}{$0$}
	\psfrag{1}{$1$}
	\psfrag{n1-2}{$n_1-2$}
	\psfrag{n1-1}{$n_1-1$}
	\psfrag{x1}{$x_1$}
	\psfrag{O}{$O$}
\begin{center}
{\includegraphics[scale = 0.3]{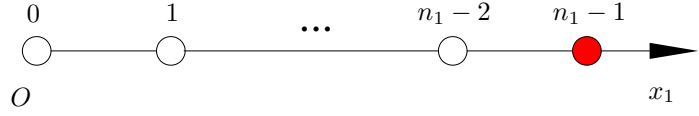}}
\end{center}
\caption{Pictorial representation of a 1D information graph.}\label{fig:1Dinformationgraph}
\end{figure}

Recall that for symmetric control (respectively, mistuning design), the stability margin scales as $O(1/n_1^2)$ (respectively, $O(1/n_1)$). Comparing with the above lemma, we can see that the PDE model provides an accurate approximation to the coupled-ODE model, and the approximation error can be ignored even for a moderate value of $n_1$, which is the number of vehicles along the $x_1$ axis of the information graph. For the ease of description, we only provide the proof for a formation with 1-dimensional information graph, i.e. the case $D=1$. Figure~\ref{fig:1Dinformationgraph} depicts a picture of the 1D information graph. The proof for higher dimensional case follows in a similar manner, upon using the closed form expressions of the eigenvalues for the discrete case~\cite{HH_PB_JV_CDC:10}.
 
\begin{proof-lemma}{\ref{lem:ssm_vs_pde}}
Given $D=1$, consider the following coupled-ODE and PDE models:
\begin{align*}
	\ddot{\tilde{p}}_{i}+b_{i}\dot{\tilde{p}}_{i}  &=-k_{(i,i-1)}(\tilde{p}_i - \tilde{p}_{i-1})  -k_{(i,i+1)}(\tilde{p}_i - \tilde{p}_{i+1}),\\
	\frac{\partial^2 \tilde{p}(x_1,t)}{\partial t^2}  + b(x_1)\frac{\partial \tilde{p}(x_1,t)}{\partial t}  & =\frac{k_1^{f-b}(x_1)}{n_1-1}\frac{\partial \tilde{p}(x_1,t)}{\partial x_1}+\frac{k_1^{f+b}(x_1)}{2{(n_1-1)}^2}\frac{\partial ^2 \tilde{p}(x_1,t)}{\partial {x_1}^2}.
\end{align*}
Using the optimal control gains given in~\eqref{eq:optimal-mistuned-gains} of Theorem~\ref{thm:mistuned}, the above models are simplified to:
\begin{align}
\ddot{\tilde{p}}_{i}+b_0\dot{\tilde{p}}_{i}  &=-(k_0-\varepsilon)(\tilde{p}_i - \tilde{p}_{i-1})  -(k_0+\varepsilon)(\tilde{p}_i - \tilde{p}_{i+1}),\label{eq:coupledode}\\
	\frac{\partial^2 \tilde{p}(x_1,t)}{\partial t^2}  + b_0\frac{\partial \tilde{p}(x_1,t)}{\partial t}  & =\varepsilon\frac{2}{n_1-1}\frac{\partial \tilde{p}(x_1,t)}{\partial x_1}+\frac{k_0}{{(n_1-1)}^2}\frac{\partial ^2 \tilde{p}(x_1,t)}{\partial {x_1}^2}\label{eq:pdemodel}.
\end{align}
Notice that when $\varepsilon=0$, it corresponds to the symmetric control case. Now consider the following discrete and continuous eigenvalue problem:
\begin{align}\label{eq:eigenvalue_problem_discrete}
1)& & -\lambda\tilde{p}_{i}  &=-(k_0-\varepsilon)(\tilde{p}_i - \tilde{p}_{i-1})  -(k_0+\varepsilon)(\tilde{p}_i - \tilde{p}_{i+1}),
\end{align}
where $i\in \{1,2,\dots,n_1-2\}$, and for the $0$-th vehicle, there is no neighbor behind it, so its equation is given by $-\lambda\tilde{p}_{0}  =-(k_0+\varepsilon)(\tilde{p}_0 - \tilde{p}_{1})$. And also, recall that the reference vehicle indexed by ``$n_1-1$'' has the property that $\tilde{p}_{n_1-1}=0$.
\begin{align}\label{eq:eigenvalue_problem_continuous}
2)& &-\mu \tilde{p}(\vec{x},t)  & =\varepsilon\frac{2}{n_1-1}\frac{\partial \tilde{p}(\vec{x},t)}{\partial x_1}+\frac{k_0}{{(n_1-1)}^2}\frac{\partial ^2 \tilde{p}(\vec{x},t)}{\partial {x_1}^2}，
\end{align}
where the boundary condition is given by $\frac{\partial  \tilde{p}(0,t)}{\partial x_1}=0, \quad \tilde{p}(1,t)=0$.

For the discrete eigenvalue problem, we can write it compactly as $\lambda \tilde{\mbf p}= L \tilde{\mbf p}$, where $\tilde{\mbf p}   \eqdef[\tilde{p}_{0},\tilde{p}_{2},\cdots,\tilde{p}_{n_1-2}]^{T}$ and $L$ is defined as follows:
\begin{align}\label{eq:grounded_laplacian}
   L=	\begin{bmatrix}
		k_0+\varepsilon & -k_0-\varepsilon& &   \\
		-k_0+\varepsilon & 2k_0&  -k_0-\varepsilon & \\
		 &\cdots  & \cdots &   \\
		& -k_0+\varepsilon & 2k_0&  -k_0-\varepsilon\\
		&  &-k_0+\varepsilon  & 2k_0
	       \end{bmatrix}.
\end{align}
For the symmetric control case ($\varepsilon=0$), the least eigenvalue of matrix $L$ is given by $4k_0\sin^2\frac{\pi}{2(2n_1-1)}$ ~\cite{yueh_tridiag}. For the case of mistuning design, under the assumption that $\varepsilon$ is small, we can use matrix perturbation method to compute the least eigenvalue of $L$ (see ~\cite{eig_perturbation_ngo:2005}). Combining the results, we have the least eigenvalue for the discrete eigenvalue problem:
\begin{align}\label{eq:least_eigenvalue_discrete1}
	\lambda=  4k_0\sin^2\frac{\pi}{2(2n_1-1)}+\varepsilon \frac{2(1+\cos\frac{\pi}{2n_1-1})}{2n_1-1}+O(\varepsilon^2).
\end{align}
By Taylor series expansion theorem, the above eigenvalue can be expressed as
\begin{align}\label{eq:least_eigenvalue_discrete}
	\lambda =  \frac{k_0\pi^2}{4(n_1-1)^2}-\frac{k_0\pi^2}{4(n_1-1)^3}+\varepsilon \frac{2}{n_1-1} -\varepsilon \frac{1}{(n_1-1)^2}+O(\varepsilon^2)+ \text{higher order terms}.
\end{align}
The continuous eigenvalue problem requires first to consider the following symmetric case ($\varepsilon=0$):
\begin{align}\label{eq:BVP}
-\mu \tilde{p}(\vec{x},t)  & =\frac{k_0}{{(n_1-1)}^2}\frac{\partial ^2 \tilde{p}(\vec{x},t)}{\partial {x_1}^2}，
\end{align}
with boundary condition $\frac{\partial  \tilde{p}(0,t)}{\partial x_1}=0, \tilde{p}(1,t)=0$, which yields the least eigenvalue $\frac{k_0\pi^2}{4(n_1-1)^2}$, which follows from straightforward algebra, see [Chapter 5]~\cite{haberman}. For the general case (mistuning design), we use the operator perturbation method [Chapter 9]~\cite{haberman}, the least eigenvalue for the continuous case is given by
\begin{align}\label{eq:least_eigenvalue_continuous}
	\mu= \frac{k_0\pi^2}{4(n_1-1)^2} +\varepsilon \frac{2}{n_1-1}+O(\varepsilon^2).
\end{align}

Comparing~\eqref{eq:least_eigenvalue_discrete} with~\eqref{eq:least_eigenvalue_continuous}, we have that for the symmetric case ($\varepsilon=0$), the eigenvalue approximation error is $O(1/(n_1-1)^3) = O(1/n_1^3)$, and for the mistuning design case, the error is $O(1/(n_1-1)^2)+O(\varepsilon^2) = O(1/n_1^2)+O(\varepsilon^2)$. Now, take Laplace transform for both~\eqref{eq:coupledode} and~\eqref{eq:pdemodel}, the characteristic equations for the coupled-ODE and PDE models are $s^2+b_0s+\lambda=0$ and $s^2+b_0s+\mu=0$ respectively, which implies that the stability margin approximation error are also $O(1/n_1^3)$ for symmetric control, and $O(1/n_1^2)+O(\varepsilon^2)$ for the mistuning design case. This completes the proof. \frQED
\end{proof-lemma}

%%%%%%%%%%%%%%%%%%%%%%%%%%%%%%%%%%%%%%%%%%%%%%%%%%%%%%%%%%%%%%%%%%%%%%
\subsection{Simulations}\label{sec:simulations}
We now present results of some time-domain simulations that show the time-domain improvements -- manifested in faster decay of initial errors -- with the mistuning-based design of control gains.  These simulations provide further corroboration of the two main conclusions of this paper:
\begin{enumerate}
\item Stability margin can be improved by using a higher-dimensional information graph with symmetric control.
\item Stability margin can be improved by using mistuned control gains for the same information graph.
\end{enumerate}

\begin{figure}[t]
  \centering
%  \psfrag{p}{$\tilde{p}=p_i-p_i^*$}
 \psfrag{p}[b][B]{$\tilde{p}_i(t) \; (\;= p_i(t) - p_i^*(t)\;)$}
 \psfrag{Time}{Time ($t$)}
  \subfigure[]{\includegraphics[scale = 0.35]{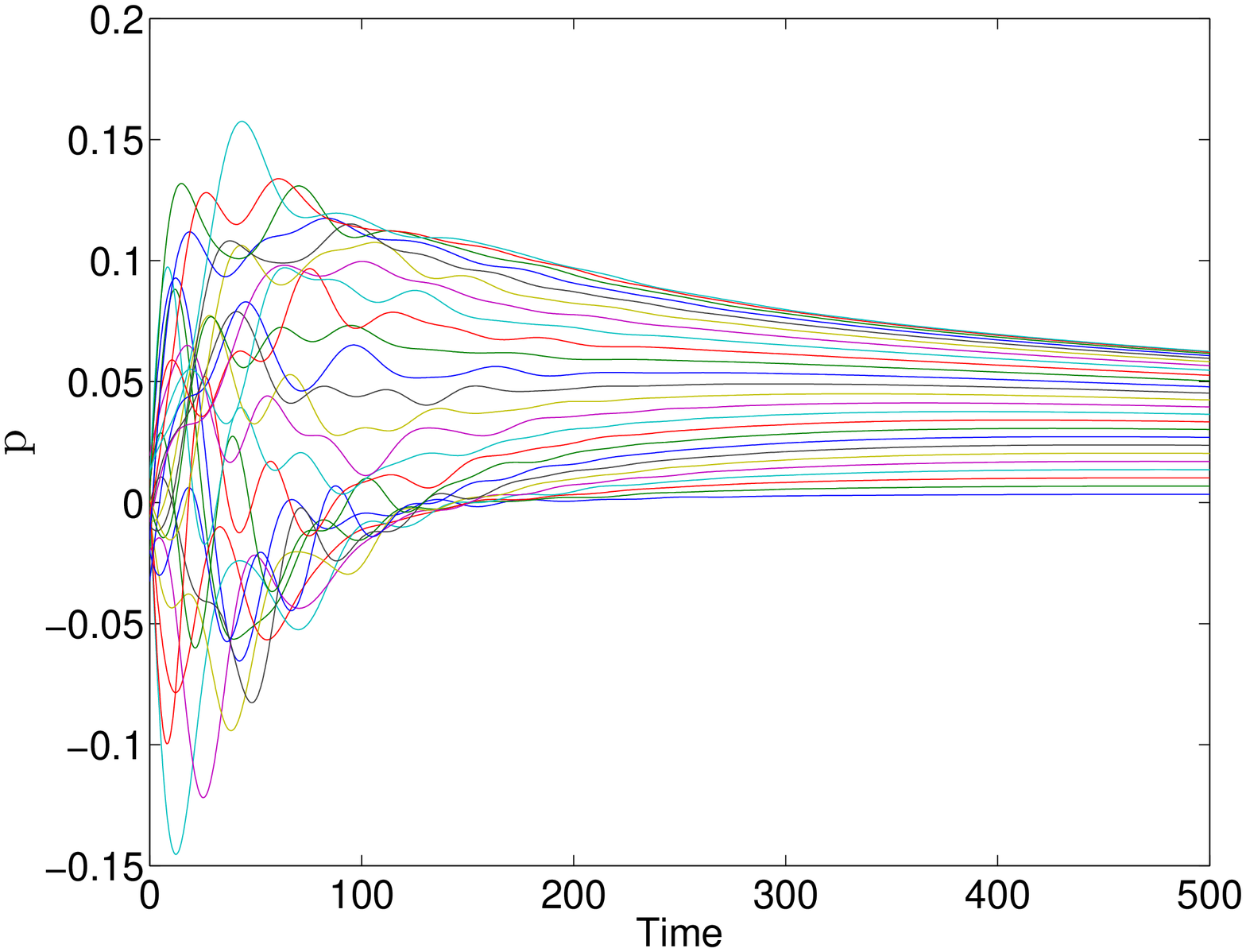}}
  \subfigure[]{\includegraphics[scale = 0.35]{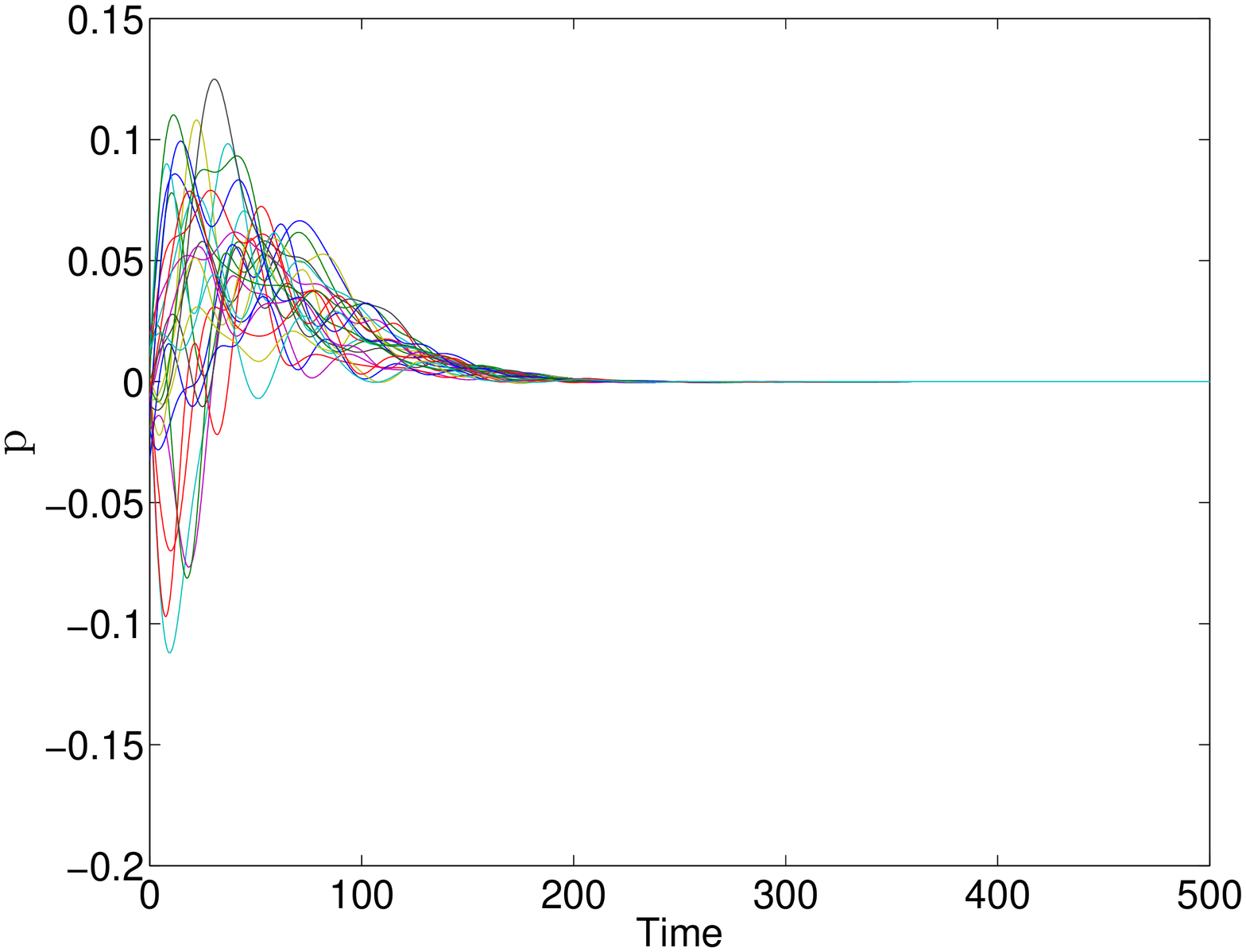}}
  \caption{Comparison of symmetric control's performance in 1D and 2D information graphs. (a) shows the relative position errors as a function of time for an $25$ vehicle platoon with a 1D lattice ($25$ vehicles and $1$ reference vehicles)  as the information graph (see Figure~\ref{fig:lattices} (a)). (b) shows the relative position errors for the same platoon (with the same initial condition) with a 2D square lattice ($25$ vehicles and $5$ reference vehicles)  as the information graph (see Figure~\ref{fig:lattices} (b)). In both cases, the gains used are $k_0 = 0.01$ and $b_0 = 0.05$, and the initial condition is such that making the position and velocity have arbitrary and small ($\leq 0.01$) perturbation from the desired position and velocity.}
  \label{fig:simulation-symmetric}
\end{figure}

For the first set of simulations, we consider $N=25$ vehicles in a one-dimensional formation ($D_s=1$). The initial position and velocity of each vehicle are randomly drawn from a uniform distribution on $[-0.01,0.01]$.  We carry out simulations for two distinct information graphs for the same physical formation which is consisted of $25$ vehicles: a $26$-node 1D lattice (including $1$ reference vehicle) and $6\times 5 $-node 2D lattice (including $5$ reference vehicles). %In the first case two vehicles that are physically adjacent are also neighbors in the information graph.
Figure~\ref{fig:simulation-symmetric} (a) and (b) show the time histories of the relative position errors of the  vehicles, for the 1D and 2D information graphs, respectively. In both cases, the control strategy is symmetric with gains $k_0=0.01$, $b_0=0.05$. On comparing Figure~\ref{fig:simulation-symmetric} (a) and (b), we see that the errors in the initial conditions are reduced faster with a two-dimensional information graph compared to the one-dimensional case.
This observation is consistent with with the result of Theorem~\ref{thm:symmetric}.

\begin{figure}[here]
  \centering
 \psfrag{p}[b][B]{$\tilde{p}_i(t) \; (\;= p_i(t) - p_i^*(t)\;)$}
 \psfrag{Time}{Time ($t$)}
  \subfigure[]{\includegraphics[scale = 0.45]{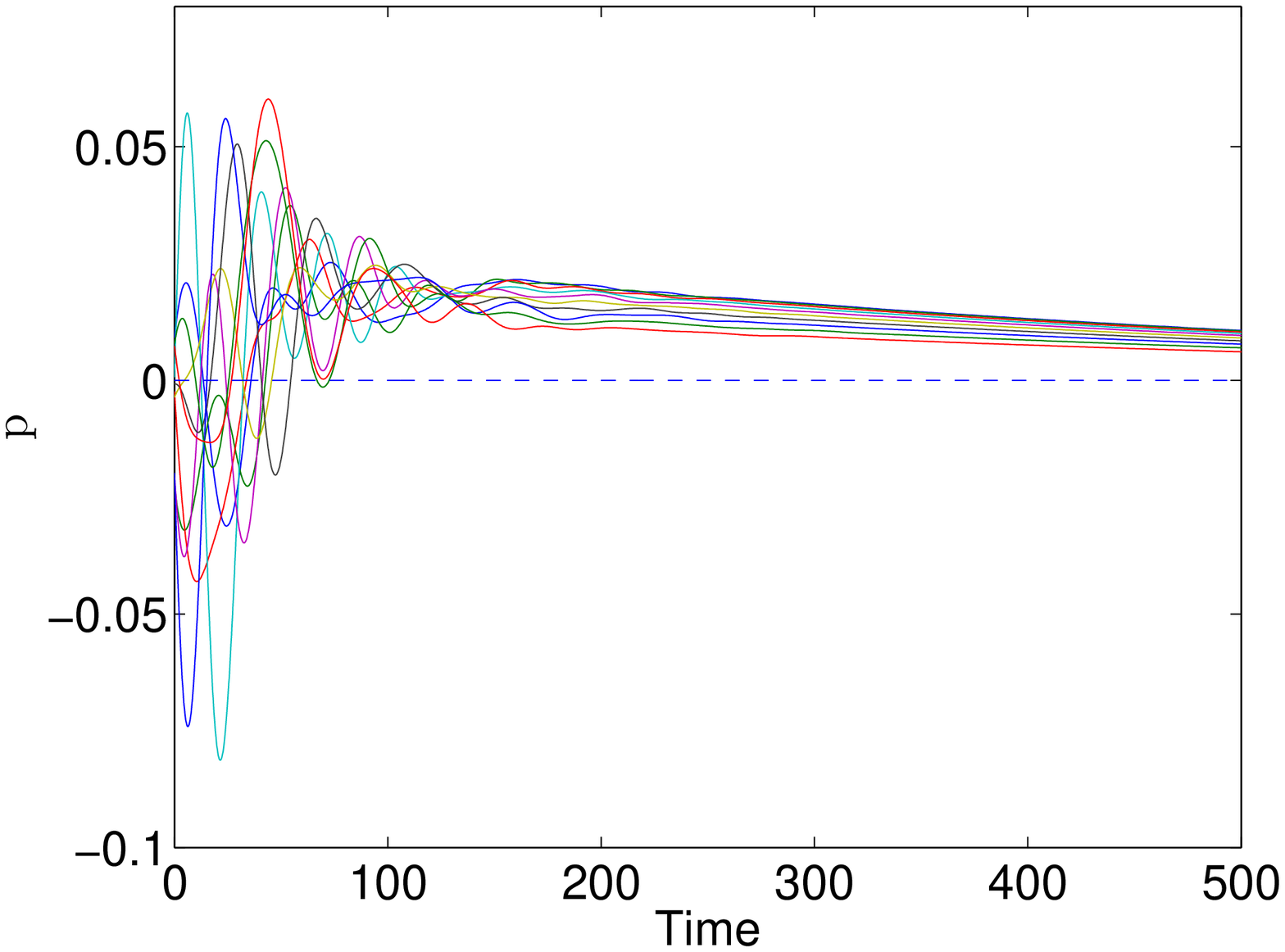}}
  \subfigure[]{\includegraphics[scale = 0.45]{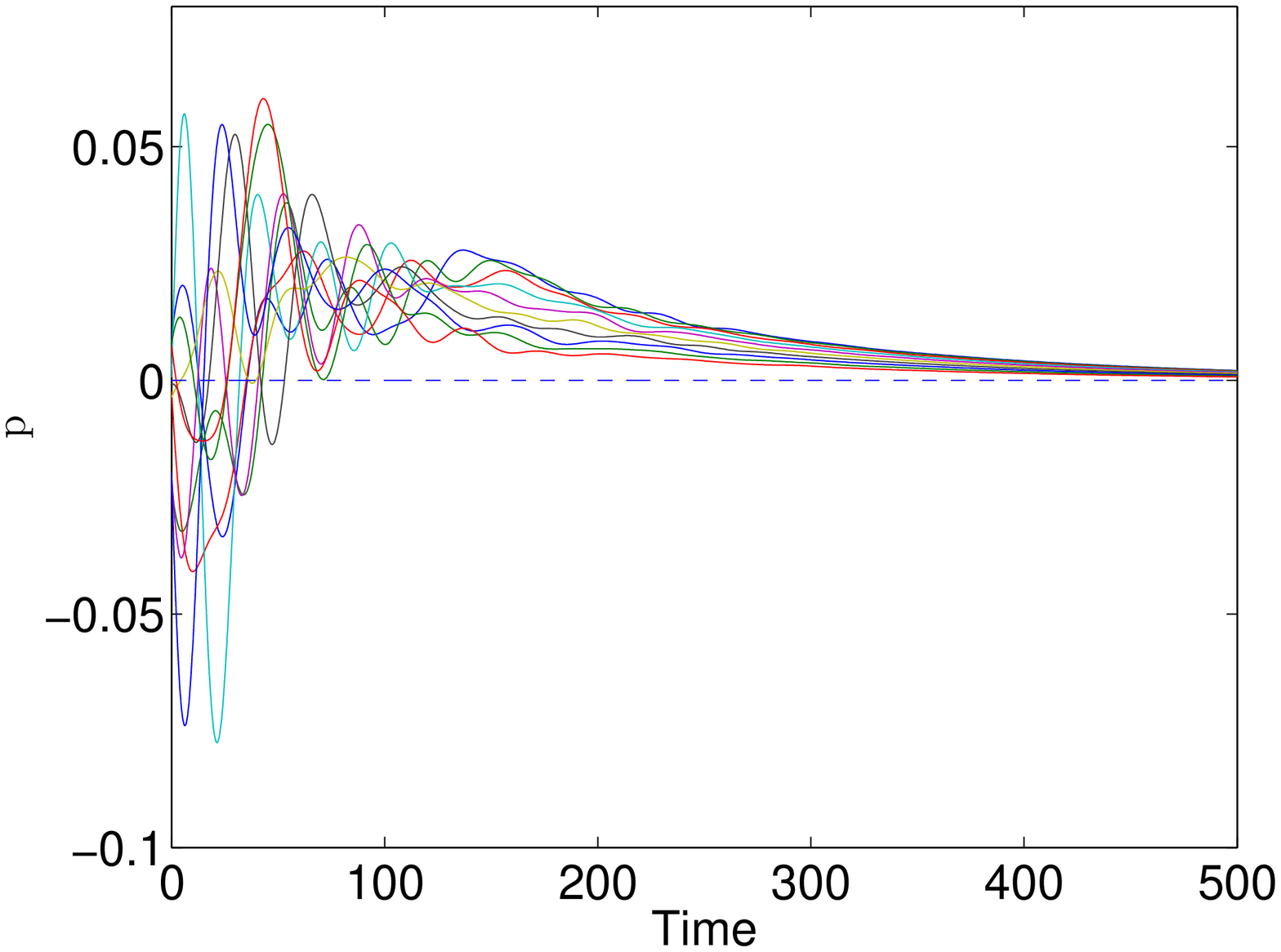}}
  \caption{Comparison of time-domain performance between symmetric and mistuned control with the same $16 \times 15$  2D square information graph. (a) shows the relative position errors as a function of time with symmetric control; (b) shows the relative position errors as a function of time for the same formation under mistuned control with control gains chosen according to the mistuned gains with parameters $k_0=0.01,b_0=0.05$, and $\epsilon = 0.001$ (i.e., $\pm 10\% $ variation from the symmetric gains). The initial condition is the similar to that as described in Figure~\ref{fig:simulation-symmetric}. Since there is a large amount of vehicles in the formation, for the purpose of showing the figure more clearly, only the first $10$ vehicles' traces are shown. The other traces have the same trend.}
  \label{fig:simulation-mistuned}
\end{figure}

The second set of simulations are carried out to test the effect of mistuning, for which we consider a formation with $225$ vehicles with a square 2D information graph -- a $16 \times 15$ lattice (including $15$ reference vehicles).  The initial position and velocity of each vehicle was again  chosen as a random small perturbation of the desired position and velocity.  Figure~\ref{fig:simulation-mistuned} (a) and (b) show the time history of the position errors with symmetric and mistuned control gains.  For the symmetric control, the control gains are $k_0=0.01$, $b_0=0.05$.  For the mistuning case $\varepsilon=0.001$, i.e., the gain $k_{(i,j)}$ is perturbed by $\pm 10\%$ from its nominal symmetric value $k_0$. On comparing Figure~\ref{fig:simulation-mistuned} (a) and (b), we see that the errors in the initial conditions are reduced faster in the mistuned case compared to the symmetric case.  This improvement is consistent with what is predicted by Theorem~\ref{thm:mistuned}.

%%%%%%%%%%%%%%%%%%%%%%%%%%%%%%%%%%%%%%%%%%%%%%%%%%%%%%%%%%%%%%%%%%%%%%
\subsection{Disturbance propagation}\label{sec:disturbance}
\begin{figure}[here]
\psfrag{N}{$N$}
\psfrag{Symmetric}{Symmetric}
\psfrag{Mistuned}{Mistuned}
\psfrag{1D information graph}{1D information graph}
\psfrag{2D infomation graph}{2D information graph}
\begin{center}
%\subfigure[]{\includegraphics[scale = 0.4]{1D_2D_h2.eps}}
%\subfigure[]{\includegraphics[scale = 0.4]{nom_mis_h2.eps}}
\subfigure[]{\includegraphics[scale = 0.35]{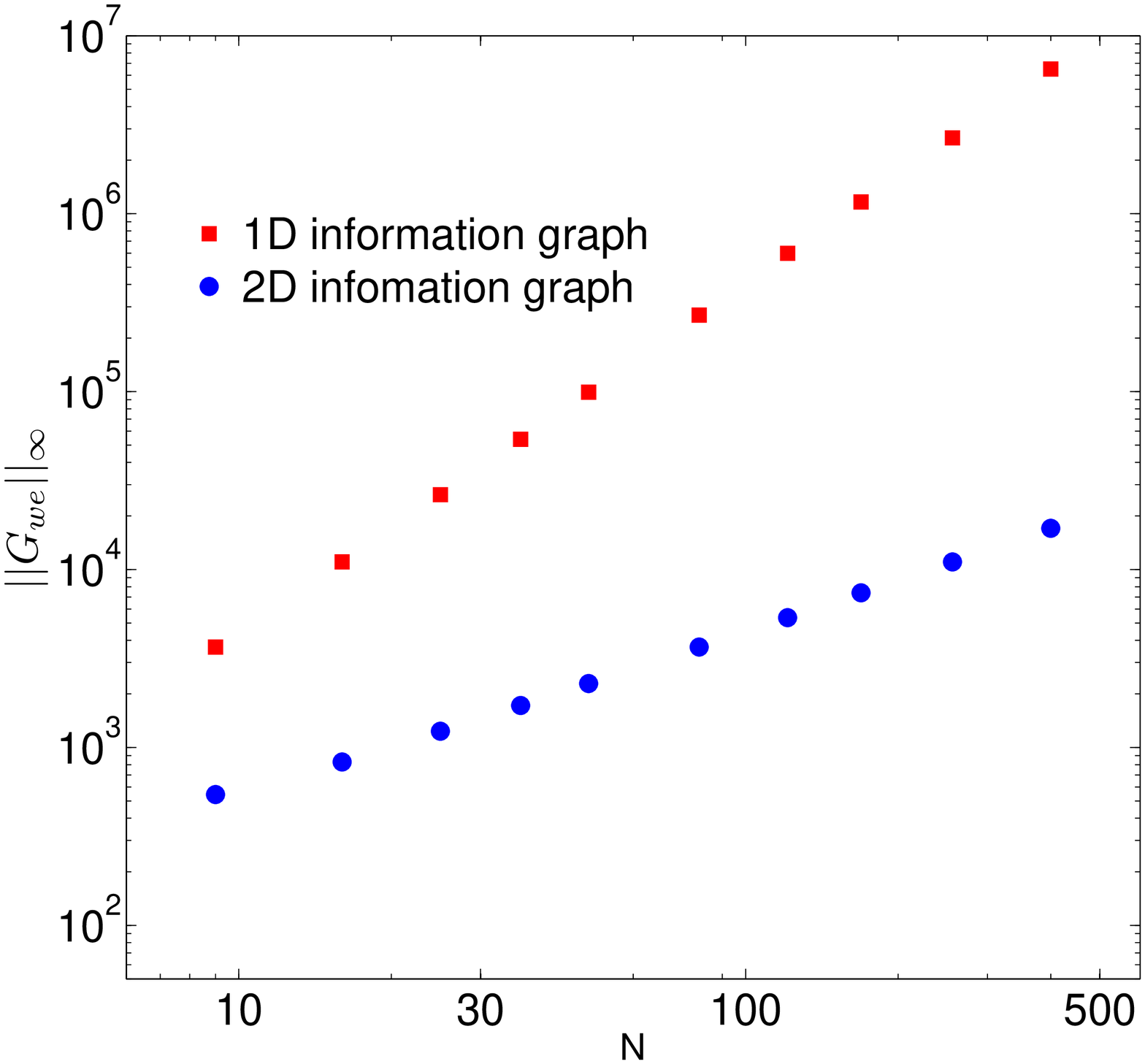}}
\subfigure[]{\includegraphics[scale = 0.35]{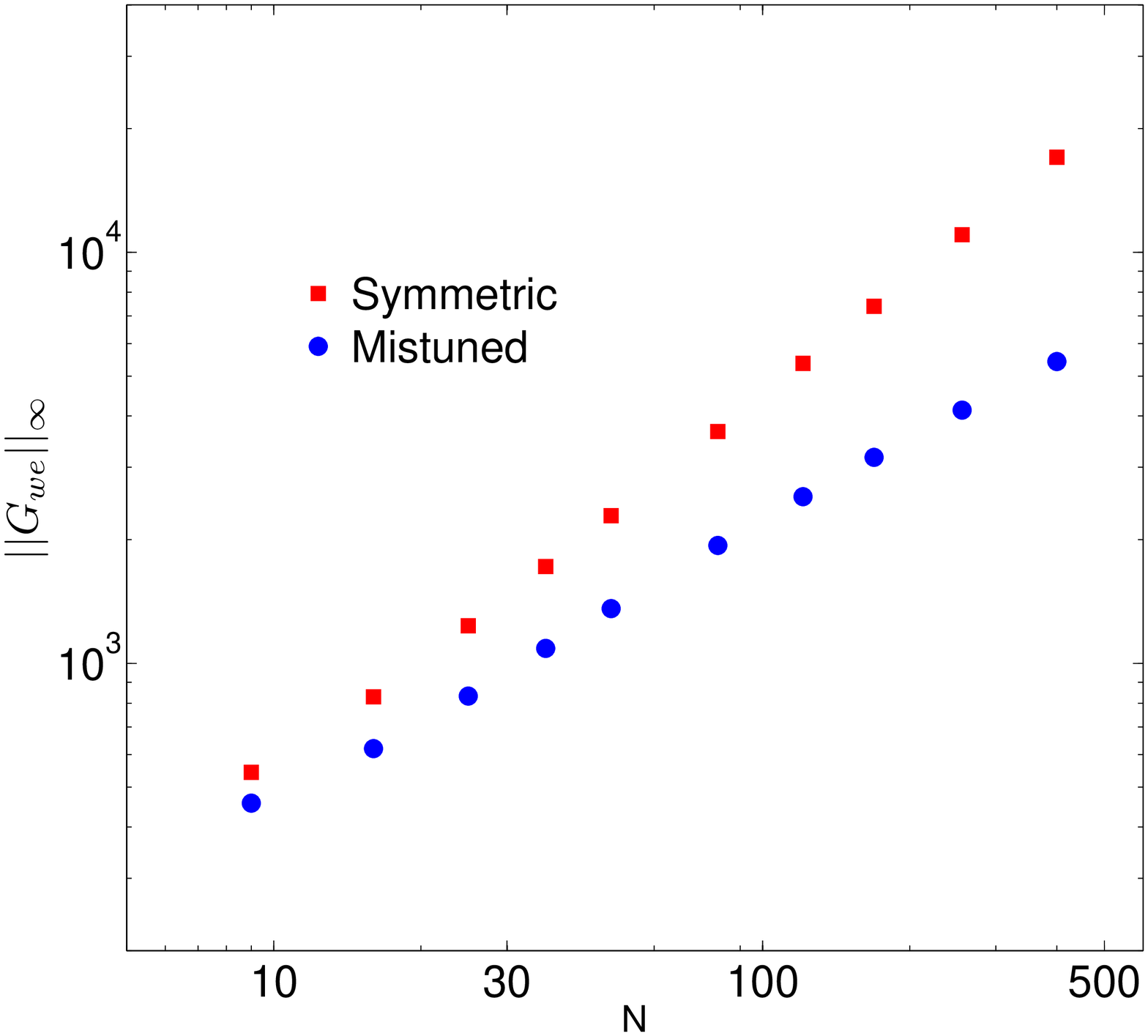}}
\caption{$H_\infty$ norm of the transfer
function $G_{we}$ from disturbance $\mbf w$ to spacing error $\mbf
{e}$. (a) compares the $H_{\infty}$ norm as a function of $N$, with 1D and 2D information graphs, with all else remaining the same. (b) compares the $H_{\infty}$ norm in symmetric control and mistuned control (with $\pm 10\%$ mistuning), when the information graph is the same (a 2D lattice). In all cases, the gains used are $k_0 = 0.01$ and $b_0 = 0.05$. The mistuned gains used are those given in Theorem~\ref{thm:mistuned}, with $\varepsilon = 0.0001$. Norms are computed using the Control Systems Toolbox in MATLAB$^\text{\copyright}$.}\label{fig:H-infinity}
\end{center}
\end{figure}

When external disturbances are present, we model the dynamics of vehicle $i$ by $\ddot{p}_i=\ddot{\tilde{p}}_i = u_i +  w_i$, where $w_i$ is the external disturbance acting on the vehicle.  Each component of the disturbance is assumed to be independent.
%
%Again, the disturbances affecting each component of the acceleration is assumed independent, so the superscript $(d)$ is omitted from the equation %above, as done in~\eqref{eq:control-law}.
%
In the $\tilde{p},\tilde{v}$ coordinates, the closed-loop dynamics of the formation is given by
\begin{align}\label{eq:error}
\dot{ \mbf \psi} & = {\mbf A}{\mbf \psi} + \underbrace{\matt{\mbf{0} \\ I}}_{\mathcal{B}}\mbf{w},
\end{align}
 where $\mbf{\psi} \eqdef [{\mbf{\tilde p}}^T, {\mbf{\tilde v}}^T]^T$ is the state vector, $\mbf{w}  \eqdef [w_1,w_2,\dots,w_N]^T$ is the vector of disturbances. We consider the vector of errors $\mbf{e} \eqdef
 [\tilde{p}_1,\dots,\tilde{p}_N]^T=\tilde{\mbf{p}}$, where $\tilde{p}_i=p_i-p_i^*$, $i=1,2,\dots,N$, as the outputs:
 \begin{align*}
\mbf{e} & = C \mbf{\psi} ,\ \ C=[I;\mbf{0}]
 \end{align*}
The $H_\infty$ norm of the transfer function $G_{we}$ from the disturbance $\mbf{w}$ to the errors $\mbf{e}$ is a measure of the closed-loop's sensitivity to external disturbance.  For one-dimensional platoons, such a norm has been used previously in~\cite{Seiler_disturb_vehicle_TAC:04,RM_JB_TAC:09,PB_PM_JH_TAC:09}.  Figure~\ref{fig:H-infinity} depicts the $H_\infty$ norm of $G_{we}$ as a function of $N$, for the two cases described in Section~\ref{sec:simulations}.  Part~(a) of the figure compares the $H_\infty$ norm of the one-dimensional and two-dimensional information graphs for the same formation with symmetric control. Part~(b) of the figure compares the $H_\infty$ norm of the symmetric and mistuned control for the two-dimensional formation.  The trends for the $H_\infty$ norm are consistent with the eigenvalue trends and the results of the time-domain simulations.  In particular,
\begin{enumerate}
\item The $H_\infty$ norm of $G_{we}$ is improved by using a higher-dimensional information graph with symmetric control.
\item For a particular information graph, the $H_\infty$ norm of $G_{we}$ is improved by using mistuned control gains over symmetric control.
\end{enumerate}
%
%
%Part~(a) of the figure shows that the $H_\infty$ norm for the two-dimensional information graph are better compared to the one-dimensional case.
%
%shows a plot of the $H_\infty$ norm of
%$G_{we}$ as a function of $N$, with and without mistuning. The
%mistuning profile used is the same as the one used for the
%eigenvalue trends reported in Figure~\ref{fig:mistuned-lseig-compare}. It is clear from the figure that: (i) for the same formation, with higher dimensional information graph can has large beneficial effect on $H_{\infty}$ norm; (ii) mistuning design improve the $H_{\infty}$ norm considerably for the same formation with the same information graph.

Analysis of these trends is beyond the scope of this work, and will be undertaken in future work.

\subsection{Other boundary conditions}
In this paper, results are derived for an arrangement of reference vehicles on one of the boundaries of a $D$-dimensional information graph (see Assumption~\ref{as:L1}).  For a one-dimensional information graph, this means there is one reference vehicle.  For a $D$-dimensional square information graph with $N$ vehicles and $D>1$, this means that there are $N^{\frac{1}{D}}$ reference vehicles.

\begin{figure}
\begin{center}
	\psfrag{x1}{$x_1$}
	\psfrag{x2}{$x_2$}
	\psfrag{O}{$O$}
\subfigure[1 Dirichlet and 3 Neumann boundaries]{\includegraphics[scale = 0.278]{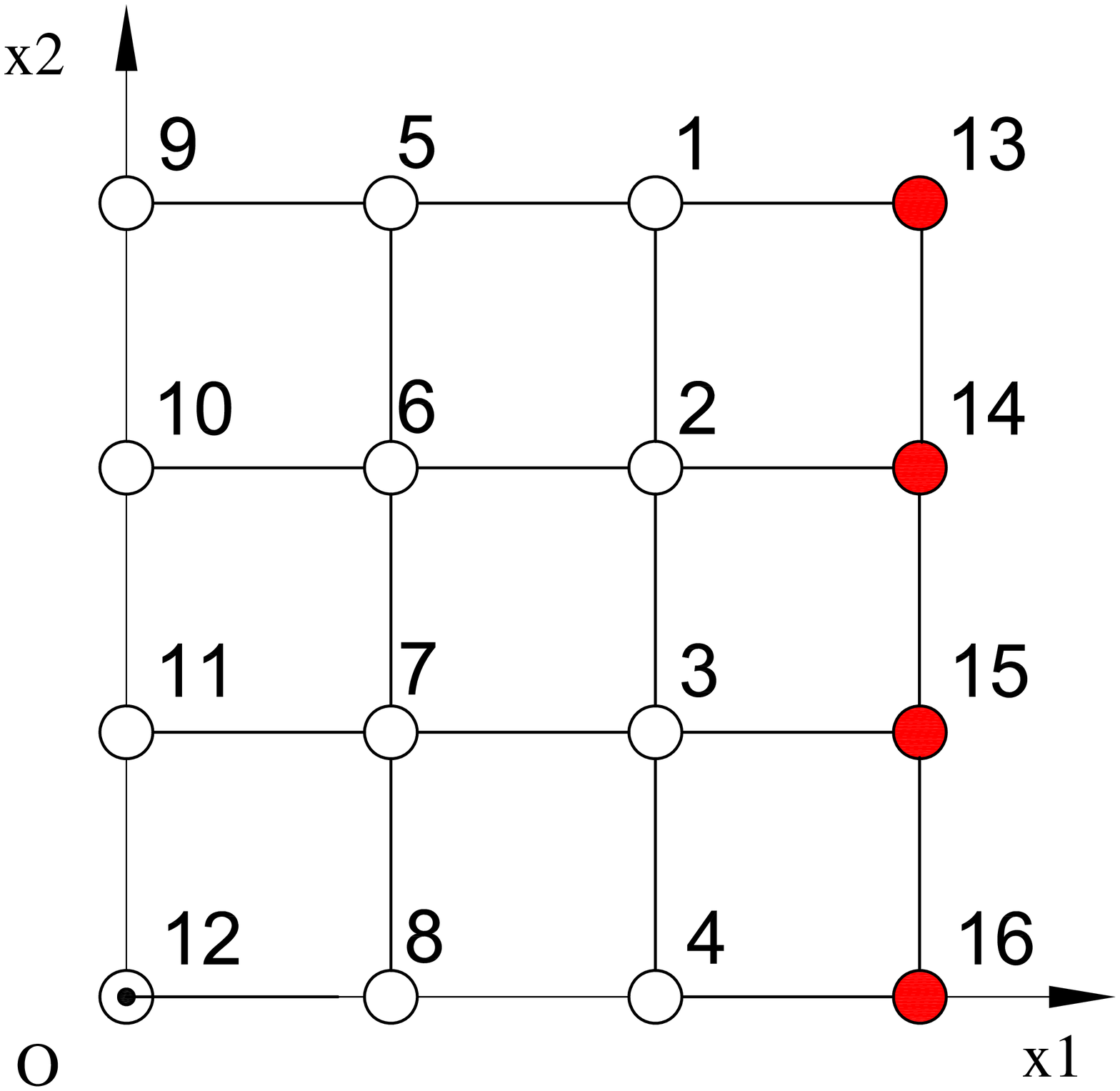}}\ \ \
\subfigure[2 Dirichlet and 2 Neumann boundaries]{\includegraphics[scale = 0.27]{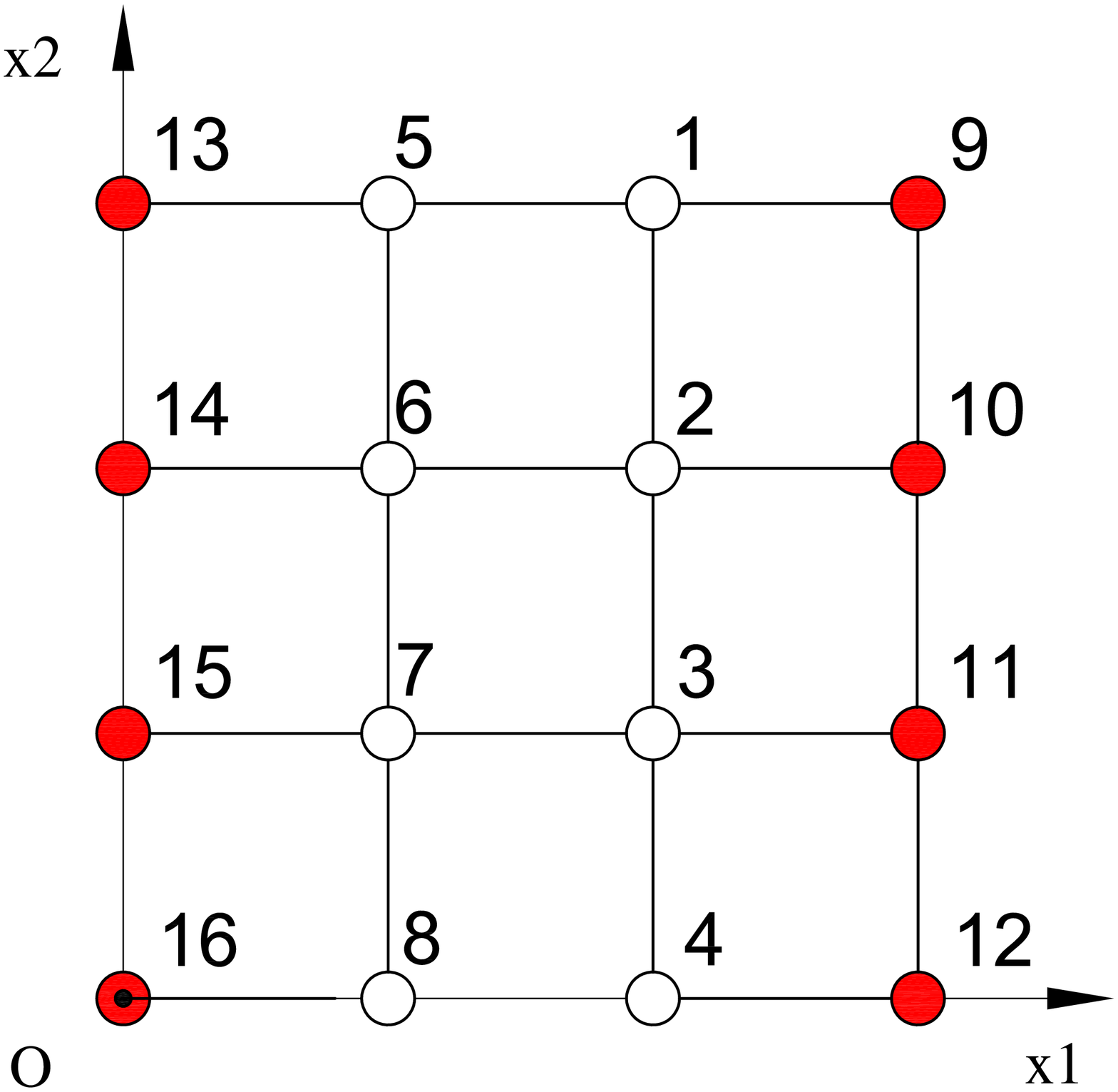}}\ \ \
\subfigure[4 Dirichlet boundaries]{\includegraphics[scale = 0.257]{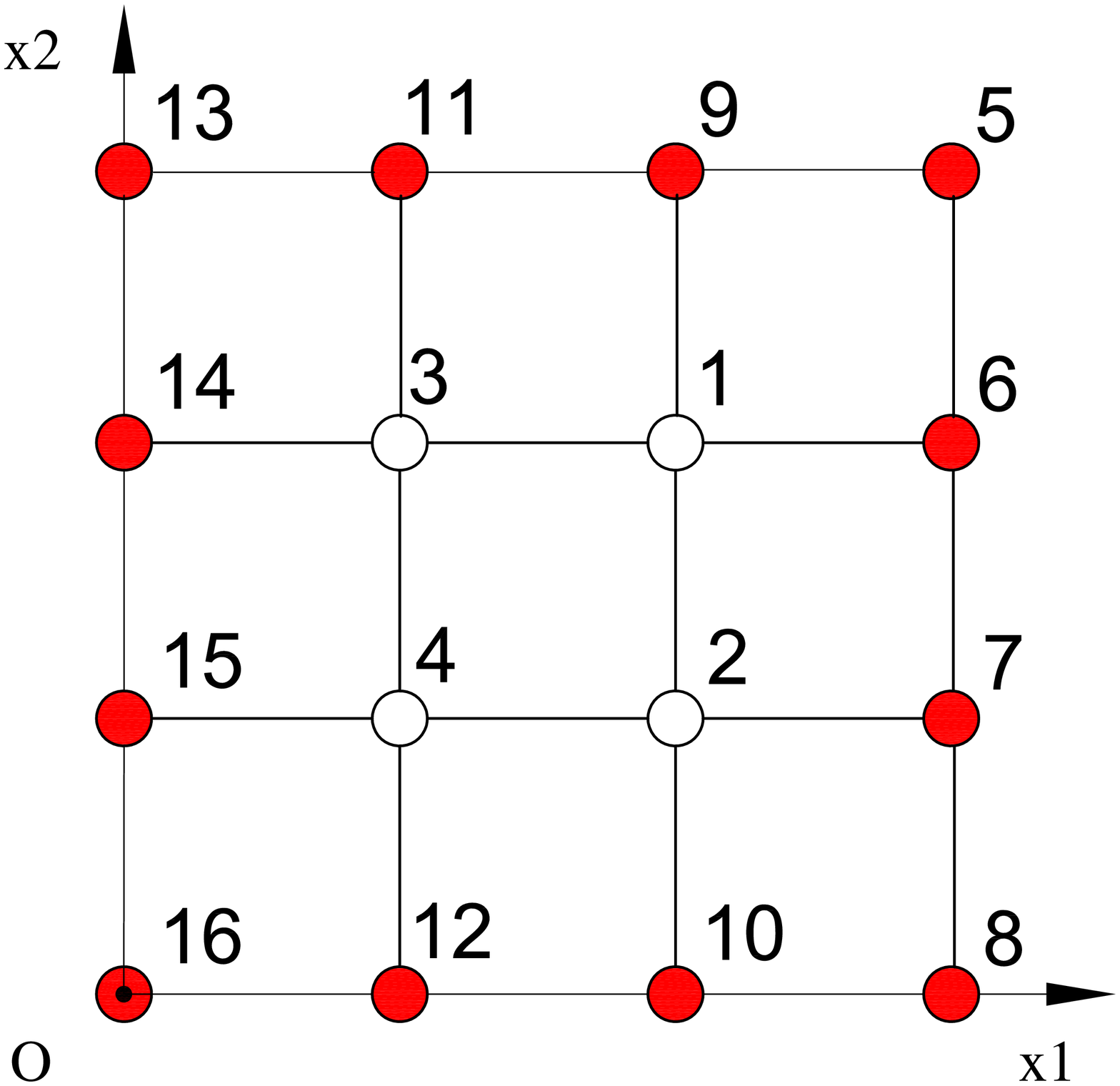}}
\end{center}
\caption{A pictorial representation of the 3 of the several possible boundary conditions for a 2D information graph.}\label{fig:bc}
\end{figure}

In terms of the methodology of this paper, the arrangement of reference vehicles affects the boundary condition in the PDE approximation but  not the PDE itself.  Under Assumption~\ref{as:L1}, the boundary condition is the Dirichlet boundary condition at $x_1 = 1$ and Neumann boundary conditions for other boundaries of $[0,1]^D$ (see~\eqref{eq:BC}).  More generally, the presence of reference vehicles on an additional boundary means that the PDE approximation will have Dirichlet boundary condition for these boundaries.  Figure~\ref{fig:bc} enumerates some of the possibilities for the two-dimensional case.

%The results presented so far are for Situation I, which refers to the case that all the fictitious leader vehicles have the same $x_1$-coordinate in the information graph. If the graph $\G$ is a D-dimensional square lattice, this corresponds to $\sqrt{N}$ vehicles having reference trajectory information. In the PDE approximation, this leads to Dirichlet boundary condition at $x_1 = 1$ and Neumann boundary conditions in all the other faces of the unit cell $[0,1]^D$. If more vehicles are provided reference trajectory information, they can be arranged in distinct manner, leading to distinct boundary conditions for the PDE.  Figure~\ref{fig:bc} shows a few possibilities.

It is straightforward to extend the analysis and the conclusions of the preceding sections to handle these more general boundary conditions.  For asymptotic trend of the stability margin with $N$ does not change with the boundary conditions.  The presence of additional reference vehicles affects only the constant in front of the asymptotic formula. Consider for example the two-dimensional square information graph with $N$ vehicles and $4\sqrt{N}-4$ reference vehicles arranged along the $4$ boundaries. The PDE approximation is again given by~\eqref{eq:PDE}.  The boundary conditions are all Dirichlet type. The scaling laws for this case are described in our conference paper~\cite{HH_PB_PM_DSCC:09}.  We summarize the results for the symmetric and the mistuned cases in the following corollaries.

\begin{corollary}[from Corollary 1 in~\cite{HH_PB_PM_DSCC:09}] \label{cor:citeHH1}
Consider an $N$-vehicle formation with dynamics~\eqref{eq:vehicle-dynamics} and control law~\eqref{eq:control-lawd} under Assumption~\ref{as:k-same}, with nominal symmetric control gains $k_0$ and $b_0$, whose information graph is a $(\sqrt{N}+2)\times(\sqrt{N}+2)$ 2D lattice. Let all the nodes on each of the $4$ faces of the information graph correspond to reference vehicles, so that the boundary conditions of the  PDE~\eqref{eq:PDE_D} are all of the Dirichlet type. %Consider a vehicular formation of $N$ vehicles  with symmetric control whose information graph is a square 2D lattice (i.e., $N=(n_1-1)^2$) with all Dirichlet boundaries. 
The closed-loop stability margin is given by
  \begin{align*}
    S = \frac{2\pi^2 k_0}{b_0}\frac{1}{N} + O(\frac{1}{N^2}) \tag*{\frqed}
  \end{align*}
\end{corollary}

\medskip

Comparing this result with Corollary~\ref{cor:symmetric-square} (when $D=2$) shows that the benefit of extra information (four times as many vehicles provided reference trajectory information) is a factor of $8$ improvement in the closed-loop stability margin.
\medskip
%%%%%%%%%%%%

%In the statement of the next corollary, $Q_k$ denotes the $k$-th quadrant of the 2D square lattice drawn in $[0,1]\times[0,1]$, which are defined with respect to a Cartesian reference frame whose axes are parallel to the original axes in $\R^2$ but with origin shifted to $(0.5,0.5)$. %We say that a node whose coordinate is $(\ell,m)$, with $|\ell|\leq 0.5$ and $|m| \leq 0.5$, lies in quadrant $Q_k$ ($k \in \{1,\dots,4\}$) if 
%%%%%%%%%%%%

\begin{corollary}[from Corollary 2 in~\cite{HH_PB_PM_DSCC:09}]
Consider the same vehicle formation with the same information graph as stated in Corollary~\ref{cor:citeHH1}. Now consider the problem of maximizing the stability margin by designing the proportional control gains $k_{(i,j)}$, where the gains are required to satisfy $| k_{(i,j)} - k_0 | \leq  \varepsilon$ for every $(i,j)\in\E$, with $\varepsilon \in ( 0, \; k_0)$ being an arbitrary pre-specified constant. For vanishingly small values of $\varepsilon$, the optimal control gains of  the $i$-th vehicle ($i=1, \dots, N$) are given by the following formula
\begin{align*}
	k_{(i,i^{1+})} & = k_0 + 2\varepsilon (H(i_1c_1-0.5) - 0.5), \\
        k_{(i,i^{1-})} & = k_0 - 2\varepsilon (H(i_1c_1-0.5) - 0.5), \\
        k_{(i,i^{2+})} & = k_0 + 2\varepsilon (H(i_2c_2-0.5) - 0.5), \\
        k_{(i,i^{2-})} & = k_0 - 2\varepsilon (H(i_2c_2-0.5) - 0.5), 
\end{align*}
where $H(x)=1$ if $x>0$, $H(x) = 0$ if $x<0$, and $H(0)=0.5$, and $c_i$ is defined in~\eqref{eq:cd_def}. The resulting stability margin is given by
\begin{align*}
    S = \frac{8\varepsilon}{b_0}\frac{1}{\sqrt{N}} +  O(\frac{1}{N} ).
\end{align*}
The formula is asymptotic in the sense that it holds when $n_1,n_2 \to \infty$ and $\varepsilon \to 0$. \frqed
 %For a vehicle formation whose information graph is a square $\sqrt{N}\times\sqrt{N}$ 2D lattice with all Dirichlet boundaries, with asymmetric gains so that $|k_{(i,j)} - k_0|< \varepsilon$, where $\varepsilon$ is a small positive number, the optimal gains so that the closed-loop stability margin is
 % \begin{align*}
  % S = \frac{8\varepsilon}{b_0}\frac{1}{\sqrt{N}} +  O(\frac{1}{N} )
 %  \end{align*}
% when $N$ is large and $\varepsilon$ is small. The optimal mistuned gains that maximize the stability margin for the same $\varepsilon$ are
% \begin{align*}
% 	k_{(i,i^{1+})} & = k_0 + 2\varepsilon (I_{Q_1\cup Q_4}(i) - 0.5), \\
%         k_{(i,i^{1-})} & = k_0 - 2\varepsilon (I_{Q_2\cup Q_3}(i) - 0.5), \\
%         k_{(i,i^{2+})} & = k_0 + 2\varepsilon (I_{Q_1\cup Q_3}(i) - 0.5), \\
%         k_{(i,i^{2-})} & = k_0 - 2\varepsilon (I_{Q_3\cup Q_4}(i) - 0.5), \\
% \end{align*}
% where $I_A(x)$ is the indicator function of the set $A$ (i.e., $I_A(x)  =1$ if $x \in A$ and $0$ otherwise)\PBcomment{This Q notation needs to be rechecked}.  \frqed
\end{corollary}

Comparing this result with corollary~\ref{cor:mistuning-square} for $D=2$ shows that with mistuning, having four times as many vehicles that have reference trajectory information results in a factor of $4$ improvement in the stability margin.

\subsection{Comparison to earlier work}
There are connections between the results of this paper and the results in~\cite{bamjovmitCDC08}. 
In~\cite{bamjovmitCDC08}, Bamieh~\textit{et. al.} proposed certain macroscopic performance measures to quantify the sensitivity to disturbances of vehicular formations. The vehicles were modeled as double integrators and the feedback control was symmetric. The information graph considered by Bamieh~\textit{et. al.}  was a $D$-dimensional torus, which is similar to a $D$-dimensional square lattice.  It was shown in~\cite{bamjovmitCDC08} that the measure of disturbance amplification proposed in~\cite{bamjovmitCDC08}  grows without bound as a function of $N$ for $D=1$ and $D=2$, but it is uniformly bounded with respect to $N$ for $D\geq 3$. In contrast, Corollary~\ref{cor:symmetric-square} shows that there is no uniform bound on stability margin in any dimension for square lattices. The scaling law for the stability margin, however, improves with increasing $D$, as well as with mistuning. In summary, the asymptotic behavior of the stability margin in dimensions $D=1$ and $D=2$ is similar to that of the macroscopic performance measure of Bamieh~\textit{et. al.} in~\cite{bamjovmitCDC08}. However, the trends are quite different in dimensions $3$ and higher.

%%%%%%%%%%%%%%%%%%%%%%%%%%%%%%%%%%%%%%%%%%%%%%%%%%%%%%%%%%%%%%%%%%%%%%
\section{Conclusion}\label{sec:conc}
%%%%%%%%%%%%%%%%%%%%%%%%%%%%%%%%%%%%%%%%%%%%%%%%%%%%%%%%%%%%%%%%%%%%%%
We studied the closed-loop stability margin with distributed control of a network of $N$ double integrator agents. Information graphs (within the class of $D$ dimensional lattices) that characterize the information exchange structure among vehicles were examined. We first examined the case of symmetric control, in which every vehicle uses the same control gains. For a square information graph, the stability margin approaches zero as $O(1/N^{2/D})$ as $N\rightarrow \infty$. Therefore, the stability margin can be improved by increasing the dimension of the information graph. For a non-square information graph, the stability margin can be made nearly independent of the number of vehicles by choosing the ``aspect ratio'' appropriately. The trade-off is that increasing the dimension of the
information graph or choosing a beneficial aspect ratio may require long range communication and/or entail an increase in the number of reference vehicles. These results are therefore useful in investigating design trade-offs between performance and the cost of designing information architectures for distributed control. 

Second, a mistuning-based approach for stability margin improvement over symmetric control is proposed that consists of making small changes to the gains over their nominal values in the symmetric case. The scaling laws for the stability margin with mistuned control showed that with arbitrarily small amount of mistuning, the stability margin can be improved significantly over symmetric control. %The information needed by the vehicles to implement the mistuned control, in addition to what is needed for  the symmetric control, is small. This makes the resulting design attractive in practice.
The mistuned control is simple to implement and therefore attractive for practical application.

A PDE approximation was derived to aid the analysis and design that was carried out in the paper.
The control design problem is much more tractable in the PDE domain than in the original state space domain. In particular, the PDE model provides insight into the effect of asymmetry in the control gains on the stability margin, which enabled the mistuning-based design. Such insight is difficult to gain by examination of the state-space model.  Although the PDE approximation is valid only for $N \to \infty$, numerical calculations using the PDE model show that accurate predictions are obtained even for small values of $N$.

The information graphs studied in this paper are limited to $D-$dimensional lattices. More complex graph structures will be explored in future work. We believe that the PDE approximation will be beneficial here, by allowing us to sample from the continuous gain functions defined over a continuous domain to assign gains to spatially discrete agents. Another future direction of research is the examination of the closed-loop's sensitivity to external disturbances. For symmetric control, this issue was investigated in~\cite{bamjovmitCDC08}. Numerical tests reported in this paper show that mistuning reduces the closed-loop's sensitivity to external disturbances. Analysis of the effect of mistuning on the closed-loop's sensitivity to external disturbances will be carried out in future work.

% The proceeding numerical computation suggested that the mistuning design method also had a beneficial effect to reduce the sensitivity to external disturbances. Analysis of mistuning on disturbance rejection is one subject of future research. In addition, distributed nonlinear control of a vehicle formation under predecessor following strategy is also a good subject which needs to be examined in the future.

%%%%%%%%%%%%%%%%%%%%%%%%%%%%%%%%%%%%%%%%%%%%%%%%%%%%%%%%%%%%%%%%%%%%%%%

\bibliographystyle{IEEEtran}
%\bibliography{../../PBbib/vehicular_platoon,../../PBbib/sensnet_bib_dbase,../../PBbib/Barooah,../../PBbib/HH}
\bibliography{../../../PBbib/vehicular_platoon,../../../PBbib/sensnet_bib_dbase,../../../PBbib/Barooah,../../../PBbib/HH}

\appendix

\begin{proof-theorem}{\ref{thm:theorem1}}
The proof proceeds by a perturbation method. Let the eigenvalues of the perturbed PDE~\eqref{eq:mistuned_PDE_D} and the Laplace transform of $\tilde{p}(\vec{x},t)$ be
\begin{align*}
s_{\vec{l}}= s^{(0)}_{\vec l} + \varepsilon s^{(\varepsilon)}_{\vec l} + O(\varepsilon^2), \quad \eta= \eta^{(0)} + \varepsilon \eta^{(\varepsilon)}+ O(\varepsilon^2)
\end{align*}
respectively, where $s^{(0)}_{\vec l}$ and $\eta^{(0)}$ are corresponding to the unperturbed PDE~\eqref{eq:PDE_D_simplified}. %, and the subscripts $(l_1,l_2,\dots, l_D)$ are suppressed.
Taking a Laplace transform of both sides of the PDE~\eqref{eq:mistuned_PDE_D} with respect to $t$, plugging in the expressions for $s$ and $\eta$, and doing an $O(1)$ balance leads to the eigenvalue equation for the unperturbed PDE:
\begin{align*}
 \scr{P}\eta^{(0)}= 0, \text { where }  \scr{P} :=  \left( (s^{(0)}_{\vec l})^2 + b_0s^{(0)}_{\vec l} - \scr{L}_0 \right)
\end{align*}
where $\scr{L}_0$ is the Laplacian operator defined in~\eqref{eq:new_laplacian}. Recall that the solution $s^{(0)}_{\vec l},\eta^{(0)}$ to this equation have been previously given. Eq.~\eqref{eq:eigenvalue} provides the formula for $s^{(0)}_{\vec l}$ (i.e $s_{\vec{l}}^+$), and $\eta^{(0)}=\sum \phi_{\vec l}(\vec x) \alpha_{\vec l}(s)$, where $\phi_{\vec l}(\vec x)$ is given by equation~\eqref{eq:lamda-symmetric}.
Next we do an $O(\varepsilon)$ balance, which leads to:
\begin{align*}
	\scr{P}\eta^{(\varepsilon)} = \Big(\sum_{d=1}^{D}{\frac{k_d^{m}(\vec{x})}{n_d-1}\frac{\partial}{\partial x_d}}  + \sum_{d=1}^{D} {\frac{k_d^{s}(\vec{x})}{2(n_d-1)^2} \frac{\partial^2}{\partial x_d^2}}  -b_0 s^{(\varepsilon)} -2 s_{\vec l}^{(0)} s^{(\varepsilon)}_{\vec l} \Big)\eta^{(0)} =: R
\end{align*}
For a solution $\eta^{(\varepsilon)}$ to exist, $R$ must lie in the range space of the operator $\scr{P}$. Since $\scr{P}$ is self-adjoint, its range space is orthogonal to its null space. Thus, we have,
\begin{align}\label{eq:inner-product}
	< R , \phi_{\vec l}(\vec x)> = 0
\end{align}
where $\phi_{\vec l}(\vec x)$ is also the $(l_1,l_2,\dots,l_D)^{\text{th}}$ basis of the null space of operator $\scr{P}$. We now have the following equation:
\begin{align*}
\int_{0}^{1} \cdots \int_0^{1} \Big( \sum_{d=1}^{D}{\frac{k_d^{m}(\vec{x})}{n_d-1}\frac{\partial \eta^{(0)}}{\partial x_d}} +  \sum_{d=1}^{D}{\frac{k_d^{s}(\vec{x})}{2(n_d-1)^2} \frac{\partial^2 \eta^{(0)}}{\partial x_d^2}}  -b_0 s^{(\varepsilon)}_{\vec l} \eta^{(0)}-2 s^{(0)}_{\vec l} s^{(\varepsilon)}_{\vec l} \eta^{(0)} \Big) \phi_{\vec l}(\vec x) dx_1 \cdots dx_D=0
\end{align*}
Following straightforward manipulations, we got:
\begin{align*}
	(b_0 + 2 s^{(0)}_{\vec l})s^{(\varepsilon)}_{\vec l} &\int_{0}^{1} \cdots \int_0^{1} (\phi_{\vec l}(\vec x))^2 dx_1 \cdots dx_D=\\
 &-\frac{(2l_1-1)\pi}{4 (n_1-1)} \int_{0}^{1} \cdots \int_0^{1}\tilde{k}_1^{m}(\vec{x})\sin \big((2l_1-1)\pi x_1 \big ) \cos ^2(l_2\pi x_2) \cdots \cos ^2(l_D\pi x_D) \ dx_1 \cdots dx_D \notag\\ & -\frac{l_2 \pi}{2 (n_2-1)} \int_{0}^{1} \cdots \int_0^{1}\tilde{k}_2^{m}(\vec{x}) \cos^2 \big(\frac{(2l_1-1)\pi x_1}{2} \big ) \sin(2l_2\pi x_2) \cdots \cos ^2(l_D\pi x_D)\ dx_1\cdots dx_D\\ &- \dots\\
 & -\frac{l_D \pi}{2 (n_D-1)} \int_{0}^{1} \cdots \int_0^{1}\tilde{k}_D^{m}(\vec{x}) \cos^2 \big(\frac{(2l_1-1)\pi x_1}{2} \big ) \cos ^2(l_2\pi x_2) \cdots \sin(2l_D\pi x_D)\ dx_1\cdots dx_D\\
&+\int_{0}^{1} \cdots \int_0^{1} \Big( \frac{\tilde{k}_1^{s}(\vec{x})}{2(n_1-1)^2} \frac{\partial^2 \eta^{(0)}}{\partial x_1^2}+ \dots+\frac{\tilde{k}_D^{s}(\vec{x})}{2(n_D-1)^2} \frac{\partial^2 \eta^{(0)}}{\partial x_D^2} \Big )  \phi_{\vec l}(\vec{x}) \ dx_1\cdots dx_D.
\end{align*}
When $n_1,\dots, n_D$ are very large, $b_0+2s^{(0)}_{\vec l} \approx b_0$. Using this, and substituting the equation above into $s_{\vec{l}}=s^{(0)}_{\vec{l}}+\varepsilon s^{(\varepsilon)}_{\vec{l}} + O(\varepsilon^2)$, we get  the following:
\begin{align}\label{eq:eig_pert}
s_{\vec{l}} & = s_{\vec{l}}^{(0)}  \notag\\  -& \frac{\varepsilon (2l_1-1)\pi}{4 b_0 (n_1-1) M} \int_{0}^{1} \dots \int_0^{1}\tilde{k}_1^{m}(\vec{x})\sin \big((2l_1-1)\pi x_1 \big ) \cos ^2(l_2\pi x_2)\cdots \cos ^2(l_D\pi x_D) \ dx_1 dx_2 \cdots dx_D \notag\\ -& \frac{\varepsilon  l_2\pi}{2 b_0 (n_2-1) M} \int_{0}^{1} \dots \int_0^{1}\tilde{k}_2^{m} (\vec{x})\cos^2 \big(\frac{(2l_1-1)\pi x_1}{2} \big ) \sin(2l_2\pi x_2)\cos ^2(l_3\pi x_3) \cdots \cos ^2(l_D\pi x_D) \ dx_1 dx_2 \cdots dx_D \notag \\-&\dots \notag\\-& \frac{\varepsilon  l_D\pi}{2 b_0 (n_D-1) M} \int_{0}^{1} \dots \int_0^{1}\tilde{k}_D^{m} (\vec{x})\cos^2 \big(\frac{(2l_1-1)\pi x_1}{2} \big ) \cdots \cos ^2(l_{(D-1)}\pi x_{(D-1)}) \sin(2l_D\pi x_D) \ dx_1 dx_2 \cdots dx_D \notag\\ +&\frac{\varepsilon}{b_0 M}\int_{0}^{1} \cdots \int_0^{1} \Big( \frac{\tilde{k}_1^{s}(\vec{x})}{2(n_1-1)^2} \frac{\partial^2 \eta^{(0)}}{\partial x_1^2}+ \dots+\frac{\tilde{k}_D^{s}(\vec{x})}{2(n_D-1)^2} \frac{\partial^2 \eta^{(0)}}{\partial x_D^2} \Big )   \phi_{\vec l}(\vec{x}) \ dx_1\cdots dx_D + O(\varepsilon^2),
\end{align}
where $M \eqdef \int_{0}^{1} \cdots \int_0^{1} (\phi_{\vec l}(\vec x))^2 dx_1 \cdots dx_D=\int_{0}^{1} \cdots \int_0^{1} \cos^2 \big( \frac{(2l_1-1)\pi x_1}{2} \big ) \cos^2(l_2\pi x_2)\cdots \cos^2(l_D\pi x_D) dx_1 \cdots dx_D$.
Without mistuning, the least stable eigenvalue is given by $s^{(0)}_{(1,0,0,\dots)}$ with an associated eigenfunction $\phi_{(1,0,0,\dots)} (\vec{x}) = \cos(\frac{\pi}{2} x_1)$, which is almost everywhere positive in $[0,\;1]^D$. As a consequence of the Sturm-Liouville theory for the elliptic boundary value problems, the possibility of ``eigenvalue cross-over'' is precluded. That is, some other eigenvalue from becoming the least stable eigenvalue in the presence of mistuning is ruled out.  The standard argument relies on the positivity of the eigenfunction corresponding to $s_{(1,0,0,\dots)}^{(0)}$; the reader is referred to~\cite{Evans:98} for the details. Thus, for vanishingly small $\varepsilon$, the least stable eigenvalue is $s_{(1,0,0,\dots)}$, even in the presence of mistuning. Setting $l_1=1$ and $l_d=0$ for $d>1$ in~\eqref{eq:eig_pert}, we obtain the result. \frQED
\end{proof-theorem}

\end{document}